\documentclass[dvipsnames]{aa}  
\usepackage{graphicx}
\usepackage{txfonts}
\usepackage{natbib}
\usepackage[colorlinks=true, linktocpage, linkcolor={blue!60!black}, citecolor={blue!60!black}, urlcolor={blue!60!black}]{hyperref}
\usepackage[dvipsnames,svgnames,x11names]{xcolor}

\usepackage{siunitx}
\usepackage{bm}

\makeatletter
\newcommand{\subalign}[1]{%
  \vcenter{%
    \Let@ \restore@math@cr \default@tag
    \baselineskip\fontdimen10 \scriptfont\tw@
    \advance\baselineskip\fontdimen12 \scriptfont\tw@
    \lineskip\thr@@\fontdimen8 \scriptfont\thr@@
    \lineskiplimit\lineskip
    \ialign{\hfil$\m@th\scriptstyle##$&$\m@th\scriptstyle{}##$\hfil\crcr
      #1\crcr
    }%
  }%
}
\makeatother

\begin{document} 


\title{The Massive and Distant Clusters of \emph{WISE} Survey}
\subtitle{SZ effect of Verification with the Atacama Compact Array -- Localization and Cluster Analysis}
\titlerunning{MaDCoWS SZ-VACA LoCA}

\author{
    Luca Di Mascolo\inst{1}\and 
    Tony Mroczkowski\inst{2}\and 
    Eugene Churazov\inst{1,3}\and
    Emily Moravec\inst{4,5}\and
    Mark Brodwin\inst{6}\and \\
    Anthony Gonzalez\inst{5}\and
    Bandon B. Decker\inst{6}\and
    Peter R. M. Eisenhardt\inst{7}\and
    Spencer A. Stanford\inst{8}\and \\
    Daniel Stern\inst{7}\and
    Rashid Sunyaev\inst{1,3}\and
    Dominika Wylezalek\inst{2}
    }
\authorrunning{L. Di Mascolo et al.}

\institute{
    Max-Planck-Institut f\"{u}r Astrophysik (MPA), Karl-Schwarzschild-Strasse 1, Garching 85741, Germany\\\email{lucadim@mpa-garching.mpg.de}
    \and
    European Southern Observatory (ESO), Karl-Schwarzschild-Strasse 2, Garching 85748, Germany
    \and
    Space Research Institute, Profsoyuznaya 84/32, Moscow 117997, Russia
    \and
    Astronomical Institute of the Czech Academy of Sciences, Prague, B\v ocn\'i II 1401/1A, 14000 Praha 4, Czech Republic
    \and
    Department of Astronomy, University of Florida, 211 Bryant Space Science Center, Gainesville, FL 32611, USA
    \and
    Department of Physics and Astronomy, University of Missouri, 5110 Rockhill Road, Kansas City, MO 64110, USA
    \and
    Jet Propulsion Laboratory, California Institute of Technology, Pasadena, CA 91109, USA
    \and
    Department of Physics, University of California, Davis, One Shields Avenue, Davis, CA 95616, USA
    }

\date{Received 25 February 2020 / Accepted 15 April 2020}

\abstract
{The Massive and Distant Clusters of WISE Survey (MaDCoWS) provides a catalog of high-redshift ($0.7\lesssim z\lesssim 1.5$) infrared-selected galaxy clusters. However, the verification of the ionized intracluster medium, indicative of a collapsed and nearly virialized system, is made challenging by the high redshifts of the sample members.}
{The main goal of this work is to test the capabilities of the Atacama Compact  Array (ACA; also known as the Morita Array) Band 3 observations, centered at about 97.5~GHz, to provide robust validation of cluster detections via the thermal Sunyaev--Zeldovich (SZ) effect.}
{Using a pilot sample that comprises ten MaDCoWS galaxy clusters, accessible to ACA and representative of the median sample richness, we infer the masses of the selected galaxy clusters and respective detection significance by means of a Bayesian analysis of the interferometric data.}
{Our test of the \textit{Verification with the ACA -- Localization and Cluster Analysis} (VACA LoCA) program demonstrates that the ACA can robustly confirm the presence of the virialized intracluster medium in galaxy clusters previously identified in full-sky surveys. In particular, we obtain a significant detection of the SZ effect for seven out of the ten VACA LoCA clusters. We note that this result is independent of the assumed pressure profile. However, the limited angular dynamic range of the ACA in Band 3 alone, short observational integration times, and possible contamination from unresolved sources limit the detailed characterization of the cluster properties and the inference of the cluster masses within scales appropriate for the robust calibration of mass--richness scaling relations.}
{}


\maketitle

\section{Introduction}\label{sec:intro}
Galaxy cluster richness has long been demonstrated to provide an observationally inexpensive proxy for cluster mass \citep[see, e.g., ][]{Rykoff2012,Andreon2015,Saro2015,Geach2017,Rettura2018,Gonzalez2019}. Being practically independent of the specific dynamical state of galaxy clusters, properly calibrated mass--richness relations play a key role in obtaining mass estimates in lieu of data that could directly probe the mass distribution of a cluster. For cluster candidates discovered through optical and infrared selection criteria such as richness, it is  essential to verify that the observed galaxy overdensities cannot be ascribed to spurious effects (e.g., line-of-sight projection of galaxies belonging to different haloes). Central to this aim is  confirming the presence of a hot X-ray emitting intracluster medium (ICM) heated by gravitational infall and nearly in virial equilibrium. X-ray confirmation, which has been the traditional tool for probing the ICM, becomes exceedingly difficult and observationally challenging at high redshift due to cosmological dimming. We note, however, that at $z\gtrsim1$ the dimming is expected to weaken due to evolution in the X-ray luminosity for a given mass \citep{Churazov2015}. 

The thermal Sunyaev--Zeldovich (SZ) effect \citep{Sunyaev1972} offers an alternative, redshift-independent way to confirm the presence of the ICM. Here we   provide a first test of the capabilities of the 7-meter Atacama Compact Array \citep[ACA;][]{Iguchi2009}, or Morita Array, in providing an SZ confirmation of cluster candidates identified in wide-field surveys. In particular, we consider a first pilot sample of the observational program, \textit{Verification with the ACA -- Localization and Cluster Analysis} (VACA LoCA), aimed at providing cluster verification and localization of the intracluster gas of galaxy clusters selected from the Massive and Distant Clusters of WISE \citep{Wright2010} Survey \citep[MaDCoWS;][]{Gonzalez2019}.

The paper is structured as follows. An overview of the observational details of the VACA LoCA cluster sample is provided in Sect.~\ref{sec:data:aca}. In Sect.~\ref{sec:analysis} we briefly discuss the modeling technique employed for inferring the cluster masses. The results of our analysis, comprising estimates of mass and detection significance for all the VACA LoCA clusters, are presented in Sect.~\ref{sec:res}. A summary of the work is then given in Sect.~\ref{sec:conclusion}.

All results discussed in this paper were derived in the framework of a spatially flat $\Lambda$CDM cosmological model, with $\Omega_{\mathrm{m}}=0.30$, $\Omega_{\Lambda}=0.70$, and $H_0=70.0~\mathrm{km\,s^{-1}\,Mpc^{-1}}$. In this cosmology, 1 arcsecond corresponds to $8.01~\mathrm{kpc}$ at the average redshift $z\simeq1$ of the VACA LoCA clusters. The best-fit estimates and uncertainties of any of the model parameters correspond respectively to the $50{\mathrm{th}}$ percentile and 68\% credibility interval of the corresponding marginalized posterior probability distributions.

\section{ACA observations}\label{sec:data:aca}
\citet{Gonzalez2019} reports a preliminary, low-scatter mass--richness scaling relation for the MaDCoWS cluster sample based on the infrared richness estimates from observations with the Infrared Array Camera \citep[IRAC;][]{Fazio2004} on the \textit{Spitzer Space Telescope} and masses derived from the SZ signal measured by the Combined Array for Research in Millimeter-wave Astronomy\footnote{\href{http://www.mmarray.org}{http://www.mmarray.org}} (CARMA; see \citealt{Brodwin2015, Gonzalez2015, Decker2019}). In order to improve the calibration of the mass--richness correlation, the VACA LoCA observations were devised to target a sample of ten MaDCoWS galaxy clusters observable by ACA and representative of the median sample richness.

The ACA observations of the selected MaDCoWS clusters were carried out between May and October 2017 as part of ALMA Cycle 4 operations (project ID: 2016.2.00014.S, PI: M. Brodwin). In order to reach a target continuum sensitivity of around $80~\mathrm{\mu Jy}$, the integration time on source for each of the pointings amounts to an average of $2.6~\mathrm{hours}$.

The overall frequency band was tuned to cover the range $89.5-105.5~\mathrm{GHz}$, using four Band 3 spectral windows in continuum mode with centers  at approximately $90.5$, $92.5$, $102.5$, and $104.5~\mathrm{GHz}$. This provided a good trade-off between probing the SZ signal spectrum near its minimum (i.e., maximum amplitude of the negative spectral distortion) and probing the largest scales accessible by ACA Band 3 data. The resulting dynamic range of $uv$ (Fourier space) distances in the ACA observations of the MaDCoWS clusters span, on average, between $2.64$ and $17.23~\mathrm{k\lambda}$, corresponding respectively to angular scales from $1.30~\mathrm{arcmin}$ to $11.97~\mathrm{arcsec}$ (we refer to Table~\ref{tab:obsdata} for further observational details).

We perform the calibration of all the data in the Common Astronomy Software Application \citep[\texttt{CASA};][]{McMullin2007} package version 4.7.2 using the standard calibration pipelines provided at data delivery. A direct inspection of the reduced data sets did not highlight any significant issue with the calibration. We hence adopt the nominal value of $5\%$ for the fiducial uncertainty on the ACA absolute calibration\footnote{\href{https://almascience.nrao.edu/documents-and-tools/cycle4/alma-technical-handbook}{https://almascience.nrao.edu/documents-and-tools/cycle4/alma-technical-handbook}}.

All the interferometric images presented in this work are generated using the \texttt{tclean} task in \texttt{CASA} version 5.6.1. To better highlight the SZ features in the maps, we do not correct for the primary beam attenuation. The fields are cut off at the standard 0.2 gain level of the ACA antenna pattern. As our study is entirely performed on the raw interferometric data, we note that the ACA maps are included  for display purposes only. No deconvolution is performed to reduce the effects of sidelobes on the reconstructed ``dirty'' maps.

\begin{figure}
    \centering
    \includegraphics[trim=0.0cm 0.25cm 0.0cm 0.35cm,clip,width=\columnwidth]{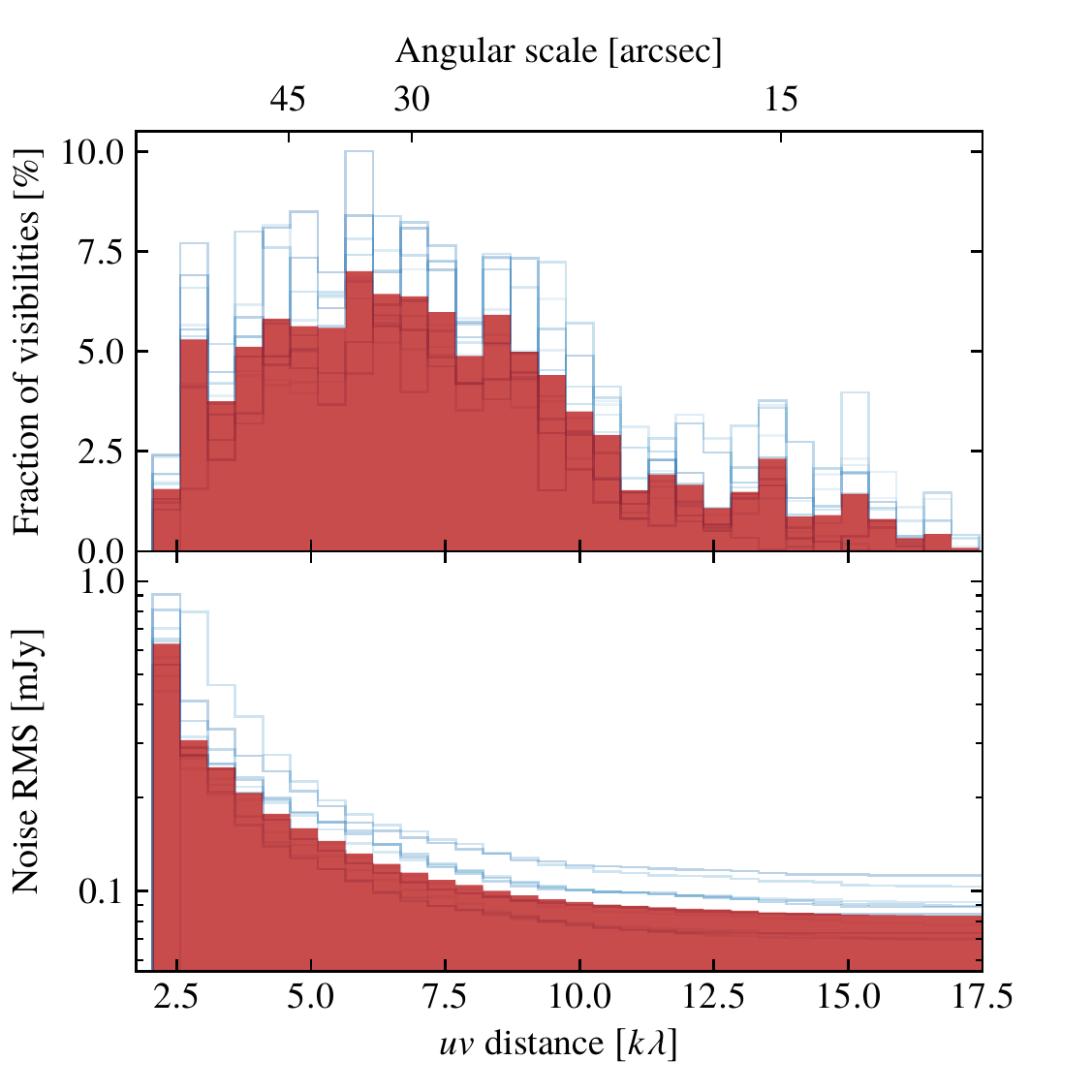}
    \caption{Fraction of visibility points for a given bin of $uv$ distances (top) and corresponding cumulative noise root mean square (RMS; bottom). The blue lines correspond to the individual fields, while the red shaded region to their average. The clear flattening of the cumulative noise curve for $uv$ distance larger than around $10~\mathrm{k\lambda}$ suggests the sensitivity budget is overall dominated by short baselines (i.e., large-scale modes).}
    \label{fig:uvdata}
\end{figure}

\begin{table*}
  \centering
  \caption{Summary of the observational properties of the VACA LoCA sample of MaDCoWS clusters. The reported noise RMS is the average noise level as measured from naturally weighted dirty images. The corresponding dirty beam is reported in the table as the nominal data resolution. The maximum recoverable scale (MRS) is instead derived from the minimum projected baseline in the full-bandwidth measurements.}
  \begin{tabular}{ccccccc}
    \hline\hline
    \noalign{\smallskip}
    Cluster ID & Obs. date & On-source time & Noise RMS &       $uv$ range      & Resolution &   MRS   \\\noalign{\vspace{1pt}}
               &           &     (hours)    &   (mJy)   & ($\mathrm{k\lambda}$) &  (arcsec)  & (arcmin)\\
    \noalign{\smallskip}
    \hline
    \noalign{\smallskip}
    \object{MOO~J0129$-$1640} & 2017-08-27 & $2.78$ & $0.061$ & $1.70-17.20$ & $17.8\times 12.4$ & $2.02$\\\noalign{\vspace{1pt}}
    \object{MOO~J0345$-$2913} & 2017-09-03 & $3.15$ & $0.055$ & $1.91-17.03$ & $18.2\times 10.6$ & $1.80$\\\noalign{\vspace{1pt}}
    \object{MOO~J0903$+$1310} & 2017-07-27 & $3.05$ & $0.087$ & $1.75-14.69$ & $17.2\times 11.4$ & $1.96$\\\noalign{\vspace{1pt}}
    \object{MOO~J0917$-$0700} & 2017-05-09 & $2.08$ & $0.069$ & $2.13-16.51$ & $19.4\times  9.5$ & $1.61$\\\noalign{\vspace{1pt}}
    \object{MOO~J1139$-$1706} & 2017-09-07 & $2.14$ & $0.081$ & $2.52-17.22$ & $19.9\times  9.6$ & $1.36$\\\noalign{\vspace{1pt}}
    \object{MOO~J1223$+$2420} & 2017-05-08 & $3.07$ & $0.061$ & $2.09-13.67$ & $16.7\times 13.2$ & $1.64$\\\noalign{\vspace{1pt}}
    \object{MOO~J1342$-$1913} & 2017-08-18 & $2.46$ & $0.071$ & $1.74-17.19$ & $17.9\times  9.9$ & $1.98$\\\noalign{\vspace{1pt}}
    \object{MOO~J1414$+$0227} & 2017-09-03 & $3.09$ & $0.066$ & $1.73-16.30$ & $17.0\times  9.6$ & $1.99$\\\noalign{\vspace{1pt}}
    \object{MOO~J2146$-$0320} & 2017-08-03 & $1.37$ & $0.101$ & $1.87-16.41$ & $19.3\times 10.9$ & $1.84$\\\noalign{\vspace{1pt}}
    \object{MOO~J2147$+$1314} & 2017-08-19 & $2.61$ & $0.062$ & $1.54-14.98$ & $18.5\times  9.6$ & $2.23$\\
    \noalign{\smallskip}
    \hline
    \noalign{\medskip}
  \end{tabular}
  \label{tab:obsdata}
\end{table*}

\section{Analysis technique}\label{sec:analysis}
\begin{figure}
    \centering
    \includegraphics[clip,trim=0.45cm 0 0.45cm 0,width=\columnwidth]{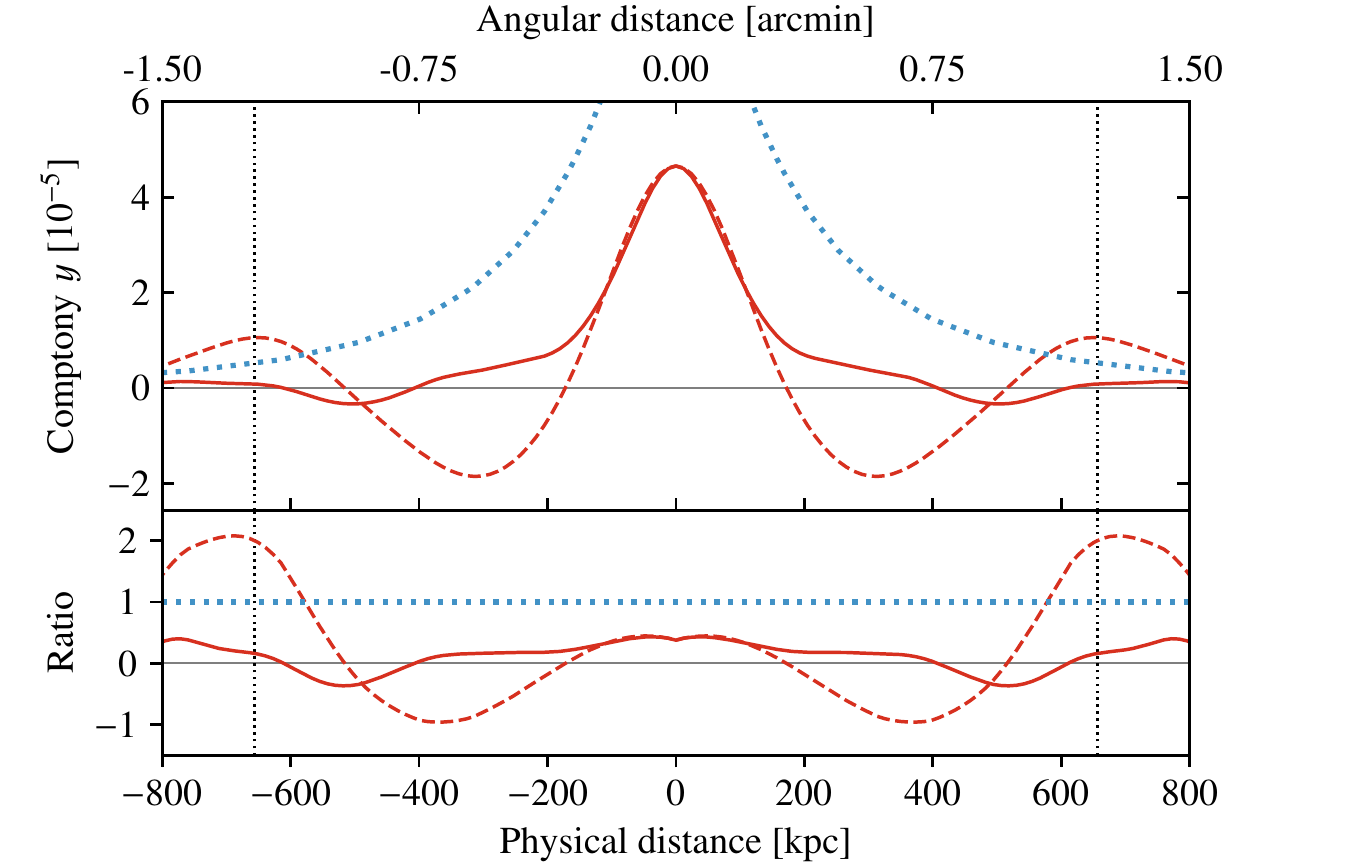}
    \caption{Simulated SZ profile for a cluster with mass of $2.5\cdot10^{14}~\mathrm{M_{\odot}}$ and redshift $z=1.00$, analogous to the  MaDCoWS targets previously reported in \citet{Gonzalez2019}. The top panel shows a comparison of the input SZ model (i.e., the true profile; dotted blue line), and the corresponding profiles after application of the interferometric transfer function (i.e., the filtered, observed profiles; red lines).  These clearly show how the fraction of missing flux is significant already well within the $r_{500}$ of the simulated cluster (see Sect.~\ref{sec:analysis:sz} for a definition; vertical lines). The two filtered profiles are measured along directions at constant right ascension or declination (respectively dashed and solid lines). Their difference reflects the asymmetry in the $uv$ coverage. The lower panel reports the ratio of the filtered and raw profiles. The line style is the same as the upper panel, corresponding to the ratio of the filtered (observed) profiles to the unfiltered (true) profile. The blue dotted line indicates unity (i.e., no filtering).}
    \label{fig:filtered}
\end{figure}

Spatial filtering due to the incomplete sampling of the Fourier modes of the observed sky may represent a severe challenge in the analysis of radio-interferometric measurements of galaxy clusters. The issue is in fact twofold: first, the sparse coverage of the Fourier plane results in poor constraints for some angular scales within the range probed by the interferometer; second, the shortest baseline achievable is essentially determined by the shadowing limit, when one antenna is in front of another as seen from the source. This sets a hard upper limit on the maximum recoverable scale (MRS) of the observation. This high-pass filtering effect is   evident in the case of astrophysical objects covering large angular scales such as galaxy clusters, whose SZ signal often extends well beyond the field of view of current millimeter and submillimeter facilities that provide subarcminute resolution (see, e.g., \citealt{Basu2016} and \citealt{DiMascolo2019b}, or \citealt{Mroczkowski2019} for a broader review). To provide a sense of the net effects of ACA filtering on the SZ signal from a galaxy cluster, in Fig.~\ref{fig:filtered} we compare the model and filtered SZ profiles for a cluster with a mass of $2.5\cdot10^{14}~\mathrm{M_{\odot}}$ at a redshift $z=1.00$.

While the missing flux issue is commonly solved by means of deconvolution techniques (at the expense of introducing correlation in the image-space noise), it is possible to include short spacings only by complementing the interferometric data with external information on larger scales (e.g., from single-dish telescopes). However, these large-scale observations should have sensitivities comparable to the corresponding interferometric measurements, a condition that is difficult to realize in the case of SZ data. 

In order to circumvent any of the above challenges, we perform a Bayesian forward-modeling analysis directly on the visibilities of the ACA MaDCoWS sample. This allows us to infer the cluster masses from the raw interferometric data, accounting for the exact sampling function of the visibility plane, and  providing a strong leverage on possible contamination from unresolved (point-like) sources. 

An extensive discussion of  the modeling methodology and the implementation is provided in \citet{DiMascolo2019a,DiMascolo2019b}.

\subsection{Estimating cluster masses}\label{sec:analysis:sz}
Hydrodynamic simulations \citep{Nagai2007} have shown that the pressure distribution of the electrons within the ICM can be reasonably described as
\begin{equation}
  P_{\mathrm{e}}(\xi) = P_{500}\times p(\xi),
  \label{eq:a10}
\end{equation}
where the scaled pressure profile $p(\xi)$ is defined by a generalized Navarro-Frenk-White (gNFW) profile,
\begin{equation}
  p(\xi) = P_{0}~\xi^{-c}\left[\left(1+\xi^a\right)\right]^{(c-b)/a}.
  \label{eq:gnfw}
\end{equation}
Here, $P_{0}$ acts as a simple normalization factor. The parameters $a$, $b$, and $c$ are respectively the radial slopes at intermediate, large, and small scales with respect to a scale radius $r_{s}$, while $\xi=r/r_{\mathrm{s}}$ is the radial distance $r$ from the pressure centroid in units of $r_{s}$.

Following \citet{Arnaud2010}, the scaling parameter $P_{500}$ is defined as
\begin{equation}
  P_{500}(M_{500},z) = 1.65\cdot10^{-3} \, E(z)^{8/3} \, \left[\frac{M_{500}}{3\cdot10^{14}M_{\odot}}\right]^{2/3+a_{\textsc{p}}(\xi)}\mathrm{keV~cm^{-3}}
  \label{eq:p500}
,\end{equation}
where $E(z)$ is the ratio of the Hubble constant at redshift $z$ to its present value $H_{0}$, and $M_{500}$ is the mass enclosed within the radius $r_{500}$ at which the average cluster density is $500\times$ the critical density $\rho_{\mathrm{c}}(z)$ of the Universe at the redshift of the cluster. Under the assumption of spherical symmetry, the radius $r_{500}$ can be easily expressed as a function of a given mass $M_{500}$ and redshift $z$ as
\begin{equation}
  r_{500}(M_{500},z) = \left[\frac{3}{4\pi}\frac{M_{500}}{500\rho_{\mathrm{c}}(z)}\right]^{1/3}.
  \label{eq:r500}
\end{equation}
This can then be related to the scale radius $r_{\mathrm{s}}$ of the normalized gNFW profile in Eq.~\eqref{eq:gnfw} by adding a concentration parameter $c_{500}$ as $r_{\mathrm{s}} = r_{500}/c_{500}$. Finally, the running slope $a_{\textsc{p}}(\xi)$ is introduced to account for any departure from self-similarity in the innermost regions of galaxy clusters, 
\begin{equation}
  a_{\textsc{p}}(\xi) = a_{0}/[1+8\xi^3]
  \label{eq:ap}
\end{equation}

In our analysis, the self-similarity deviation parameter $a_{0}$, the pressure normalization $P_{0}$, the concentration parameter $c_{500}$, and the gNFW slopes $a$, $b$, and $c$ are kept fixed. The parameters $a_{0}$, $P_{0}$, and $c_{500}$ are degenerate with the mass parameter $M_{500}$, and unconstrained fittings would significantly affect the recovery of the cluster masses. On the other hand, test fittings with free gNFW slopes have shown that all  three parameters $a$, $b$, and $c$ would remain entirely unconstrained, and would undergo strong degeneracies with any of the other gNFW parameters and mass $M_{500}$. This is a direct consequence of the limited dynamic range of angular scales probed by ACA which, in combination with the modest sensitivity of the VACA LoCA observations, limits the information available for reconstructing pressure profiles for the individual fields.

Hence, we set the above gNFW parameters alternatively to the best-fit values reported in \citet{Arnaud2010} for the universal pressure  profile, or for the  subsamples of cool-core and morphologically disturbed clusters. For a comparison, we additionally consider gNFW parameters derived in \citet{Planck2013V} from the joint fit of the \textit{XMM-Newton}- and \textit{Planck}-selected sample of galaxy clusters, as well as the high-redshift \textit{Chandra} gNFW model by \citet[][]{McDonald2014}. The different set of parameters adopted in our analysis is summarized in Table~\ref{tab:gnfw}.

\begin{table}
  \centering
  \caption{Best-fit parameters of the gNFW pressure models from \citet{Arnaud2010}, \citet{Planck2013V}, and \citet[][referred to here as MD14]{McDonald2014}.}
  \begin{tabular}{cccccc}
    \hline\hline
    \noalign{\smallskip}
              & universal & cool core & disturbed & \textit{Planck} & MD14\\
    \noalign{\smallskip}
    \hline
    \noalign{\smallskip}
    $P_{0}$  \hspace{-5pt} &  $8.40$  &  $3.25$  &  $3.20$  & $6.41$ & $3.47\substack{+1.09\\-0.67}$ \\\noalign{\vspace{1pt}}
    $c_{500}$\hspace{-5pt} &  $1.18$  &  $1.13$  &  $1.08$  & $1.81$ & $2.59\substack{+0.37\\-0.38}$ \\\noalign{\vspace{1pt}}
    $a$      \hspace{-5pt} &  $1.05$  &  $1.22$  &  $1.41$  & $1.33$ & $2.27\substack{+0.89\\-0.40}$ \\\noalign{\vspace{1pt}}
    $b$      \hspace{-5pt} &  $5.49$  &  $5.49$  &  $5.49$  & $4.13$ & $3.48\substack{+0.60\\-0.39}$ \\\noalign{\vspace{1pt}}
    $c$      \hspace{-5pt} &  $0.31$  &  $0.77$  &  $0.38$  & $0.31$ & $0.15\substack{+0.13\\-0.15}$ \\\noalign{\vspace{1pt}}
    $a_{0}$  \hspace{-5pt} &  $0.22$  &  $0.22$  &  $0.22$  & $0.00$ & $0.22$ \\
    \noalign{\smallskip}
    \hline
    \noalign{\medskip}
  \end{tabular}
  \label{tab:gnfw}
\end{table}

At each iteration of the posterior sampling, we then compute the expected thermal SZ signal by integrating the pressure model defined in Eq.~\eqref{eq:a10} for a given value of $M_{500}$ and redshift $z$. The resulting variation in the CMB surface brightness in a direction $\bm{x}$ on the plane of the sky and at a frequency $\nu$ is \citep{Sunyaev1972}
\begin{equation}
  \delta i_{\mathrm{t\textsc{sz}}}(\bm{x},\nu) \propto g_{\mathrm{t\textsc{sz}}}(\nu) \, {\textstyle\int} P_{\mathrm{e}}(\bm{x},\ell)\,\mathrm{d}\ell.
  \label{eq:tsz}
\end{equation}
The integral along the line-of-sight coordinate $\ell$ is computed from $0$ up to the fiducial value of $5 r_{500}$ \citep{Arnaud2010}. The factor $g_{\mathrm{t\textsc{sz}}}(\nu,T_{\mathrm{e}})$ represents the frequency scaling of the non-relativistic thermal SZ effect \citep{Sunyaev1972}. For simplicity, we neglect any temperature-dependent corrections arising from the fully relativistic treatment of the thermal SZ effect. The ACA observations cover a frequency band that is not   broad enough, or deep enough, to constrain any relativistic contribution to the SZ spectrum, and hence to get direct constraints on the average temperature of the electron populations within the observed clusters \citep{Challinor1998,Itoh1998,Sazonov1998}. Further, the correction to the non-relativistic thermal signal for the ACA MaDCoWS clusters is expected\footnote{The average relativistic correction reported in the text is computed employing the formulation by \citet{Itoh2004}. The average electron temperature is inferred from the core-excised temperature-redshift-mass scaling relation in \citet{Bulbul2019}. To keep a conservative upper limit, we consider an extreme case of a galaxy cluster with the same mass as the most massive object identified in the MaDCoWS survey \citep{Ruppin2020} at a redshift equal to the highest value in the ACA sample.} to be on average less than $\sim5\%$. Although this will bias the reconstructed masses systematically to lower values, the effect will be at most on the same order as the flux uncertainties discussed in Sect.~\ref{sec:data:aca}, and well within the modeling statistical uncertainties (see Sect.~\ref{sec:res} below).

Similarly, the ACA frequency coverage is not wide enough to   retrieve any information about the bulk velocities of the observed clusters (or parts of them). Therefore, we assume any contributions from a possible kinetic SZ component \citep{Sunyaev1980} to be subdominant with respect to the thermal effect, and we neglect it in our analysis.

\subsection{Unresolved sources}\label{sec:analysis:pts}
Contamination from point-like radio sources may limit and significantly affect the reconstruction of a cluster model from SZ observations \citep{Gobat2019,Mroczkowski2019}. In order to assess the level at which the unresolved flux might have contributed to the estimates of the masses of VACA LoCA clusters, we perform blind searches of point-like components over the entire fields of view of the ACA observations and simultaneously with the SZ analysis. We assume the unresolved components to be described by a Dirac-$\delta$ model with flat spectrum over the entire ACA band. The long-baseline data range, most sensitive to the signal from compact sources, is the least densely parsed region of the visibility plane (see Fig.~\ref{fig:uvdata}). This results in high noise on the smaller angular scales, hence limiting the possibility of constraining the spectral properties of the unresolved sources in the observed fields. The point-like model thus simplifies to \citep{DiMascolo2019a}
\begin{equation}
    V(\bm{u},\nu) = i_{\textsc{ps}}e^{2\pi j\bm{u}\cdot\bm{x}_{\textsc{ps}}},
  \label{eq:pts}
\end{equation}
given a set of interferometric data with visibility coordinates $\bm{u}$. The position $\bm{x}_{\textsc{ps}}$ and the source flux $i_{\textsc{ps}}$ are left free to vary. 

Due to the limited information  provided by the ACA data about the population of unresolved sources in the VACA LoCA fields, here we   consider them as nuisance model components and generally marginalize over them. A future analysis with higher resolution, multi-frequency observations will be key for their proper characterization.

\subsection{Parameter priors}\label{sec:analysis:prior}
The comparison of cluster positions identified through the MaDCoWS search and the galaxy distribution centroids measured by \textit{Spitzer} are found to deviate by $\sigma_{\mathrm{RA}}=14.3~\mathrm{arcsec}$ in right ascension and $\sigma_{\mathrm{Dec}}=15~\mathrm{arcsec}$ in declination \citep{Gonzalez2019}. We thus assume normal priors with standard deviations of $\sigma_{\mathrm{ra}}$ and $\sigma_{\mathrm{Dec}}$ on the right ascension and declination coordinates of the cluster centroids, respectively.

The mass parameter $M_{500}$ and the redshift $z$ are heavily degenerate as they both enter in the determination of the pressure model through the pressure normalization $P_{500}$ and the scale radius $r_{500}$. In order to alleviate the degeneracy, we adopt split-normal priors \citep{Wallis2014} on the redshifts $z$ based on the photometric constraints on the cluster members from \textit{Spitzer} \citep{Gonzalez2019}. When fitting the gNFW profile from \citet{McDonald2014}, the gNFW parameters were also assigned split-normal priors, with standard deviations given by the respective parameter uncertainties in Table~\ref{tab:gnfw}.

To account for the ACA flux uncertainties in the recovered masses (Sect.~\ref{sec:data:aca}), we introduce a normalization hyperparameter, as detailed in \citet{DiMascolo2019a}. In particular, we consider a scaling parameter characterized by a normal prior distribution with unitary mean value and standard deviation equal to the inherent calibration uncertainty.

Finally, we assume wide uninformative priors on all the point source parameters apart from the position. For the blind search, this is bound to vary uniformly within the region defined by the first null of the ACA primary beam. 
Data-free runs for each of the analyzed data sets \citep{DiMascolo2019a,DiMascolo2019b} have shown no biases in the parameter inference related to choice in the prior distributions.

\section{Results and discussion}\label{sec:res}
\begin{table*}
    \centering
    \caption{Inferred quantities for the VACA LoCA sample clusters under the assumption of a universal pressure profile \citep{Arnaud2010}. See Sect.~\ref{sec:res} for more details about the effective significance estimate $\sigma_{\mathrm{eff}}$. The photometric redshift $z_{\mathrm{phot}}$ and infrared richness $\lambda$ are taken from \citet{Gonzalez2019}.}
    \begin{tabular}{ccccccccc}
        \hline\hline
        \noalign{\smallskip}
        Cluster ID & $z_{\mathrm{phot}}$ & $\lambda$ & $r_{500}$ &  $\theta_{500}$ & $Y_{\mathrm{sph}}(<r_{500})$ & $Y_{\mathrm{cyl}}(<r_{500})$ &       $M_{500}$       & $\sigma_{\mathrm{eff}}$\\
                   &          --         &       --       &   (Mpc)   &    (arcmin)     &   $10^{-5}~\mathrm{Mpc^2}$  &   $10^{-5}~\mathrm{Mpc^2}$  & $(10^{14}~M_{\odot})$ &           --           \\
        \noalign{\smallskip}
        \hline
        \noalign{\smallskip}
        \multicolumn{5}{l}{\textit{Significant detection}}\\\noalign{\vspace{1pt}}
        \object{MOO~J0129$-$1640}\hspace{5pt} & $1.05\substack{+0.04\\-0.05}$ & $49\pm7$ & $0.67\substack{+0.09\\-0.08}$ & $1.38\substack{+0.03\\-0.21}$ & $2.13\substack{+0.01\\-0.66}$ & $2.34\substack{+0.01\\-0.72}$ & $2.57\substack{+0.30\\-0.30}$ & $7.77$ \\\noalign{\vspace{1pt}}
        \object{MOO~J0345$-$2913}\hspace{5pt} & $1.08\substack{+0.03\\-0.04}$ & $53\pm7$ & $0.57\substack{+0.02\\-0.03}$ & $1.17\substack{+0.05\\-0.07}$ & $1.04\substack{+0.25\\-0.26}$ & $1.15\substack{+0.27\\-0.29}$ & $1.78\substack{+0.20\\-0.29}$ & $5.32$ \\\noalign{\vspace{1pt}}
        \object{MOO~J0917$-$0700}\tablefootmark{a} & $1.10\substack{+0.05\\-0.05}$ & $58\pm7$ & $0.55\substack{+0.04\\-0.05}$ & $1.11\substack{+0.08\\-0.08}$ & $0.97\substack{+0.40\\-0.33}$ & $1.07\substack{+0.44\\-0.37}$ & $1.66\substack{+0.31\\-0.38}$ & $4.26$ \\\noalign{\vspace{1pt}}
                         &                               &          & $0.58\substack{+0.04\\-0.04}$ & $1.19\substack{+0.08\\-0.09}$ & $1.48\substack{+0.54\\-0.49}$ & $1.63\substack{+0.59\\-0.53}$ & $2.13\substack{+0.40\\-0.49}$ &        \\\noalign{\vspace{1pt}}
        \object{MOO~J1139$-$1706}\hspace{5pt} & $1.31\substack{+0.03\\-0.05}$ & $53\pm7$ & $0.56\substack{+0.02\\-0.03}$ & $1.10\substack{+0.05\\-0.05}$ & $1.51\substack{+0.38\\-0.25}$ & $1.66\substack{+0.41\\-0.27}$ & $2.24\substack{+0.36\\-0.52}$ & $3.81$ \\\noalign{\vspace{1pt}}
        \object{MOO~J1342$-$1913}\hspace{5pt} & $1.08\substack{+0.04\\-0.05}$ & $41\pm6$ & $0.59\substack{+0.03\\-0.03}$ & $1.20\substack{+0.08\\-0.07}$ & $1.22\substack{+0.39\\-0.29}$ & $1.34\substack{+0.42\\-0.32}$ & $1.95\substack{+0.31\\-0.31}$ & $4.53$ \\\noalign{\vspace{1pt}}
        \object{MOO~J1414$+$0227}\hspace{5pt} & $1.02\substack{+0.07\\-0.06}$ & $41\pm7$ & $0.67\substack{+0.04\\-0.04}$ & $1.40\substack{+0.09\\-0.08}$ & $2.29\substack{+0.50\\-0.43}$ & $2.52\substack{+0.55\\-0.48}$ & $2.75\substack{+0.32\\-0.32}$ & $6.99$ \\\noalign{\vspace{1pt}}
        \object{MOO~J2146$-$0320}\tablefootmark{a} & $1.16\substack{+0.05\\-0.05}$ & $50\pm7$ & $0.57\substack{+0.05\\-0.05}$ & $0.94\substack{+0.11\\-0.05}$ & $1.19\substack{+0.45\\-0.40}$ & $1.32\substack{+0.49\\-0.44}$ & $1.86\substack{+0.34\\-0.52}$ & $5.35$ \\\noalign{\vspace{1pt}}
                         &                               &          & $0.55\substack{+0.05\\-0.05}$ & $1.09\substack{+0.06\\-0.05}$ & $1.43\substack{+0.64\\-0.56}$ & $1.58\substack{+0.71\\-0.62}$ & $2.04\substack{+0.42\\-0.56}$ &        \\\noalign{\medskip}

        \multicolumn{5}{l}{\textit{Non-detection}}\\\noalign{\vspace{1pt}}
        \object{MOO~J0903$+$1310}\hspace{5pt} & $1.26\substack{+0.05\\-0.08}$ & $29\pm5$ & $0.30^{+0.05}$ & $0.59^{+0.10}$ & $0.04^{+0.06}$ & $0.05^{+0.07}$ & $0.30^{+0.18}$ & -- \\\noalign{\vspace{1pt}}
        \object{MOO~J1223$+$2420}\tablefootmark{b} & $1.09\substack{+0.04\\-0.04}$ & $51\pm7$ & $0.49\substack{+0.04\\-0.07}$ & $0.99\substack{+0.09\\-0.14}$ & $0.49\substack{+0.18\\-0.29}$ & $0.54\substack{+0.19\\-0.32}$ & $1.17\substack{+0.27\\-0.38}$ & $2.40$ \\\noalign{\vspace{1pt}}
        \object{MOO~J2147$+$1314}\hspace{5pt} & $1.01\substack{+0.06\\-0.07}$ & $38\pm6$ & $0.59\substack{+0.05\\-0.04}$ & $1.22\substack{+0.08\\-0.12}$ & $1.01\substack{+0.40\\-0.43}$ & $1.11\substack{+0.44\\-0.48}$ & $1.82\substack{+0.34\\-0.42}$ & $1.26$ \\\noalign{\vspace{1pt}} 
        \noalign{\smallskip}
        \hline
        \noalign{\medskip}
    \end{tabular}
    \tablefoot{
        \tablefoottext{a}{The two mass values provided for \object{MOO~J0917$-$0700} and \object{MOO~J2146$-$0320} correspond to the masses of each of the individual SZ components detected.} 
        \tablefoottext{b}{The SZ signal from \object{MOO~J1223$+$2420} has significant contamination from an FR II radio galaxy at the center of the cluster (see Sect.~\ref{sec:res:pts} for a discussion).}
    }
    \label{tab:masses}
\end{table*}

A summary of the masses of the VACA LoCA pilot sample is presented in Table~\ref{tab:masses}. The results presented in previous MaDCoWS papers \citep{Brodwin2015,Decker2019,Gonzalez2019} were derived adopting the universal profile by \citet{Arnaud2010} to describe the electron pressure distribution. For consistency, we  report here only the masses estimated under the same assumption. A discussion of the impact of model choice on the inferred masses is  presented in Sect.~\ref{sec:res:mass}.

We quantify the detection significance of the SZ signal in the VACA LoCA observations by comparing the log-evidence of the full modeling runs $\mathcal{Z}_1$ with those considering only the point source model component $\mathcal{Z}_0$ by means of the Jeffreys scale\footnote{As a reference, we consider a cluster to be significantly detected if the corresponding model has a Bayes factor $\mathcal{Z}_1/\mathcal{Z}_0$ higher than 100.} \citep{Jeffreys1961}. To get a more immediate handle on the significance of each detection, we report in Table~\ref{tab:masses} the number of effective standard deviations $\sigma_{\mathrm{eff}}$ between the model with and without an SZ component. This can be computed as $\sigma_{\mathrm{eff}}\simeq\sqrt{2\Delta\log{\mathcal{Z}}}$ given a log-Bayes factor $\Delta\log{\mathcal{Z}}=\log{(\mathcal{Z}_1/\mathcal{Z}_0)}$ \citep{Trotta2008}. This differs from the approach taken in the CARMA SZ follow-up papers \citep{Brodwin2015, Gonzalez2015, Decker2019} in that it is more statistically robust, as it properly accounts for the change between different models in the number of parameters and respective priors. We note that $\sigma_{\mathrm{eff}}$ is to be interpreted in a merely heuristic manner, as we are not accounting for the different degrees of freedom or prior volumes between the models. According to the Jeffreys criterion, introduced above, a value of $\sigma_{\mathrm{eff}}\gtrsim3$ is indicative of a robust detection.

Overall, we significantly detect the SZ effect toward seven out of the ten clusters of the VACA LoCA sample, while the presence of SZ signal is only weakly favored for \object{MOO~J1223$+$2420} and \object{MOO~J2147$+$1314}. Conservatively, we consider these two clusters as being undetected. Nevertheless, it is worth noting that, according to the Jeffreys criterion, the detection significance of \object{MOO~J1223$+$2420} provides moderately significant statistical support for the presence of SZ signal \citep{Jeffreys1961,Trotta2008}. For the single case of \object{MOO~J0903$+$1310}, the analysis favors the model without the SZ component, thus resulting in a clear non-detection. Reported in Table~\ref{tab:masses} is the estimated upper limit for the cluster mass.

\begin{figure}
    \centering
    \includegraphics[clip,trim=0 0.50cm 0 1.50cm,width=\columnwidth]{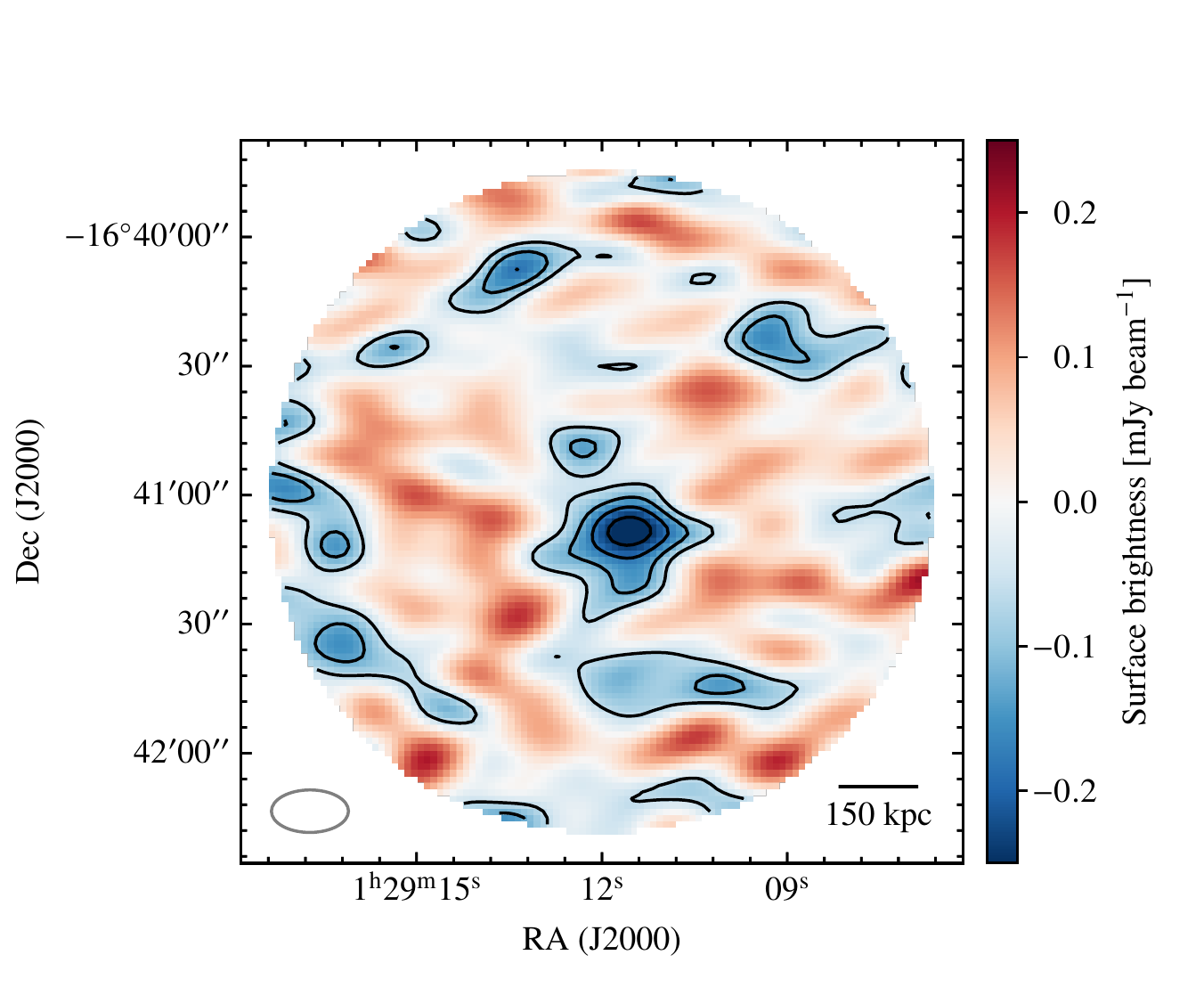}
    \caption{Dirty image of \object{MOO~J0129$-$1640} generated from point source-subtracted visibilities. The point source components are identified as the peaks in the joint posterior probability distribution point source position parameters. The contours correspond to the $1\sigma$, $2\sigma$,  $3\sigma$, and $4\sigma$ significance levels of the SZ signal, with $\sigma=0.061~\mathrm{mJy beam^{-1}}$. Although the integrated SZ decrement is detected at $\sigma_{\mathrm{eff}}=7.77$ (Table~\ref{tab:masses}), the peak SZ amplitude has a significance only slightly higher than $4\sigma$ when measured in image space.}
    \label{fig:imdirt:mooj0129}
\end{figure}

As for the analyses presented in \cite{Brodwin2015}, \cite{Gonzalez2015}, and \cite{Decker2019} of the CARMA data, the fitting of the ACA data is performed entirely in $uv$-space. Along with the advantages discussed in Sect.~\ref{sec:analysis}, this provides an approach to cluster detection in interferometric SZ data that avoids the drawbacks of image-space analysis, in particular the biased reconstructions produced by the \texttt{CLEAN} algorithm. For a comparison, we show in Fig.~\ref{fig:imdirt:mooj0129} the dirty image of the most significant detection in our sample, \object{MOO~J0129$-$1640}.
The peak of the SZ decrement has an amplitude of $-0.24~\mathrm{mJy~beam^{-1}}$, corresponding to a statistical significance of $4.06\sigma$, which is lower than the cluster detection $\sigma_{\mathrm{eff}}=7.77$ (Table~\ref{tab:masses}). This is not surprising, as $\sigma_{\mathrm{eff}}$ is a measure of the significance of the total SZ signal. However, in addition to the resolved SZ signal, the reason for this discrepancy also resides in the fact that the interferometric images are affected by heavily correlated noise. As a consequence, the resulting fluctuations may attenuate the measured signal and limit the confidence of its detection. On the other hand, side lobes further contaminate interferometric images. 
This is generally solved by applying \texttt{CLEAN}-like deconvolution techniques to the data \citep{Hoegbom1974,Thompson1986}. However, these techniques are specifically devised to reduce the effects of the incomplete sampling of the visibility plane on the overall quality of the reconstructed image, and would not provide any serious improvement in the significance of the observed SZ signal. It is worth noting that any deconvolved image would still provide a heavily high-pass filtered view of the very core of a galaxy cluster, as ACA does not measure the SZ signal on scales larger than the maximum recovered scale (Table~\ref{tab:obsdata}).

Another important remark is that the dirty map shown in Fig.~\ref{fig:imdirt:mooj0129} is generated only after subtraction from the visibility data of the most significant point-like sources detected by our modeling algorithm, allowing for a cleaner identification of the SZ signal in the cluster image. In fact, the presence of very bright compact sources may completely hide any SZ effect component, as either their signal would be superimposed on the one from the galaxy cluster or the side lobes would be blended with the SZ feature.

\subsection{Mass--richness relation}\label{sec:res:rich}
\begin{figure}
    \centering
    \includegraphics[clip,trim=0.4cm 0.35cm 0.4cm 1.35cm,width=\columnwidth]{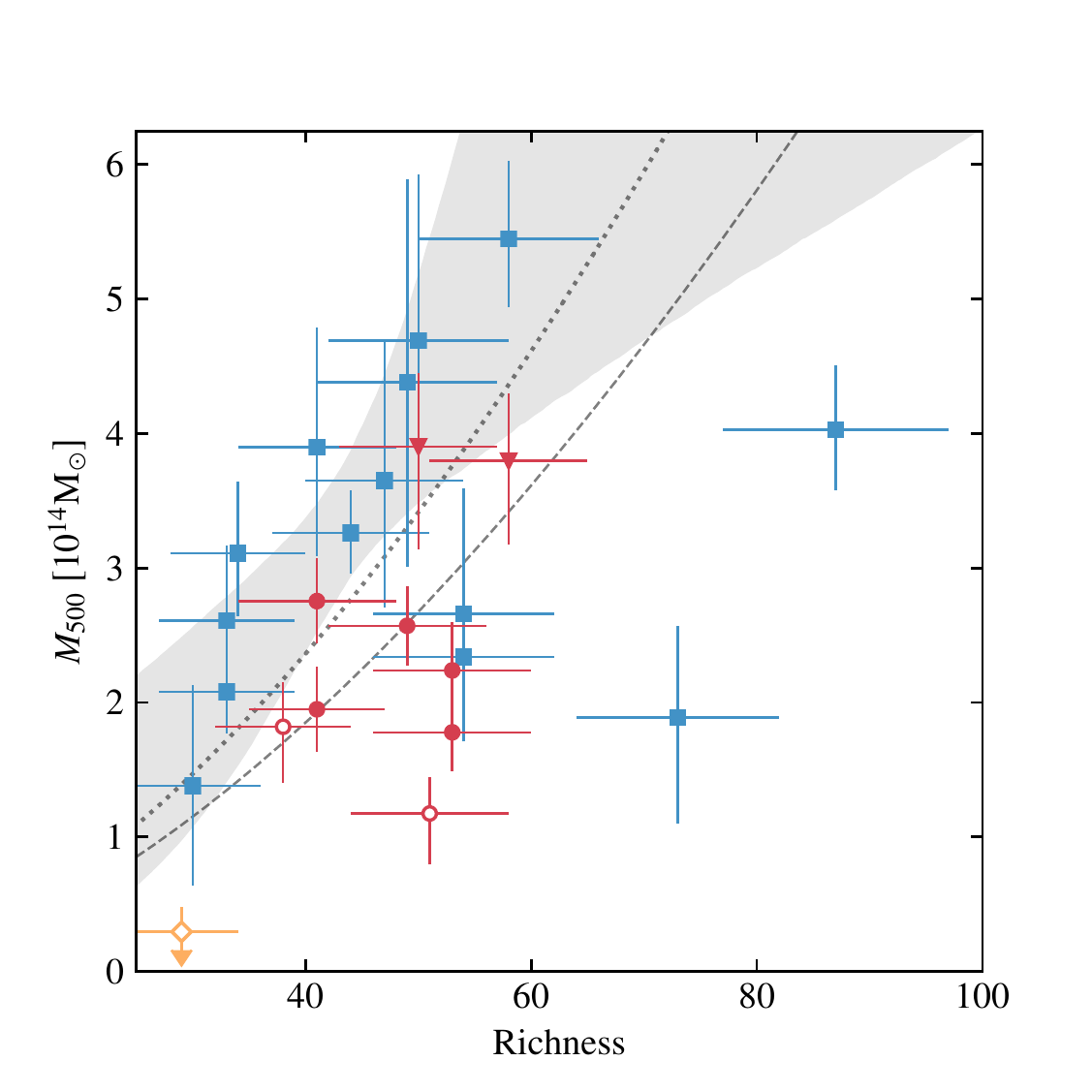}
    \caption{Mass vs. richness relation for all the MaDCoWS clusters with SZ-based mass estimates. The blue squares correspond to the CARMA MaDCoWS cluster sample from \citet{Gonzalez2019}. In solid red are the clusters from this work that have been significantly detected, while open red points denote the clusters \object{MOO~J1223$+$2420} and \object{MOO~J2147$+$1314}, with only weak statistical support for the presence of SZ signal. The upper limit for non-detected \object{MOO~J0903$+$1310} is denoted with a yellow open diamond.
    We use circles and triangles for the clusters with single or double SZ features, respectively. In the latter case, the plot reports the sum of the masses of the individual SZ components. The shaded region is the 68\% confidence interval for the mass-scaling relation reported in \citet{Gonzalez2019}. The VACA LoCA distribution is observed to lie below the mass--richness relation previously reported, highlighting potential systematics in the mass reconstruction from either or both the CARMA and ACA observations. The dashed and dotted lines correspond to the mass--richness scaling derived in Sect.~\ref{sec:res:rich} respectively from the VACA LoCA points only (excluding the non-detection) and from the joint modeling of the VACA LoCA and CARMA measurements.}
    \label{fig:massrich}
\end{figure}

Figure~\ref{fig:massrich} shows a comparison of the mass--richness scaling for the VACA LoCA sample and the CARMA measurements previously reported by \citet{Gonzalez2019}. Although there is good consistency between our estimates and the MaDCoWS mass--richness scaling, the VACA LoCA mass--richness distribution is systematically below the expected correlation. We quantify the average scaling by fitting the VACA LoCA data points with a linear function, with the slope constrained to that of the mass--richness scaling in \citet{Gonzalez2019} but with a free normalization parameter (which translates to an offset in the logarithmic relation). We find that the VACA LoCA cluster masses are downscaled by a factor of $0.56\substack{+0.13\\-0.05}$ with respect to the CARMA-derived mass--richness scaling when considering all the VACA LoCA clusters. The resulting scatter of the ACA masses with respect to the reconstructed relation is $\sigma\subalign{&\,\textsc{aca}\\&\log{M}|\lambda}=0.25\substack{+0.06\\-0.02}$, broader than the scatter observed in the CARMA measurements. However, if we exclude all the non-detections from the analysis, the scatter decreases to a value comparable with the CARMA measurement, $\sigma\subalign{&\,\textsc{aca}\\&\log{M}|\lambda}=0.14\substack{+0.02\\-0.01}$, while the relative normalization remains statistically consistent with the previous estimate ($0.61\substack{+0.15\\-0.06}$).
This suggests that the non-detections are major actors in the increase of the measured scatter, possibly representing outliers from the mass--richness relation (either due to properties inherent to the clusters themselves or as a result of modeling issues). The observed deviation from the nominal mass--richness relation may imply an overall systematic in the cluster mass estimates. On the other hand, a joint fit of   the CARMA and VACA LoCA samples provides an overall scatter of $\sigma\subalign{&\,\textsc{joint}\\&\log{M}|\lambda}=0.16\substack{+0.04\\-0.02}$. It may thus be possible that a scatter intrinsic to the mass--richness distribution or arising due to the limited size of studied sample may dominate the calibration of the mass--richness relation. Further observations of the SZ footprint of galaxy clusters spanning a broader richness range will be key to improving the current constraints on the MaDCoWS mass--richness scaling relation.

\begin{figure}
    \centering
    \includegraphics[clip,trim=0.25cm 0.12cm 1.00cm 0,width=\columnwidth]{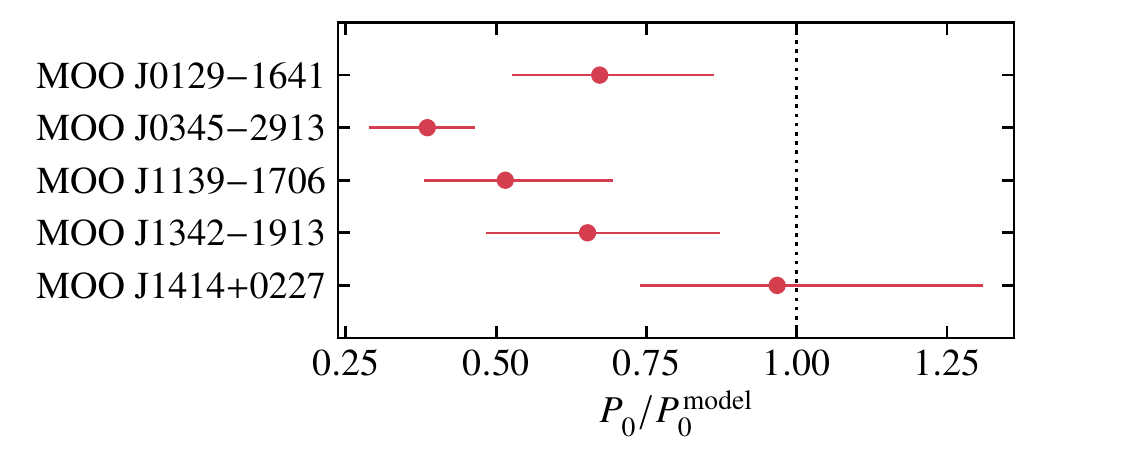}
    \caption{Inferred pressure normalization $P_0$ when assuming a cluster mass derived using the mass--richness relation from \citet{Gonzalez2019} and a universal pressure profile \citep{Arnaud2010}. The ratios reported here are normalized by the nominal value for $P_{0}$ given in Table~\ref{tab:gnfw}. As discussed in Sect.~\ref{sec:res:rich}, the observed scatter indicates a true discrepancy, which could be due to deviations from the \citet{Gonzalez2019} mass--richness scaling or to deviations from the \citet{Arnaud2010} ensemble-average pressure profile. If ACA filtering were driving the mass reconstruction, we would expect a uniformly low value for the ratio, which is not observed. The error bars for each of the points incorporates both statistical uncertainties and scatter intrinsic to MaDCoWS mass--richness scaling relation.}
    \label{fig:pnorm}
\end{figure}

The fact that ACA can only provide a high-pass filtered view of the SZ signal (coupled with possible deviations from the fiducial average pressure profile; see discussion below) may be among the main causes of the slight discrepancy of the VACA LoCA masses with respect to the scaling relation obtained using SZ measurements from CARMA. In fact, CARMA probed the SZ signal out to scales larger the $r_{500}$ values of the observed clusters, hence accessing spatial information crucial to mass determination within a cosmologically relevant overdensity. In contrast, though the ACA observations have improved sensitivity on subarcminute scales, the reconstructed masses are derived by extrapolating the assumed pressure profile from the very core regions of the clusters. In order to assess whether filtering effects play a major role in biasing the cluster masses low, we re-run the modeling by forcing the model mass $M_{500}$ to be equal to the value expected from the mass--richness relation by \citet{Gonzalez2019}, and fit for the normalization $P_0$ by assuming a wide uninformative prior. Once again, to be consistent with previous studies, we only consider the universal profile case. If the mass--richness relation provides an unbiased estimate of the cluster masses for the measured richnesses, we should then expect the respective SZ model to  describe the ACA $uv$ data well, and the inferred estimates of $P_0$ to be consistent with the nominal value in Table~\ref{tab:gnfw}. To facilitate interpretation, we  limit the analysis here to the clusters with a single SZ feature with a strong significance. As shown in Fig.~\ref{fig:pnorm}, the results are in qualitative agreement with the overall low-mass trend observed in the mass--richness distribution of the VACA LoCA sample cluster. Nevertheless, it is not possible to highlight any evident systematic effect common to all the data points. We thus conclude that the interferometric filtering is unlikely to play a major role in biasing our mass reconstruction to lower masses.

This of course presumes that the universal pressure model by \citet{Arnaud2010} can successfully describe the electron pressure distribution of such systems. However, departures from self-similarity, for example due to  an actual evolution of the average pressure profile with the cluster redshift \citep{McDonald2014} or the disturbed state of any of the studied clusters, may significantly affect the mass reconstruction \citep[see][for a cosmological application]{Ruppin2019b}.

\subsection{Multiple SZ features}\label{sec:res:multi}
\begin{figure}
    \centering
    \includegraphics[clip,trim=0.30cm 2.25cm 0.50cm 3.50cm,width=\columnwidth]{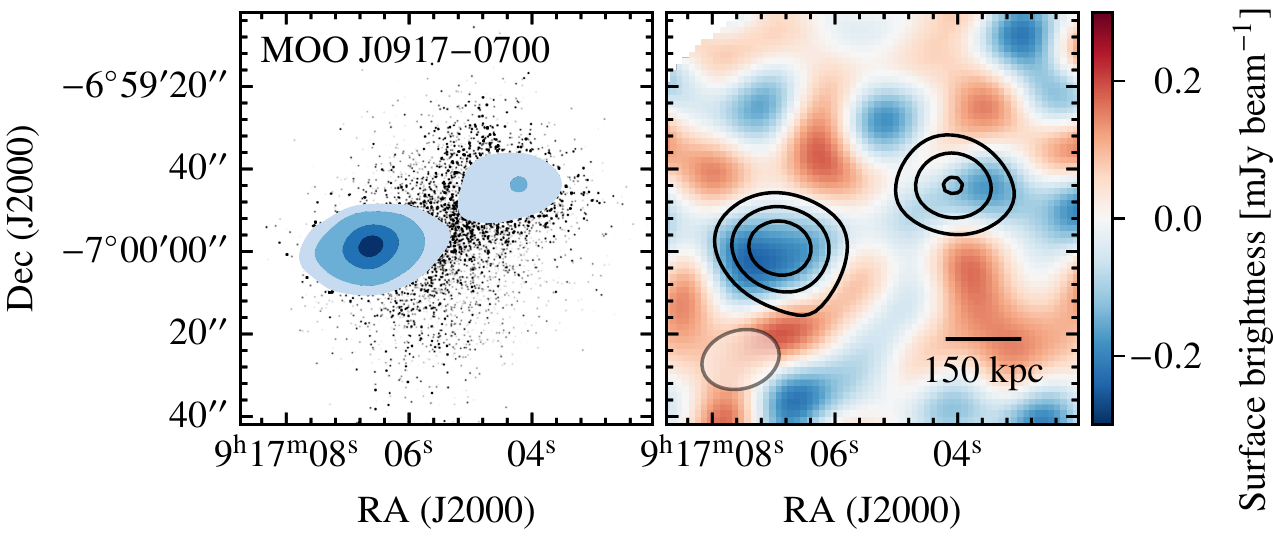}\vspace{10pt}\\
    \includegraphics[clip,trim=0.30cm 2.25cm 0.50cm 3.50cm,width=\columnwidth]{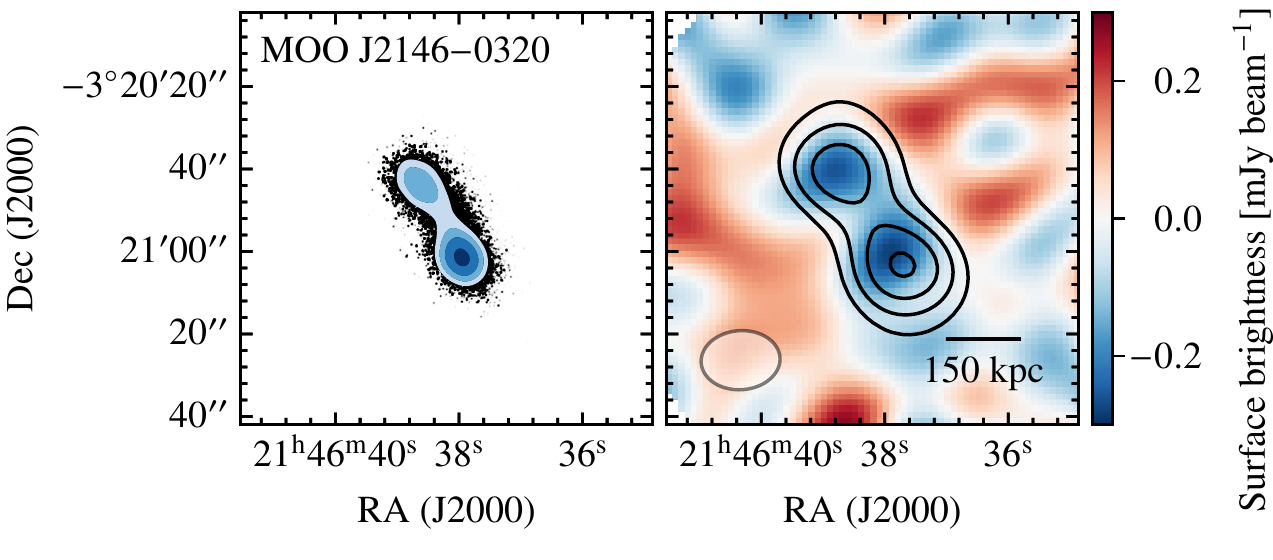}
    \caption{Marginalized posterior for the cluster centroids and dirty images (left and right panels, respectively) of the two VACA LoCA clusters characterized by multiple SZ features, \object{MOO~J0917$-$0700} and \object{MOO~J2146$-$0320} (top and bottom panels, respectively). The posterior contours correspond (from  innermost to  outermost) to 38\%, 68\%, 87\%, and 95\% credibility levels. Contours in the right panels show the $0.5\sigma$, $1\sigma$, $1.5\sigma$, and $2\sigma$ statistical significance levels of the filtered model with respect to the map noise RMS. To better highlight the SZ effect, we subtract from the visibility data the most significant point-like sources, as in Fig.~\ref{fig:imdirt:mooj0129}, and apply a $10~\mathrm{k\lambda}$ taper to the data.}
    \label{fig:bimodal}
\end{figure}

The marginalized posterior distribution for the centroids of the galaxy clusters \object{MOO~J0917$-$0700} and \object{MOO~J2146$-$0320} manifest a clear bimodal behavior (top and bottom panels of Fig.~\ref{fig:bimodal}, respectively).
We   checked that they are  dependent neither on the assumed prior on the position of the cluster centroid nor on the inclusion of point-like model components.

It may be possible that the individual posterior modes are actually related to distinct SZ components. These may arise, for example, due to the presence of a cluster pair in an early to mid merging phase (the situation is similar,  in terms of cluster masses and of separation, to the merging system 1E~2216.0-0401/1E~2215.7-0404; \citealt{Akamatsu2016}). In such cases, however, the electron pressure distribution will deviate significantly from the average gNFW models adopted in our analysis \citep{Wik2008,Sembolini2014,Yu2015,Ruppin2019c}. In particular, we expect the resulting pressure distribution to be shallower than for the case of a relaxed cluster, hence resulting in an SZ signal more affected by the interferometric short-spacing filtering (see discussion in Sect.~\ref{sec:analysis}). Similarly, non-thermal effects may play a central role in providing pressure support to the system \citep[e.g.,][]{Battaglia2012,Shi2015,Biffi2016,Ansarifard2020}. As a consequence, we might expect the reconstructed masses to be greatly biased toward values lower than the true ones. 

On the other hand, the elongation observed in the marginalized posterior probability for the cluster centroid may be a consequence of the combined effect of an elliptical geometry of the core region of the ICM and residual contributions from unresolved sources (see discussion in Sect.~\ref{sec:res:pts} below). In any case, the non-regular electron pressure distribution would indicate that the clusters may be highly disturbed, again inducing a potential bias in the reconstructed masses.

\begin{figure}
    \centering
    \includegraphics[clip,trim=0.30cm 0.50cm 0.60cm 1.40cm,width=\columnwidth]{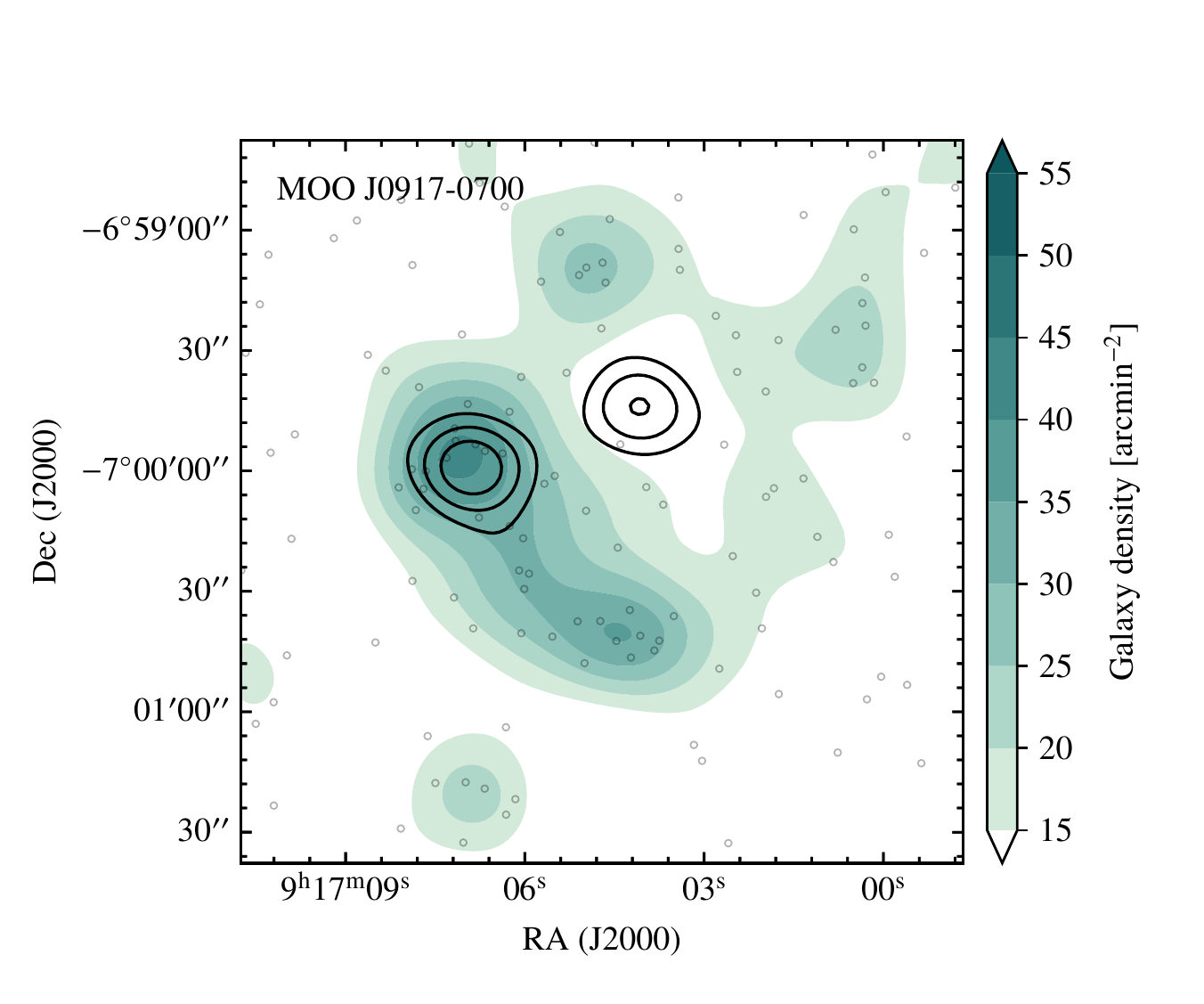}\vspace{10pt}\\
    \includegraphics[clip,trim=0.30cm 0.50cm 0.60cm 1.40cm,width=\columnwidth]{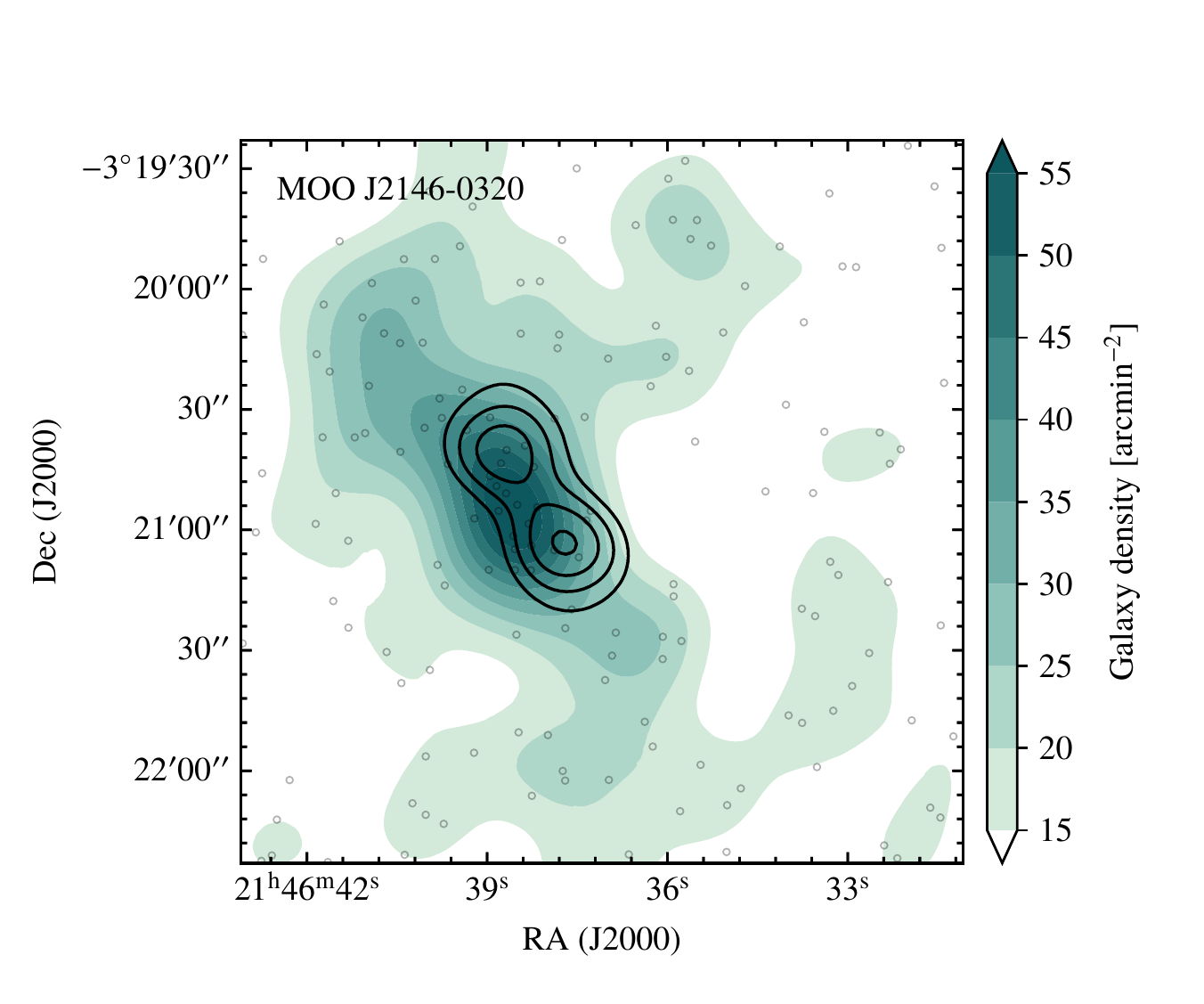}
    \caption{Maps of the color-selected galaxy overdensities around \object{MOO~J0917$-$0700} and \object{MOO~J2146$-$0320}, as measured by \textit{Spitzer}/IRAC. Overplotted are the contours of the SZ models of the two clusters, as in Figure~\ref{fig:bimodal}. In both cases the elongated morphology of the galaxy density distribution may support the merger scenario. The light gray points denote the positions of the individual IRAC-selected galaxies. For display purposes, the galaxy overdensity maps have been preliminary smoothed by a Gaussian kernel with standard deviation of $\sim12~\mathrm{arcsec}$.}
    \label{fig:irac}
\end{figure}

Additional hints about the potential presence of merger activity in the two clusters come from the analysis of the distribution of their member galaxies. In Figure~\ref{fig:irac}, we show \textit{Spitzer}/IRAC color-selected galaxy overdensities (a description of the specific color-selection strategy employed here can be found in \citealt{Gonzalez2019}, and is based on the works by \citealt{Wylezalek2013,Wylezalek2014}). For both clusters a significant elongation is observed in the galaxy density distribution. In the specific case of \object{MOO~J0917$-$0700}, the eastern component, the broader and more massive of the two SZ features, sits on the most prominent peak of the galaxy distribution, while the second SZ feature is significantly displaced with respect to the main galaxy overdensity structure. This may be related to the case in which a subcluster (in this case, the western one) has undergone an off-axis collision at a large impact parameter with the main cluster (the eastern SZ component). The two mass components should then be interpreted as highlighting a disturbed and elongated cluster morphology, rather than the presence of separate subclusters. On the other hand, the agreement in both the orientation and position of the SZ model and galaxy overdensity in \object{MOO~J2146$-$0320} supports the original hypothesis of an early to mid-stage, almost edge-on merger, in which the subclusters are still able to partially retain the respective intracluster gas in their potential well.

Unfortunately, the low spatial resolution of the available ACA data does not allow for a proper characterization of the dynamical state of the two systems, and the above arguments can only provide a speculative scenario that may explain the inferred SZ morphology. Additional observations will thus be necessary to shed a definitive light on cluster structures and to help understand whether the multiple SZ features belong to separate physical components.

\subsection{Unresolved sources}\label{sec:res:pts}
As already briefly mentioned in the previous section, a possible systematic effect that may prevent the proper estimation of the cluster masses from the modeling of ACA data is any residual contamination from emissions that have not been accounted for. Along with radio synchrotron sources, we expect dusty galaxies to contribute to the overall confusion noise \citep[see discussion in][]{DiMascolo2019a}.  However, although we were able to locate and constrain a number of unresolved components, the lack of high-resolution data (e.g., from the main 12-meter array) may in fact have limited the identification to the brightest end of the source population contaminating the SZ signal.

On the other hand, the poor resolution and sensitivity do not allow us to  separate with reasonable confidence the SZ effect from any possible diffuse radio components. Studies by \citet{Moravec2019,Moravec2020} show that a large fraction of the sources belonging to the population of radio-loud AGNs within the MaDCoWS clusters exhibit extended morphologies. In this regard, external data may be key to complementing information about radio contaminants. Unfortunately, the available radio surveys offer only partial coverage of the VACA LoCA sample. 

In particular, we first checked the NRAO VLA Sky Survey \citep[NVSS;][]{Condon1998} for possible radio components. The inspection of the NVSS images of the VACA LoCA fields, however,  does not  highlight any significant radio sources,  point-like or diffuse.

We further inspect the VLA Faint Images of the Radio Sky at Twenty-Centimeters \citep[FIRST;][]{Becker1995} survey, which provides coverage for only five sources from the VACA LoCA pilot sample. One low-significance source is found in each of the \object{MOO~J0917$-$0700} and \object{MOO~J1414$+$0227} fields, both coinciding with bright point-like sources identified by the blind search over the ACA data (Sect.~\ref{sec:analysis:pts}).

\begin{figure}
    \centering
    \includegraphics[clip,trim=0 0.50cm 0 1.40cm,width=\columnwidth]{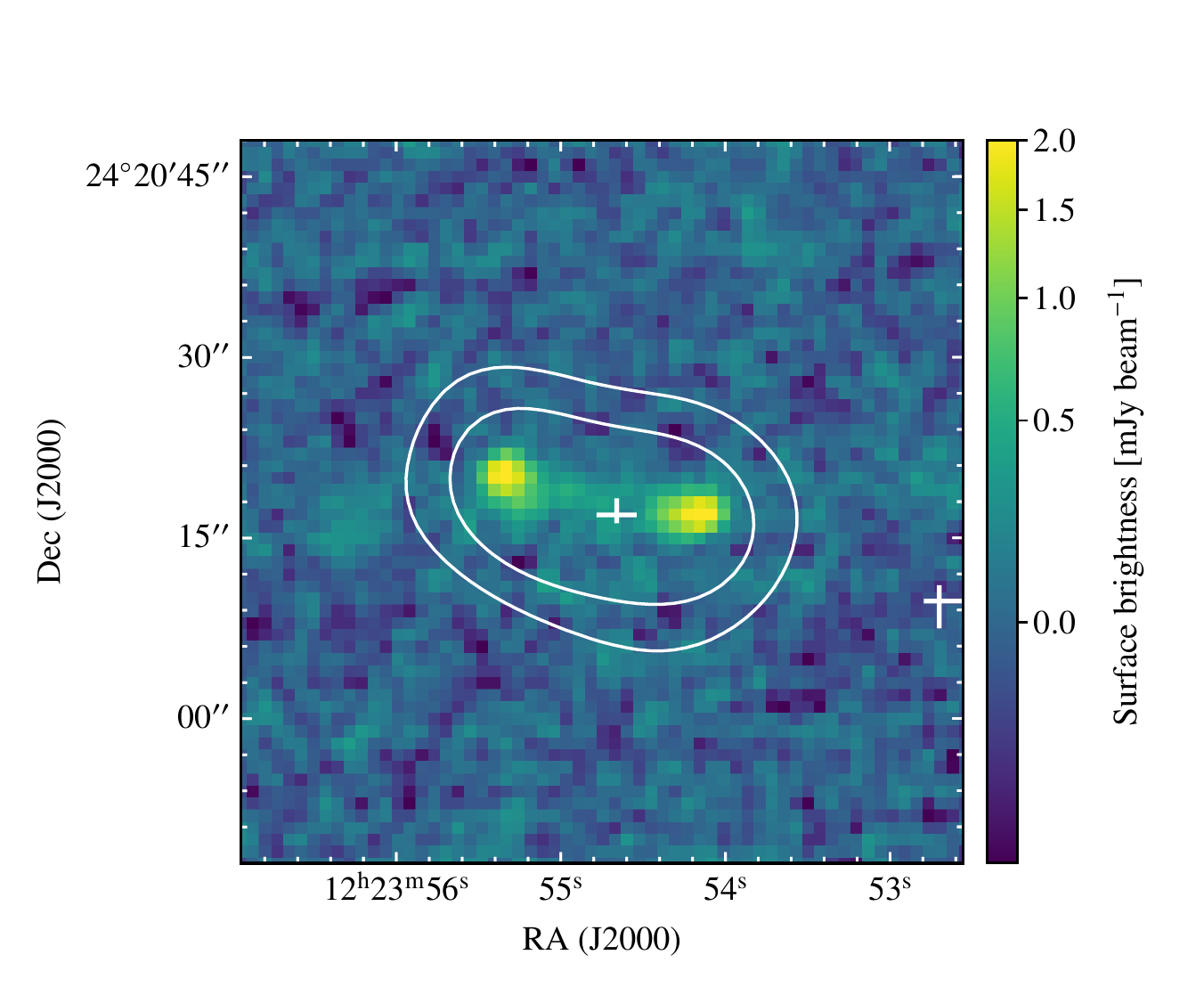}
    \caption{VLASS map of the radio structure in \object{MOO~J1223$+$2420}. Overplotted are the contours from the same image smoothed to ACA resolution (the levels correspond to the arbitrary values of 0.10 and 0.05 mJy beam$^{-1}$). The white crosses denote the position of the most significant point sources and respective uncertainties from the 68\% credibility interval around each posterior peak. Regardless of the accuracy in the determination of the position of any point-like sources, the low resolution of ACA does not   resolve the possible different contributions from the jets and the central galaxy.}
    \label{fig:vlass}
\end{figure}

More recently, the Very Large Array Sky Survey \citep[VLASS;][]{Lacy2020} completed its first epoch of observations covering the entire sky north of $\delta=-40$ in S Band (2-4~GHz), which offers the advantage of sharing the same sky coverage as the full MaDCoWS sample but at much higher resolution than NVSS. The first epoch maps reach a depth of $\approx 120~\mu$Jy RMS on average. Surprisingly, the only sources we were able to confirm at $>3\sigma$ significance are those also seen in the FIRST data on \object{MOO~J1414$+$0227} and \object{MOO~J1223$+$2420}.

The FIRST and VLASS images of the \object{MOO~J1223$+$2420} field are the only ones to show the presence of a clear radio source within any of the VACA LoCA clusters. This exhibits a double-lobed feature, identified by \citet{vanVelzen2015} as belonging to the radio jets from a central FR II radio galaxy. It corresponds to a strong radio source detected near the center of the \object{MOO~J1223$+$2420} ACA field, which may be the cause of the inferred mass well below the value predicted from MaDCoWS mass--richness scaling relation. Although we model all possible contributions from unresolved radio emission from the central regions of the cluster, any residual contributions (e.g., from extended structures) may still limit our ability to retrieve an accurate estimate of the cluster mass. Unfortunately, the resolution of the ACA observation of \object{MOO~J1223$+$2420} does not allow us to determine whether the model describes the signal from the central galaxy, the radio lobes, or a blend of these two forms of emission (see Fig.~\ref{fig:vlass}).

To get a sense of whether the contamination from the extended radio lobes may contribute substantially to limiting the statistical significance of the SZ signal from \object{MOO~J1223$+$2420}, we scale the integrated flux measured in the VLASS image to the central frequency of the VACA LoCA data, $\nu_{\textsc{aca}}=97.5~\mathrm{GHz}$. We assume the radio lobes exhibit a typical synchrotron spectrum with average spectral index of $\alpha=-0.85$, appropriate for this redshift range \citep[see][]{vanVelzen2015}, and account for both the VLASS and ACA broadband spectral coverage. The fluxes integrated over the eastern and western lobes in the VLASS data are respectively $i_{\textsc{vlass}}=6.35~\mathrm{mJy}$ and $i_{\textsc{vlass}}=6.11~\mathrm{mJy}$, corresponding to $i_{\textsc{aca}}=0.32~\mathrm{mJy}$ and $i_{\textsc{aca}}=0.31~\mathrm{mJy}$ at the central frequency of our ACA observations. As indicated by the contours in Fig.~\ref{fig:vlass}, the ACA observations do not resolve the individual radio lobes, and these will manifest in the interferometric image of \object{MOO~J1223$+$2420} as an unresolved emission with a flux density of $0.60~\mathrm{mJy}$. On the other hand, the peak of the SZ signal expected in the case that the cluster mass exactly follows the MaDCoWS mass--richness relation would be around $-0.45~\mathrm{mJy~beam^{-1}}$. Furthermore, the sum of the extrapolated lobe flux and the SZ signal is consistent with the amplitude of the point source component identified by the blind search, $i_{\textsc{ps}}=0.20\substack{+0.06\\-0.04}~\mathrm{mJy}$. Under the assumption of a spatial correspondence of the radio source and the SZ centroid, this implies that most of the SZ signal from the cluster core could be entirely dominated by the emission of the radio lobes and, in turn, dim the measured flux from the radio source. Since the ACA is capable of probing the SZ signal from only the innermost radii of \object{MOO~J1223$+$2420} (see Fig.~\ref{fig:filtered}), the above estimates further confirm that the radio contamination may have been critical in biasing the mass reconstruction low. This combines with the fact that a higher mass would correspond to a more extended and hence more severely filtered SZ signal. We note, however, that the above discussion is only approximate, as we do not have information on the real spectral index over the range of frequencies covered by the ACA observation. Again, the proper characterization of the radio source within \object{MOO~J1223$+$2420} with data at higher angular resolution will be crucial to improving the constraints on the cluster mass.

\subsection{Dependence on the assumed pressure model}\label{sec:res:mass}
\begin{figure}
    \centering
    \includegraphics[clip,trim=0 1.25cm 0 2.74cm,width=\columnwidth]{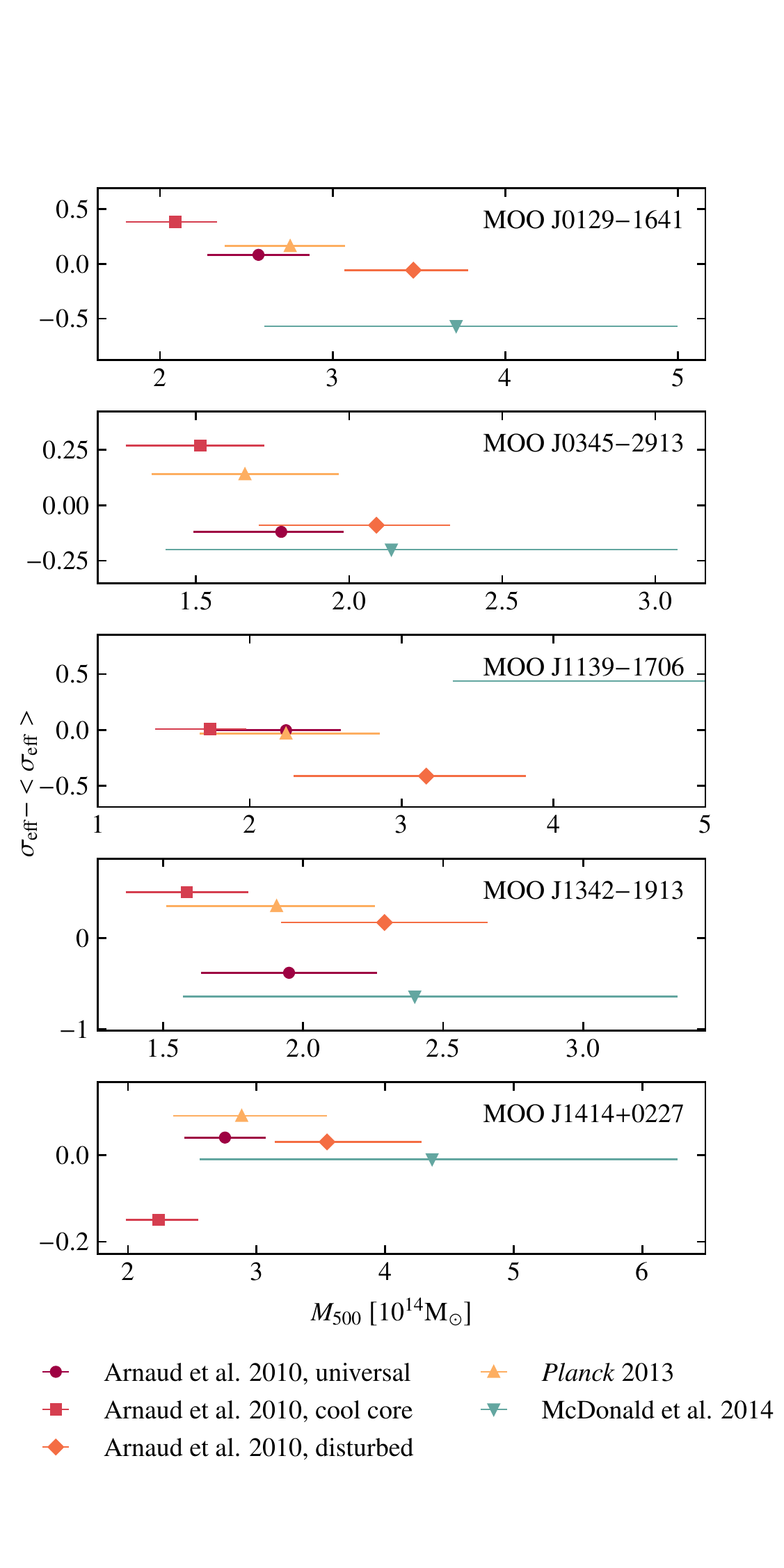}
    \caption{Deviations from average significance levels for the best-detected galaxy clusters of the VACA LoCA pilot sample. The points correspond to the mass estimates obtained by assuming different versions of the gNFW pressure profile. The variations in $\sigma_{\mathrm{eff}}$ are always less than 3 (see Sect.~\ref{sec:res}), which, according to the Jeffreys model selection criterion, implies that no pressure model is strongly favored over the others for any of the VACA LoCA clusters.
    The large uncertainties on the masses derived assuming the profile by \citet{McDonald2014} are due to the large uncertainties on the respective best-fit gNFW parameters.}
    \label{fig:masssigma}
\end{figure}

As already mentioned in Sect.~\ref{sec:analysis:sz}, we further test the mass reconstruction against different versions of the gNFW pressure profiles. In Fig.~\ref{fig:masssigma}, we provide a direct comparison of the masses and respective effective significance levels for the VACA LoCA clusters with significant detections. The full list with the estimates of the cluster masses for all the profiles considered in Table~\ref{tab:gnfw} can be found in Table~\ref{tab:app:masses}.

\begin{figure}
    \centering
    \includegraphics[clip,trim=0 0.25cm 0 0,width=\columnwidth]{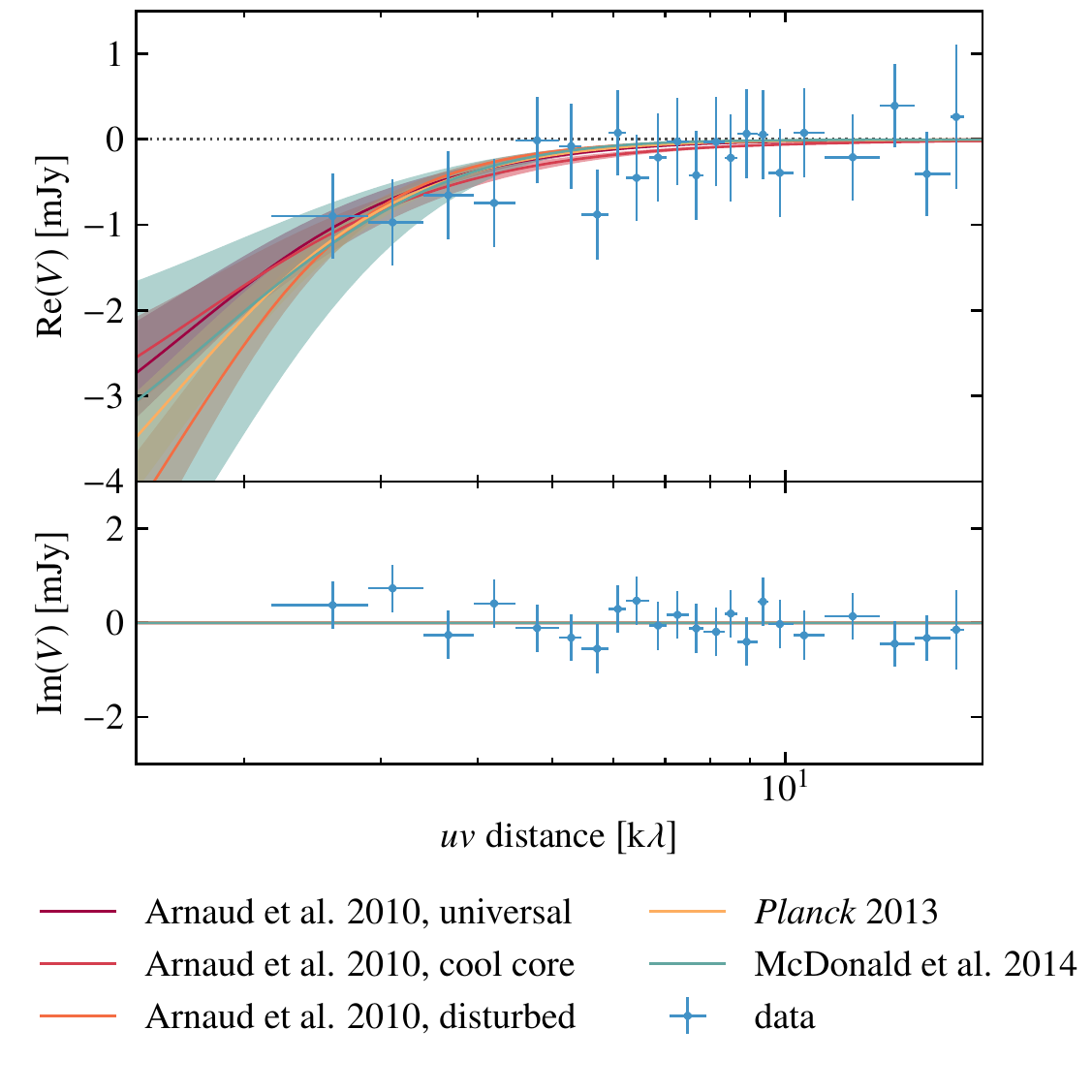}
    \caption{Comparison of the real (top) and imaginary (bottom) parts of ACA point source-subtracted visibilities $V=V(u,v)$ for \object{MOO~J0129$-$1640}, the most significant detection, and the $uv$ radial profiles for the different flavors of gNFW (Table~\ref{tab:gnfw}). The data are binned so that each bin contains the same number of visibilities (here set to 2500 for plotting purposes). Before averaging, we shifted the phase center to the position of the cluster centroid to minimize the ringing effect due to non-zero phases. As a result, the imaginary part of the visibilities are overall consistent with zero. Any significant deviations would be symptomatic of residual off-center point-like sources or asymmetries in the cluster SZ signal that are unaccounted for in the analysis, for example.}
    \label{fig:uvprof:mooj0129}
\end{figure}

Not unexpectedly, the specific value for the cluster mass is highly dependent on the specific profile assumed to describe the pressure distribution. As shown in the $uv$ radial profile of Fig.~\ref{fig:uvprof:mooj0129}, most of the SZ flux is not probed by ACA, making it sensitive only to the pressure distribution within the inner region of galaxy clusters. This can also be inferred by comparing the values for the MRS in each observation to $2 \times \theta_{500}$ using Tables \ref{tab:obsdata} and \ref{tab:masses}, respectively. This effect couples with the primary beam attenuation of the edges of the ACA fields, which drives the characteristic radius of the gNFW profile to be on the same of order of the antenna pattern half width half maximum, and thus affecting the mass reconstruction. As a result, the model based on the gNFW parameterization from \citet{McDonald2014} and for the morphologically disturbed sample in \citet{Arnaud2010} present masses systematically higher than the other profiles, as a direct consequence of their flatter radial trend at small radii. Conversely, the strongly peaked cool-core profile by \citet{Arnaud2010} allows us to  easily fit low-mass (and, then, very compact) cluster models to the observed SZ signal.
Nevertheless, as we were not able to infer any of the parameters defining the gNFW pressure profile in Eq.~\eqref{eq:gnfw}, the small scatter in the effective significance for each of the different pressure models does not allow us to  select or rule out any of the mass estimates. 

\begin{figure}
    \centering
    \includegraphics[clip,trim=0.30cm 1.10cm 0.30cm 1.4cm,width=\columnwidth]{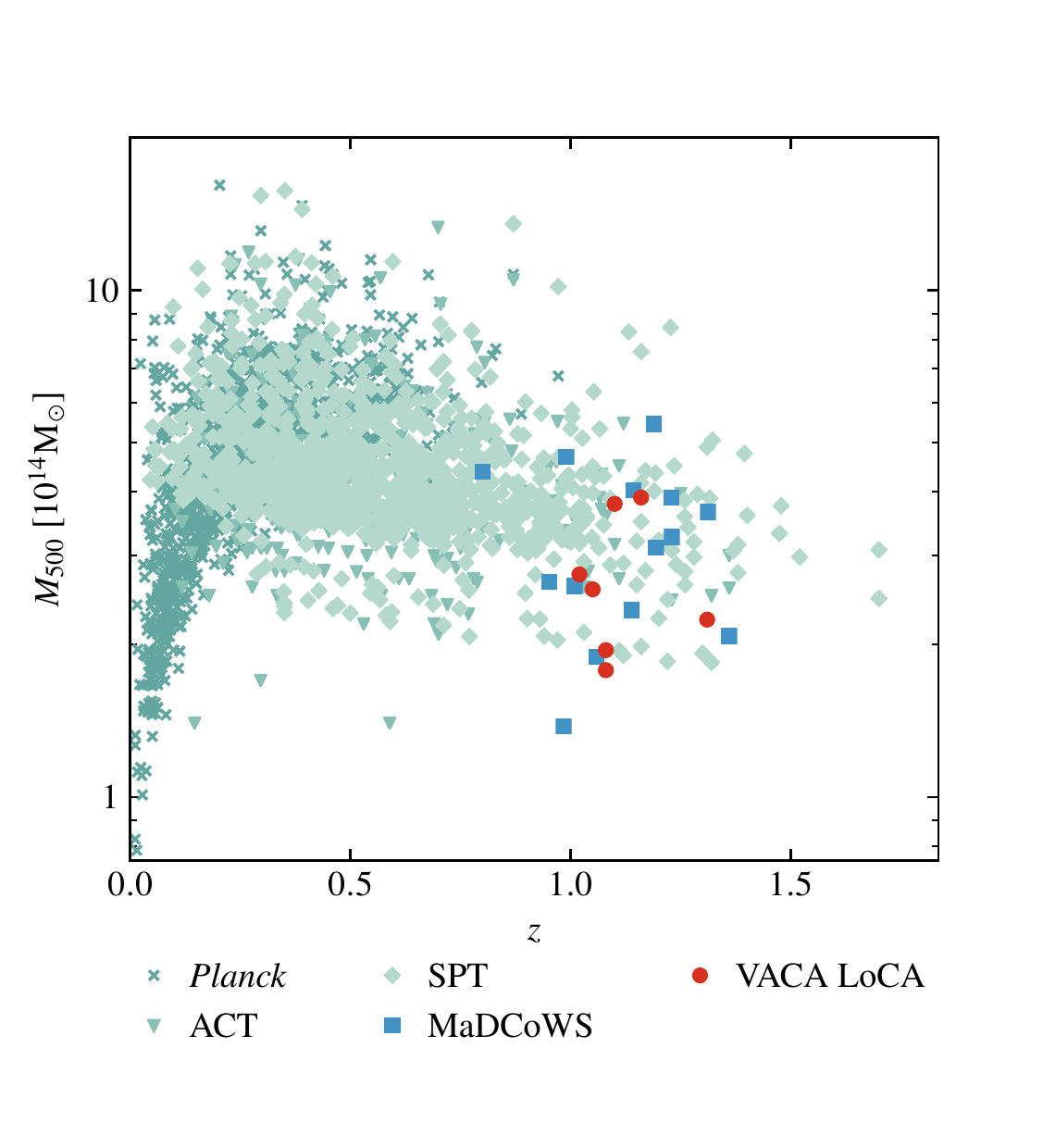}
    \caption{Mass vs. redshift distribution of galaxy clusters in the VACA LoCA pilot sample (red circles). The red points denote the VACA LoCA clusters with significant detections. For comparison, we include the mass estimates of previously reported MaDCoWS clusters \citep[blue squares;][]{Gonzalez2019} and the samples from the   SZ surveys of \textit{Planck} \citep[green crosses;][]{Planck2016XXVII}, ACT \citep[green triangles;][]{Marriage2011,Hasselfield2013,Hilton2018}, and SPT \citep[green diamonds;][]{Bocquet2019,Huang2019,Bleem2020}. All the MaDCoWS clusters with SZ data are found to be comparable in mass with the clusters detected by these surveys over the same range of redshifts.}
    \label{fig:massredshift}
\end{figure}

The impossibility of discriminating between different gNFW scenarios is an immediate consequence of the limited sensitivity of the ACA observations we are analyzing, along with the lack of information on large angular scales. Figure~\ref{fig:uvprof:mooj0129} shows the $uv$ radial plot for the different gNFW best-fit models for the most significantly detected cluster of the VACA LoCA sample, \object{MOO~J0129$-$1640}. Although they all succeed in describing the long baseline data, they also present a non-negligible scatter over angular scales larger than the maximum recovered scale in the observation.
As discussed in \citet{DiMascolo2019a} and \citet{Perrott2019}, the joint analysis of interferometric measurements and lower-resolution, single-dish observations provides a straightforward solution for improving the reconstruction of models of the SZ signal from galaxy clusters. Cosmic microwave background experiments designed to detect clusters at arcminute resolution, such as the Atacama Cosmology Telescope \citep[ACT; see, e.g.,][]{Hilton2018} or the South Pole Telescope \citep[SPT; see, e.g.,][]{Bleem2015,Bleem2020} could fulfill the needs of complementary large-scale data, and the VACA LoCA clusters are comparable in mass to some of the high-redshift systems detected by those surveys (see Fig.~\ref{fig:massredshift}). However, the publicly available data do not cover the portion of the sky comprising the VACA LoCA fields. 

Additionally, \cite{Gonzalez2019} compares the \textit{Planck} mass-redshift relation to the masses inferred for the entire MaDCoWS sample, and finds that they predominantly lie below the mass selection function of \textit{Planck}.
For the VACA LoCA sample of MaDCoWS clusters, we find that no useful constraint on the integrated Compton parameter $Y$ can be obtained from the \textit{Planck} maps, due to beam dilution and limited sensitivity.  
In all but the most extreme case the integrated SZ signal for each clusters would fall within a single 10\arcmin\ resolution element of \textit{Planck}. Extrapolating the fits to the VACA LoCA sample, each member should have an average Compton $Y$ value $\langle Y \rangle \lesssim 1.6\times10^{-6}$ over an area of 100 square arcminutes, while the RMS noise level in the \textit{Planck} maps is $\approx 1.7 \times 10^{-6}$ on average \citep{Planck2015_XXII}. This indicates that the most massive clusters in VACA LoCA may be on the order of $1\sigma$ significance in the \textit{Planck} maps, while the rest are well below that, and even a stacked measurement using the ten members of the pilot sample would be marginal.

It is worth noting that the small range of inferred $\sigma_{\mathrm{eff}}$ implies that ACA is able to provide robust detections of the SZ signal from the VACA LoCA sample clusters independent of assumptions about the underlying pressure electron distribution. 

\section{Conclusions}\label{sec:conclusion}
In this work, we analyze a pilot sample of ACA observations of ten high-redshift galaxy clusters representative of the typical richness of the MaDCoWS catalog. This has been mainly aimed at directly testing the capability of the ACA in ALMA Band 3 for measurements of the SZ signal from high-redshift systems. In summary, our main findings are the following:
\begin{itemize}
    \item The ACA can provide robust and relatively straightforward validation of galaxy cluster identifications through the detection of the SZ signal from the intracluster gas. We note that the on-source integration times are typically $\lesssim 3$~hours per target. More importantly, the detection significance is not affected by the specific choice of pressure distribution model.\\
    \item The limited sensitivity and angular dynamic range probed by the observations limit the accuracy of the mass estimates. The mass estimates within $r_{500}$ are strongly dependent on the specific choice of pressure model, as the maximum recoverable scale in the observations is smaller than the typical radius within which one would like to probe the integrated SZ signal, $\theta_{500}$ (see Tables \ref{tab:obsdata} and \ref{tab:masses}, respectively).\\
    \item Related to the point above, a thorough characterization of the cluster dynamical state cannot be achieved, as the ACA angular resolution and limited sensitivity do not constrain small-scale features in the ICM within the observed galaxy clusters.  However, the analysis does reveal two potentially exciting merging cluster candidates that merit more detailed follow-up.\\
    \item The $uv$-space analysis of ACA data is crucial for separating the SZ signal from unresolved sources of contamination. However, the reconstruction of a proper and exhaustive model is limited by the above-mentioned sensitivity, resolution, and maximum recoverable scale.
\end{itemize}

Data at higher angular resolution than ACA (e.g., from the ALMA 12-meter array) would provide a dramatic improvement in the identification and characterization of point-like sources populating the cluster fields. Furthermore, multi-frequency coverage of the cluster fields would provide fundamental insight into, and better constraints on, the spectral properties of any contaminant source, and thus better disentangle its signal from the underlying SZ effect.

On the other hand, as discussed in Sect.~\ref{sec:res:mass}, the possibility of complementing interferometric observations with single-dish measurements of the same targets will be key to gaining a better description of the electron pressure distribution out to large scales, and hence a more accurate reconstruction of the cluster masses. 
However, the scales recoverable with current SZ survey instruments are limited to greater than 1 arcminute at 90~GHz, leaving a gap with little to  no overlap in Fourier modes probed in such joint analyses \citep[see discussion in][]{DiMascolo2019a}. An exciting advance could be obtained using a future large single-dish telescope with at least three times the size of the ACA and ALMA primary mirrors and a wide ($>1^\circ$) field of view, such as the Atacama Large Aperture Submillimeter Telescope \citep[AtLAST;][]{Klaassen2019}.

Finally, ALMA Bands 1 \citep[35-51~GHz;][]{DiFrancesco2013,Huang2016}
and 2 \citep[67-116~GHz;][]{Yagoubov2020} will further increase  the maximum recoverable scale over the next
few years, and thus the sensitivity of ALMA and the ACA to diffuse, low surface brightness signals on arcminute scales.

\section*{Acknowledgments}
We thank the anonymous referee for the thoughtful suggestions and comments that helped improving this work.
This paper makes use of the following ALMA data: ADS/JAO.ALMA\#2016.2.00014.S. ALMA is a partnership of ESO (representing its member states), NSF (USA) and NINS (Japan), together with NRC (Canada), MOST and ASIAA (Taiwan), and KASI (Republic of Korea), in cooperation with the Republic of Chile. The Joint ALMA Observatory is operated by ESO, AUI/NRAO and NAOJ. EC and RS acknowledge partial support by the Russian Science Foundation grant 19-12-00369. The work of PRME and DS was carried out at the Jet Propulsion Laboratory, California Institute of Technology, under a contract with NASA.

\bibliographystyle{aa}
\bibliography{madcows}

\begin{thebibliography}{68}
\expandafter\ifx\csname natexlab\endcsname\relax\def\natexlab#1{#1}\fi

\bibitem[{{Akamatsu} {et~al.}(2016){Akamatsu}, {Gu}, {Shimwell}, {Mernier},
  {Mao}, {Urdampilleta}, {de Plaa}, {R{\"o}ttgering}, \&
  {Kaastra}}]{Akamatsu2016}
{Akamatsu}, H., {Gu}, L., {Shimwell}, T.~W., {et~al.} 2016, \aap, 593, L7

\bibitem[{{Andreon}(2015)}]{Andreon2015}
{Andreon}, S. 2015, \aap, 582, A100

\bibitem[{{Ansarifard} {et~al.}(2020){Ansarifard}, {Rasia}, {Biffi}, {Borgani},
  {Cui}, {De Petris}, {Dolag}, {Ettori}, {Movahed}, {Murante}, \&
  {Yepes}}]{Ansarifard2020}
{Ansarifard}, S., {Rasia}, E., {Biffi}, V., {et~al.} 2020, \aap, 634, A113

\bibitem[{{Arnaud} {et~al.}(2010)}]{Arnaud2010}
{Arnaud}, M. {et~al.} 2010, \aap, 517, A92

\bibitem[{{Basu} {et~al.}(2016){Basu}, {Sommer}, {Erler}, {Eckert}, {Vazza},
  {Magnelli}, {Bertoldi}, \& {Tozzi}}]{Basu2016}
{Basu}, K., {Sommer}, M., {Erler}, J., {et~al.} 2016, \apjl, 829, L23

\bibitem[{{Battaglia} {et~al.}(2012){Battaglia}, {Bond}, {Pfrommer}, \&
  {Sievers}}]{Battaglia2012}
{Battaglia}, N., {Bond}, J.~R., {Pfrommer}, C., \& {Sievers}, J.~L. 2012, \apj,
  758, 74

\bibitem[{{Becker} {et~al.}(1995){Becker}, {White}, \& {Helfand}}]{Becker1995}
{Becker}, R.~H., {White}, R.~L., \& {Helfand}, D.~J. 1995, \apj, 450, 559

\bibitem[{{Biffi} {et~al.}(2016){Biffi}, {Borgani}, {Murante}, {Rasia},
  {Planelles}, {Granato}, {Ragone-Figueroa}, {Beck}, {Gaspari}, \&
  {Dolag}}]{Biffi2016}
{Biffi}, V., {Borgani}, S., {Murante}, G., {et~al.} 2016, \apj, 827, 112

\bibitem[{{Bleem} {et~al.}(2020){Bleem}, {Bocquet}, {Stalder}, {Gladders},
  {Ade}, {Allen}, {Anderson}, {Annis}, {Ashby}, {Austermann}, {Avila}, {Avva},
  {Bayliss}, {Beall}, {Bechtol}, {Bender}, {Benson}, {Bertin}, {Bianchini},
  {Blake}, {Brodwin}, {Brooks}, {Buckley-Geer}, {Burke}, {Carlstrom}, {Rosell},
  {Carrasco Kind}, {Carretero}, {Chang}, {Chiang}, {Citron}, {Moran},
  {Costanzi}, {Crawford}, {Crites}, {da Costa}, {de Haan}, {De Vicente},
  {Desai}, {Diehl}, {Dietrich}, {Dobbs}, {Eifler}, {Everett}, {Flaugher},
  {Floyd}, {Frieman}, {Gallicchio}, {Garc{\'\i}a-Bellido}, {George}, {Gerdes},
  {Gilbert}, {Gruen}, {Gruendl}, {Gschwend}, {Gupta}, {Gutierrez}, {Halverson},
  {Harrington}, {Henning}, {Heymans}, {Holder}, {Hollowood}, {Holzapfel},
  {Honscheid}, {Hrubes}, {Huang}, {Hubmayr}, {Irwin}, {James}, {Jeltema},
  {Joudaki}, {Khullar}, {Klein}, {Knox}, {Kuropatkin}, {Lee}, {Li}, {Lidman},
  {Lowitz}, {MacCrann}, {Mahler}, {Maia}, {Marshall}, {McDonald}, {McMahon},
  {Melchior}, {Menanteau}, {Meyer}, {Miquel}, {Mocanu}, {Mohr}, {Montgomery},
  {Nadolski}, {Natoli}, {Nibarger}, {Noble}, {Novosad}, {Padin}, {Palmese},
  {Parkinson}, {Patil}, {Paz-Chinch{\'o}n}, {Plazas}, {Pryke}, {Ramachandra},
  {Reichardt}, {Remolina Gonz{\'a}lez}, {Romer}, {Roodman}, {Ruhl}, {Rykoff},
  {Saliwanchik}, {Sanchez}, {Saro}, {Sayre}, {Schaffer}, {Schrabback},
  {Serrano}, {Sharon}, {Sievers}, {Smecher}, {Smith}, {Soares-Santos}, {Stark},
  {Story}, {Suchyta}, {Tarle}, {Tucker}, {Vanderlinde}, {Veach}, {Vieira},
  {Wang}, {Weller}, {Whitehorn}, {Wu}, {Yefremenko}, \& {Zhang}}]{Bleem2020}
{Bleem}, L.~E., {Bocquet}, S., {Stalder}, B., {et~al.} 2020, \apjs, 247, 25

\bibitem[{{Bleem} {et~al.}(2015){Bleem}, {Stalder}, {de Haan}, {Aird}, {Allen},
  {Applegate}, {Ashby}, {Bautz}, {Bayliss}, {Benson}, {Bocquet}, {Brodwin},
  {Carlstrom}, {Chang}, {Chiu}, {Cho}, {Clocchiatti}, {Crawford}, {Crites},
  {Desai}, {Dietrich}, {Dobbs}, {Foley}, {Forman}, {George}, {Gladders},
  {Gonzalez}, {Halverson}, {Hennig}, {Hoekstra}, {Holder}, {Holzapfel},
  {Hrubes}, {Jones}, {Keisler}, {Knox}, {Lee}, {Leitch}, {Liu}, {Lueker},
  {Luong-Van}, {Mantz}, {Marrone}, {McDonald}, {McMahon}, {Meyer}, {Mocanu},
  {Mohr}, {Murray}, {Padin}, {Pryke}, {Reichardt}, {Rest}, {Ruel}, {Ruhl},
  {Saliwanchik}, {Saro}, {Sayre}, {Schaffer}, {Schrabback}, {Shirokoff},
  {Song}, {Spieler}, {Stanford}, {Staniszewski}, {Stark}, {Story}, {Stubbs},
  {Vand erlinde}, {Vieira}, {Vikhlinin}, {Williamson}, {Zahn}, \&
  {Zenteno}}]{Bleem2015}
{Bleem}, L.~E., {Stalder}, B., {de Haan}, T., {et~al.} 2015, \apjs, 216, 27

\bibitem[{{Bocquet} {et~al.}(2019){Bocquet}, {Dietrich}, {Schrabback}, {Bleem},
  {Klein}, {Allen}, {Applegate}, {Ashby}, {Bautz}, {Bayliss}, {Benson},
  {Brodwin}, {Bulbul}, {Canning}, {Capasso}, {Carlstrom}, {Chang}, {Chiu},
  {Cho}, {Clocchiatti}, {Crawford}, {Crites}, {de Haan}, {Desai}, {Dobbs},
  {Foley}, {Forman}, {Garmire}, {George}, {Gladders}, {Gonzalez}, {Grandis},
  {Gupta}, {Halverson}, {Hlavacek-Larrondo}, {Hoekstra}, {Holder}, {Holzapfel},
  {Hou}, {Hrubes}, {Huang}, {Jones}, {Khullar}, {Knox}, {Kraft}, {Lee}, {von
  der Linden}, {Luong-Van}, {Mantz}, {Marrone}, {McDonald}, {McMahon}, {Meyer},
  {Mocanu}, {Mohr}, {Morris}, {Padin}, {Patil}, {Pryke}, {Rapetti},
  {Reichardt}, {Rest}, {Ruhl}, {Saliwanchik}, {Saro}, {Sayre}, {Schaffer},
  {Shirokoff}, {Stalder}, {Stanford}, {Staniszewski}, {Stark}, {Story},
  {Strazzullo}, {Stubbs}, {Vanderlinde}, {Vieira}, {Vikhlinin}, {Williamson},
  \& {Zenteno}}]{Bocquet2019}
{Bocquet}, S., {Dietrich}, J.~P., {Schrabback}, T., {et~al.} 2019, \apj, 878,
  55

\bibitem[{{Brodwin} {et~al.}(2015){Brodwin}, {Greer}, {Leitch}, {Stanford},
  {Gonzalez}, {Gettings}, {Abdulla}, {Carlstrom}, {Decker}, {Eisenhardt},
  {Lin}, {Mantz}, {Marrone}, {McDonald}, {Stalder}, {Stern}, \&
  {Wylezalek}}]{Brodwin2015}
{Brodwin}, M., {Greer}, C.~H., {Leitch}, E.~M., {et~al.} 2015, \apj, 806, 26

\bibitem[{{Bulbul} {et~al.}(2019){Bulbul}, {Chiu}, {Mohr}, {McDonald},
  {Benson}, {Bautz}, {Bayliss}, {Bleem}, {Brodwin}, {Bocquet}, {Capasso},
  {Dietrich}, {Forman}, {Hlavacek-Larrondo}, {Holzapfel}, {Khullar}, {Klein},
  {Kraft}, {Miller}, {Reichardt}, {Saro}, {Sharon}, {Stalder}, {Schrabback}, \&
  {Stanford}}]{Bulbul2019}
{Bulbul}, E., {Chiu}, I.~N., {Mohr}, J.~J., {et~al.} 2019, \apj, 871, 50

\bibitem[{{Challinor} \& {Lasenby}(1998)}]{Challinor1998}
{Challinor}, A. \& {Lasenby}, A. 1998, \apj, 499, 1

\bibitem[{{Churazov} {et~al.}(2015){Churazov}, {Vikhlinin}, \&
  {Sunyaev}}]{Churazov2015}
{Churazov}, E., {Vikhlinin}, A., \& {Sunyaev}, R. 2015, \mnras, 450, 1984

\bibitem[{{Condon} {et~al.}(1998){Condon}, {Cotton}, {Greisen}, {Yin},
  {Perley}, {Taylor}, \& {Broderick}}]{Condon1998}
{Condon}, J.~J., {Cotton}, W.~D., {Greisen}, E.~W., {et~al.} 1998, \aj, 115,
  1693

\bibitem[{{Decker} {et~al.}(2019){Decker}, {Brodwin}, {Abdulla}, {Gonzalez},
  {Marrone}, {O'Donnell}, {Stanford}, {Wylezalek}, {Carlstrom}, {Eisenhardt},
  {Mantz}, {Mo}, {Moravec}, {Stern}, {Aldering}, {Ashby}, {Boone}, {Hayden},
  {Gupta}, \& {McDonald}}]{Decker2019}
{Decker}, B., {Brodwin}, M., {Abdulla}, Z., {et~al.} 2019, \apj, 878, 72

\bibitem[{{Di Francesco} {et~al.}(2013){Di Francesco}, {Johnstone}, {Matthews},
  {Bartel}, {Bronfman}, {Casassus}, {Chitsazzadeh}, {Chou}, {Cunningham},
  {Duchene}, {Geisbuesch}, {Hales}, {Ho}, {Houde}, {Iono}, {Kemper}, {Kepley},
  {Koch}, {Kohno}, {Kothes}, {Lai}, {Lin}, {Liu}, {Mason}, {Maccarone},
  {Mizuno}, {Morata}, {Schieven}, {Scaife}, {Scott}, {Shang}, {Shimojo}, {Su},
  {Takakuwa}, {Wagg}, {Wootten}, \& {Yusef-Zadeh}}]{DiFrancesco2013}
{Di Francesco}, J., {Johnstone}, D., {Matthews}, B.~C., {et~al.} 2013, arXiv
  e-prints, arXiv:1310.1604

\bibitem[{{Di Mascolo} {et~al.}(2019{\natexlab{a}}){Di Mascolo}, {Churazov}, \&
  {Mroczkowski}}]{DiMascolo2019a}
{Di Mascolo}, L., {Churazov}, E., \& {Mroczkowski}, T. 2019{\natexlab{a}},
  \mnras, 487, 4037

\bibitem[{{Di Mascolo} {et~al.}(2019{\natexlab{b}}){Di Mascolo}, {Mroczkowski},
  {Churazov}, {Markevitch}, {Basu}, {Clarke}, {Devlin}, {Mason}, {Randall},
  {Reese}, {Sunyaev}, \& {Wik}}]{DiMascolo2019b}
{Di Mascolo}, L., {Mroczkowski}, T., {Churazov}, E., {et~al.}
  2019{\natexlab{b}}, Astronomy and Astrophysics, 628, A100

\bibitem[{{Fazio} {et~al.}(2004){Fazio}, {Hora}, {Allen}, {Ashby}, {Barmby},
  {Deutsch}, {Huang}, {Kleiner}, {Marengo}, {Megeath}, {Melnick}, {Pahre},
  {Patten}, {Polizotti}, {Smith}, {Taylor}, {Wang}, {Willner}, {Hoffmann},
  {Pipher}, {Forrest}, {McMurty}, {McCreight}, {McKelvey}, {McMurray}, {Koch},
  {Moseley}, {Arendt}, {Mentzell}, {Marx}, {Losch}, {Mayman}, {Eichhorn},
  {Krebs}, {Jhabvala}, {Gezari}, {Fixsen}, {Flores}, {Shakoorzadeh}, {Jungo},
  {Hakun}, {Workman}, {Karpati}, {Kichak}, {Whitley}, {Mann}, {Tollestrup},
  {Eisenhardt}, {Stern}, {Gorjian}, {Bhattacharya}, {Carey}, {Nelson},
  {Glaccum}, {Lacy}, {Lowrance}, {Laine}, {Reach}, {Stauffer}, {Surace},
  {Wilson}, {Wright}, {Hoffman}, {Domingo}, \& {Cohen}}]{Fazio2004}
{Fazio}, G.~G., {Hora}, J.~L., {Allen}, L.~E., {et~al.} 2004, \apjs, 154, 10

\bibitem[{{Geach} \& {Peacock}(2017)}]{Geach2017}
{Geach}, J.~E. \& {Peacock}, J.~A. 2017, Nature Astronomy, 1, 795

\bibitem[{{Gobat} {et~al.}(2019){Gobat}, {Daddi}, {Coogan}, {Le Brun},
  {Bournaud}, {Melin}, {Riechers}, {Sargent}, {Valentino}, {Hwang},
  {Finoguenov}, \& {Strazzullo}}]{Gobat2019}
{Gobat}, R., {Daddi}, E., {Coogan}, R.~T., {et~al.} 2019, \aap, 629, A104

\bibitem[{{Gonzalez} {et~al.}(2015){Gonzalez}, {Decker}, {Brodwin},
  {Eisenhardt}, {Marrone}, {Stanford}, {Stern}, {Wylezalek}, {Aldering},
  {Abdulla}, {Boone}, {Carlstrom}, {Fagrelius}, {Gettings}, {Greer}, {Hayden},
  {Leitch}, {Lin}, {Mantz}, {Muchovej}, {Perlmutter}, \&
  {Zeimann}}]{Gonzalez2015}
{Gonzalez}, A.~H., {Decker}, B., {Brodwin}, M., {et~al.} 2015, \apjl, 812, L40

\bibitem[{{Gonzalez} {et~al.}(2019){Gonzalez}, {Gettings}, {Brodwin},
  {Eisenhardt}, {Stanford}, {Wylezalek}, {Decker}, {Marrone}, {Moravec},
  {O'Donnell}, {Stalder}, {Stern}, {Abdulla}, {Brown}, {Carlstrom}, {Chambers},
  {Hayden}, {Lin}, {Magnier}, {Masci}, {Mantz}, {McDonald}, {Mo}, {Perlmutter},
  {Wright}, \& {Zeimann}}]{Gonzalez2019}
{Gonzalez}, A.~H., {Gettings}, D.~P., {Brodwin}, M., {et~al.} 2019, \apjs, 240,
  33

\bibitem[{{Hasselfield} {et~al.}(2013){Hasselfield}, {Hilton}, {Marriage},
  {Addison}, {Barrientos}, {Battaglia}, {Battistelli}, {Bond}, {Crichton},
  {Das}, {Devlin}, {Dicker}, {Dunkley}, {D{\"u}nner}, {Fowler}, {Gralla},
  {Hajian}, {Halpern}, {Hincks}, {Hlozek}, {Hughes}, {Infante}, {Irwin},
  {Kosowsky}, {Marsden}, {Menanteau}, {Moodley}, {Niemack}, {Nolta}, {Page},
  {Partridge}, {Reese}, {Schmitt}, {Sehgal}, {Sherwin}, {Sievers}, {Sif{\'o}n},
  {Spergel}, {Staggs}, {Swetz}, {Switzer}, {Thornton}, {Trac}, \&
  {Wollack}}]{Hasselfield2013}
{Hasselfield}, M., {Hilton}, M., {Marriage}, T.~A., {et~al.} 2013, \jcap, 2013,
  008

\bibitem[{{Hilton} {et~al.}(2018){Hilton}, {Hasselfield}, {Sif{\'o}n},
  {Battaglia}, {Aiola}, {Bharadwaj}, {Bond}, {Choi}, {Crichton}, {Datta},
  {Devlin}, {Dunkley}, {D{\"u}nner}, {Gallardo}, {Gralla}, {Hincks}, {Ho},
  {Hubmayr}, {Huffenberger}, {Hughes}, {Koopman}, {Kosowsky}, {Louis},
  {Madhavacheril}, {Marriage}, {Maurin}, {McMahon}, {Miyatake}, {Moodley},
  {N{\ae}ss}, {Nati}, {Newburgh}, {Niemack}, {Oguri}, {Page}, {Partridge},
  {Schmitt}, {Sievers}, {Spergel}, {Staggs}, {Trac}, {van Engelen},
  {Vavagiakis}, \& {Wollack}}]{Hilton2018}
{Hilton}, M., {Hasselfield}, M., {Sif{\'o}n}, C., {et~al.} 2018, \apjs, 235, 20

\bibitem[{{H{\"o}gbom}(1974)}]{Hoegbom1974}
{H{\"o}gbom}, J.~A. 1974, \aaps, 15, 417

\bibitem[{{Huang} {et~al.}(2020){Huang}, {Bleem}, {Stalder}, {Ade}, {Allen},
  {Anderson}, {Austermann}, {Avva}, {Beall}, {Bender}, {Benson}, {Bianchini},
  {Bocquet}, {Brodwin}, {Carlstrom}, {Chang}, {Chiang}, {Citron}, {Moran},
  {Crawford}, {Crites}, {Haan}, {Dobbs}, {Everett}, {Floyd}, {Gallicchio},
  {George}, {Gilbert}, {Gladders}, {Guns}, {Gupta}, {Halverson}, {Harrington},
  {Henning}, {Hilton}, {Holder}, {Holzapfel}, {Hrubes}, {Hubmayr}, {Irwin},
  {Khullar}, {Knox}, {Lee}, {Li}, {Lowitz}, {McDonald}, {McMahon}, {Meyer},
  {Mocanu}, {Montgomery}, {Nadolski}, {Natoli}, {Nibarger}, {Noble}, {Novosad},
  {Padin}, {Patil}, {Pryke}, {Reichardt}, {Ruhl}, {Saliwanchik}, {Saro},
  {Sayre}, {Schaffer}, {Sharon}, {Sievers}, {Smecher}, {Stark}, {Story},
  {Tucker}, {Vanderlinde}, {Veach}, {Vieira}, {Wang}, {Whitehorn}, {Wu}, \&
  {Yefremenko}}]{Huang2019}
{Huang}, N., {Bleem}, L.~E., {Stalder}, B., {et~al.} 2020, \aj, 159, 110

\bibitem[{{Huang} {et~al.}(2016){Huang}, {Morata}, {Koch}, {Kemper}, {Hwang},
  {Chiong}, {Ho}, {Chu}, {Huang}, {Liu}, {Hsieh}, {Tseng}, {Weng}, {Ho},
  {Chiang}, {Wu}, {Chang}, {Jian}, {Lee}, {Lee}, {Iguchi}, {Asayama}, {Iono},
  {Gonzalez}, {Effland}, {Saini}, {Pospieszalski}, {Henke}, {Yeung}, {Finger},
  {Tapia}, \& {Reyes}}]{Huang2016}
{Huang}, Y. D.~T., {Morata}, O., {Koch}, P.~M., {et~al.} 2016, Society of
  Photo-Optical Instrumentation Engineers (SPIE) Conference Series, Vol. 9911,
  {The Atacama Large Millimeter/sub-millimeter Array band-1 receiver}, 99111V

\bibitem[{{Iguchi} {et~al.}(2009){Iguchi}, {Morita}, {Sugimoto}, {Vilar{\'o}},
  {Saito}, {Hasegawa}, {Kawabe}, {Tatematsu}, {Sakamoto}, {Kiuchi}, {Okumura},
  {Kosugi}, {Inatani}, {Takakuwa}, {Iono}, {Kamazaki}, {Ogasawara}, \&
  {Ishiguro}}]{Iguchi2009}
{Iguchi}, S., {Morita}, K.-I., {Sugimoto}, M., {et~al.} 2009, \pasj, 61, 1

\bibitem[{{Itoh} {et~al.}(1998){Itoh}, {Kohyama}, \& {Nozawa}}]{Itoh1998}
{Itoh}, N., {Kohyama}, Y., \& {Nozawa}, S. 1998, \apj, 502, 7

\bibitem[{{Itoh} \& {Nozawa}(2004)}]{Itoh2004}
{Itoh}, N. \& {Nozawa}, S. 2004, \aap, 417, 827

\bibitem[{Jeffreys(1961)}]{Jeffreys1961}
Jeffreys, H. 1961, Theory of Probability, 3rd edn. (Oxford, England: Oxford)

\bibitem[{{Klaassen} {et~al.}(2019){Klaassen}, {Mroczkowski}, {Bryan},
  {Groppi}, {Basu}, {Cicone}, {Dannerbauer}, {De Breuck}, {Fischer}, {Geach},
  {Hatziminaoglou}, {Holland}, {Kawabe}, {Sehgal}, {Stanke}, \& {van
  Kampen}}]{Klaassen2019}
{Klaassen}, P., {Mroczkowski}, T., {Bryan}, S., {et~al.} 2019, in \baas,
  Vol.~51, 58

\bibitem[{{Lacy} {et~al.}(2020){Lacy}, {Baum}, {Chandler}, {Chatterjee},
  {Clarke}, {Deustua}, {English}, {Farnes}, {Gaensler}, {Gugliucci},
  {Hallinan}, {Kent}, {Kimball}, {Law}, {Lazio}, {Marvil}, {Mao}, {Medlin},
  {Mooley}, {Murphy}, {Myers}, {Osten}, {Richards}, {Rosolowsky}, {Rudnick},
  {Schinzel}, {Sivakoff}, {Sjouwerman}, {Taylor}, {White}, {Wrobel},
  {Andernach}, {Beasley}, {Berger}, {Bhatnager}, {Birkinshaw}, {Bower},
  {Brandt}, {Brown}, {Burke-Spolaor}, {Butler}, {Comerford}, {Demorest}, {Fu},
  {Giacintucci}, {Golap}, {G{\"u}th}, {Hales}, {Hiriart}, {Hodge}, {Horesh},
  {Ivezi{\'c}}, {Jarvis}, {Kamble}, {Kassim}, {Liu}, {Loinard}, {Lyons},
  {Masters}, {Mezcua}, {Moellenbrock}, {Mroczkowski}, {Nyland},
  {O{\textquoteright}Dea}, {O{\textquoteright}Sullivan}, {Peters}, {Radford},
  {Rao}, {Robnett}, {Salcido}, {Shen}, {Sobotka}, {Witz}, {Vaccari}, {van
  Weeren}, {Vargas}, {Williams}, \& {Yoon}}]{Lacy2020}
{Lacy}, M., {Baum}, S.~A., {Chandler}, C.~J., {et~al.} 2020, \pasp, 132, 035001

\bibitem[{{Marriage} {et~al.}(2011){Marriage}, {Acquaviva}, {Ade}, {Aguirre},
  {Amiri}, {Appel}, {Barrientos}, {Battistelli}, {Bond}, {Brown}, {Burger},
  {Chervenak}, {Das}, {Devlin}, {Dicker}, {Bertrand Doriese}, {Dunkley},
  {D{\"u}nner}, {Essinger-Hileman}, {Fisher}, {Fowler}, {Hajian}, {Halpern},
  {Hasselfield}, {Hern{\'a}ndez-Monteagudo}, {Hilton}, {Hilton}, {Hincks},
  {Hlozek}, {Huffenberger}, {Handel Hughes}, {Hughes}, {Infante}, {Irwin},
  {Baptiste Juin}, {Kaul}, {Klein}, {Kosowsky}, {Lau}, {Limon}, {Lin},
  {Lupton}, {Marsden}, {Martocci}, {Mauskopf}, {Menanteau}, {Moodley},
  {Moseley}, {Netterfield}, {Niemack}, {Nolta}, {Page}, {Parker}, {Partridge},
  {Quintana}, {Reese}, {Reid}, {Sehgal}, {Sherwin}, {Sievers}, {Spergel},
  {Staggs}, {Swetz}, {Switzer}, {Thornton}, {Trac}, {Tucker}, {Warne},
  {Wilson}, {Wollack}, \& {Zhao}}]{Marriage2011}
{Marriage}, T.~A., {Acquaviva}, V., {Ade}, P. A.~R., {et~al.} 2011, \apj, 737,
  61

\bibitem[{{McDonald} {et~al.}(2014){McDonald}, {Benson}, {Vikhlinin}, {Aird},
  {Allen}, {Bautz}, {Bayliss}, {Bleem}, {Bocquet}, {Brodwin}, {Carlstrom},
  {Chang}, {Cho}, {Clocchiatti}, {Crawford}, {Crites}, {de Haan}, {Dobbs},
  {Foley}, {Forman}, {George}, {Gladders}, {Gonzalez}, {Halverson},
  {Hlavacek-Larrondo}, {Holder}, {Holzapfel}, {Hrubes}, {Jones}, {Keisler},
  {Knox}, {Lee}, {Leitch}, {Liu}, {Lueker}, {Luong-Van}, {Mantz}, {Marrone},
  {McMahon}, {Meyer}, {Miller}, {Mocanu}, {Mohr}, {Murray}, {Padin}, {Pryke},
  {Reichardt}, {Rest}, {Ruhl}, {Saliwanchik}, {Saro}, {Sayre}, {Schaffer},
  {Shirokoff}, {Spieler}, {Stalder}, {Stanford}, {Staniszewski}, {Stark},
  {Story}, {Stubbs}, {Vanderlinde}, {Vieira}, {Williamson}, {Zahn}, \&
  {Zenteno}}]{McDonald2014}
{McDonald}, M., {Benson}, B.~A., {Vikhlinin}, A., {et~al.} 2014, \apj, 794, 67

\bibitem[{{McMullin} {et~al.}(2007){McMullin}, {Waters}, {Schiebel}, {Young},
  \& {Golap}}]{McMullin2007}
{McMullin}, J.~P., {Waters}, B., {Schiebel}, D., {Young}, W., \& {Golap}, K.
  2007, in Astronomical Society of the Pacific Conference Series, Vol. 376,
  Astronomical Data Analysis Software and Systems XVI, ed. R.~A. {Shaw},
  F.~{Hill}, \& D.~J. {Bell}, 127

\bibitem[{{Moravec} {et~al.}(2019){Moravec}, {Gonzalez}, {Stern}, {Brodwin},
  {Clarke}, {Decker}, {Eisenhardt}, {Mo}, {O'Donnell}, {Pope}, {Stanford}, \&
  {Wylezalek}}]{Moravec2019}
{Moravec}, E., {Gonzalez}, A.~H., {Stern}, D., {et~al.} 2019, \apj, 871, 186

\bibitem[{{Moravec} {et~al.}(2020){Moravec}, {Gonzalez}, {Stern}, {Clarke},
  {Brodwin}, {Decker}, {Eisenhardt}, {Mo}, {Pope}, {Stanford}, \&
  {Wylezalek}}]{Moravec2020}
{Moravec}, E., {Gonzalez}, A.~H., {Stern}, D., {et~al.} 2020, \apj, 888, 74

\bibitem[{{Mroczkowski} {et~al.}(2019){Mroczkowski}, {Nagai}, {Basu}, {Chluba},
  {Sayers}, {Adam}, {Churazov}, {Crites}, {Di Mascolo}, {Eckert},
  {Macias-Perez}, {Mayet}, {Perotto}, {Pointecouteau}, {Romero}, {Ruppin},
  {Scannapieco}, \& {ZuHone}}]{Mroczkowski2019}
{Mroczkowski}, T., {Nagai}, D., {Basu}, K., {et~al.} 2019, \ssr, 215, 17

\bibitem[{{Nagai} {et~al.}(2007){Nagai}, {Kravtsov}, \&
  {Vikhlinin}}]{Nagai2007}
{Nagai}, D., {Kravtsov}, A.~V., \& {Vikhlinin}, A. 2007, \apj, 668, 1

\bibitem[{{Perrott} {et~al.}(2019){Perrott}, {Javid}, {Carvalho}, {Elwood},
  {Hobson}, {Lasenby}, {Olamaie}, \& {Saunders}}]{Perrott2019}
{Perrott}, Y.~C., {Javid}, K., {Carvalho}, P., {et~al.} 2019, \mnras, 486, 2116

\bibitem[{{Planck Collaboration} {et~al.}(2013){Planck Collaboration}, {Ade},
  {Aghanim}, {Arnaud}, {Ashdown}, {Atrio-Barandela}, {Aumont}, {Baccigalupi},
  {Balbi}, {Banday}, {Barreiro}, {Bartlett}, {Battaner}, {Benabed},
  {Beno{\^\i}t}, {Bernard}, {Bersanelli}, {Bhatia}, {Bikmaev}, {Bobin},
  {B{\"o}hringer}, {Bonaldi}, {Bond}, {Borgani}, {Borrill}, {Bouchet},
  {Bourdin}, {Brown}, {Burenin}, {Burigana}, {Cabella}, {Cardoso}, {Carvalho},
  {Castex}, {Catalano}, {Cay{\'o}n}, {Chamballu}, {Chiang}, {Chon},
  {Christensen}, {Churazov}, {Clements}, {Colafrancesco}, {Colombi}, {Colombo},
  {Comis}, {Coulais}, {Crill}, {Cuttaia}, {Da Silva}, {Dahle}, {Danese},
  {Davis}, {de Bernardis}, {de Gasperis}, {de Zotti}, {Delabrouille},
  {D{\'e}mocl{\`e}s}, {D{\'e}sert}, {Diego}, {Dolag}, {Dole}, {Donzelli},
  {Dor{\'e}}, {D{\"o}rl}, {Douspis}, {Dupac}, {Efstathiou}, {En{\ss}lin},
  {Eriksen}, {Finelli}, {Flores-Cacho}, {Forni}, {Fosalba}, {Frailis},
  {Franceschi}, {Frommert}, {Galeotta}, {Ganga}, {G{\'e}nova-Santos}, {Giard},
  {Giraud-H{\'e}raud}, {Gonz{\'a}lez-Nuevo}, {G{\'o}rski}, {Gregorio},
  {Gruppuso}, {Hansen}, {Harrison}, {Hempel}, {Henrot-Versill{\'e}},
  {Hern{\'a}ndez-Monteagudo}, {Herranz}, {Hildebrandt}, {Hivon}, {Hobson},
  {Holmes}, {Hurier}, {Jaffe}, {Jaffe}, {Jagemann}, {Jones}, {Juvela},
  {Keih{\"a}nen}, {Khamitov}, {Kisner}, {Kneissl}, {Knoche}, {Knox}, {Kunz},
  {Kurki-Suonio}, {Lagache}, {L{\"a}hteenm{\"a}ki}, {Lamarre}, {Lasenby},
  {Lawrence}, {Le Jeune}, {Leonardi}, {Liddle}, {Lilje}, {L{\'o}pez-Caniego},
  {Luzzi}, {Mac{\'\i}as-P{\'e}rez}, {Maino}, {Mandolesi}, {Maris}, {Marleau},
  {Marshall}, {Mart{\'\i}nez-Gonz{\'a}lez}, {Masi}, {Massardi}, {Matarrese},
  {Mazzotta}, {Mei}, {Melchiorri}, {Melin}, {Mendes}, {Mennella}, {Mitra},
  {Miville-Desch{\^e}nes}, {Moneti}, {Montier}, {Morgante}, {Mortlock},
  {Munshi}, {Murphy}, {Naselsky}, {Nati}, {Natoli}, {N{\o}rgaard-Nielsen},
  {Noviello}, {Novikov}, {Novikov}, {Osborne}, {Pajot}, {Paoletti}, {Pasian},
  {Patanchon}, {Perdereau}, {Perotto}, {Perrotta}, {Piacentini}, {Piat},
  {Pierpaoli}, {Piffaretti}, {Plaszczynski}, {Pointecouteau}, {Polenta},
  {Ponthieu}, {Popa}, {Poutanen}, {Pratt}, {Prunet}, {Puget}, {Rachen},
  {Reach}, {Rebolo}, {Reinecke}, {Remazeilles}, {Renault}, {Ricciardi},
  {Riller}, {Ristorcelli}, {Rocha}, {Roman}, {Rosset}, {Rossetti},
  {Rubi{\~n}o-Mart{\'\i}n}, {Rusholme}, {Sandri}, {Savini}, {Scott}, {Smoot},
  {Starck}, {Sudiwala}, {Sunyaev}, {Sutton}, {Suur-Uski}, {Sygnet}, {Tauber},
  {Terenzi}, {Toffolatti}, {Tomasi}, {Tristram}, {Tuovinen}, {Valenziano}, {Van
  Tent}, {Varis}, {Vielva}, {Villa}, {Vittorio}, {Wade}, {Wandelt}, {Welikala},
  {White}, {White}, {Yvon}, {Zacchei}, \& {Zonca}}]{Planck2013V}
{Planck Collaboration}, {Ade}, P.~A.~R., {Aghanim}, N., {et~al.} 2013, \aap,
  550, A131

\bibitem[{{Planck Collaboration} {et~al.}(2016{\natexlab{a}}){Planck
  Collaboration}, {Ade}, {Aghanim}, {Arnaud}, {Ashdown}, {Aumont},
  {Baccigalupi}, {Banday}, {Barreiro}, {Barrena}, {Bartlett}, {Bartolo},
  {Battaner}, {Battye}, {Benabed}, {Beno{\^\i}t}, {Benoit-L{\'e}vy}, {Bernard},
  {Bersanelli}, {Bielewicz}, {Bikmaev}, {B{\"o}hringer}, {Bonaldi}, {Bonavera},
  {Bond}, {Borrill}, {Bouchet}, {Bucher}, {Burenin}, {Burigana}, {Butler},
  {Calabrese}, {Cardoso}, {Carvalho}, {Catalano}, {Challinor}, {Chamballu},
  {Chary}, {Chiang}, {Chon}, {Christensen}, {Clements}, {Colombi}, {Colombo},
  {Combet}, {Comis}, {Couchot}, {Coulais}, {Crill}, {Curto}, {Cuttaia},
  {Dahle}, {Danese}, {Davies}, {Davis}, {de Bernardis}, {de Rosa}, {de Zotti},
  {Delabrouille}, {D{\'e}sert}, {Dickinson}, {Diego}, {Dolag}, {Dole},
  {Donzelli}, {Dor{\'e}}, {Douspis}, {Ducout}, {Dupac}, {Efstathiou},
  {Eisenhardt}, {Elsner}, {En{\ss}lin}, {Eriksen}, {Falgarone}, {Fergusson},
  {Feroz}, {Ferragamo}, {Finelli}, {Forni}, {Frailis}, {Fraisse}, {Franceschi},
  {Frejsel}, {Galeotta}, {Galli}, {Ganga}, {G{\'e}nova-Santos}, {Giard},
  {Giraud-H{\'e}raud}, {Gjerl{\o}w}, {Gonz{\'a}lez-Nuevo}, {G{\'o}rski},
  {Grainge}, {Gratton}, {Gregorio}, {Gruppuso}, {Gudmundsson}, {Hansen},
  {Hanson}, {Harrison}, {Hempel}, {Henrot-Versill{\'e}},
  {Hern{\'a}ndez-Monteagudo}, {Herranz}, {Hildebrandt}, {Hivon}, {Hobson},
  {Holmes}, {Hornstrup}, {Hovest}, {Huffenberger}, {Hurier}, {Jaffe}, {Jaffe},
  {Jin}, {Jones}, {Juvela}, {Keih{\"a}nen}, {Keskitalo}, {Khamitov}, {Kisner},
  {Kneissl}, {Knoche}, {Kunz}, {Kurki-Suonio}, {Lagache}, {Lamarre}, {Lasenby},
  {Lattanzi}, {Lawrence}, {Leonardi}, {Lesgourgues}, {Levrier}, {Liguori},
  {Lilje}, {Linden-V{\o}rnle}, {L{\'o}pez-Caniego}, {Lubin},
  {Mac{\'\i}as-P{\'e}rez}, {Maggio}, {Maino}, {Mak}, {Mandolesi}, {Mangilli},
  {Martin}, {Mart{\'\i}nez-Gonz{\'a}lez}, {Masi}, {Matarrese}, {Mazzotta},
  {McGehee}, {Mei}, {Melchiorri}, {Melin}, {Mendes}, {Mennella}, {Migliaccio},
  {Mitra}, {Miville-Desch{\^e}nes}, {Moneti}, {Montier}, {Morgante},
  {Mortlock}, {Moss}, {Munshi}, {Murphy}, {Naselsky}, {Nastasi}, {Nati},
  {Natoli}, {Netterfield}, {N{\o}rgaard-Nielsen}, {Noviello}, {Novikov},
  {Novikov}, {Olamaie}, {Oxborrow}, {Paci}, {Pagano}, {Pajot}, {Paoletti},
  {Pasian}, {Patanchon}, {Pearson}, {Perdereau}, {Perotto}, {Perrott},
  {Perrotta}, {Pettorino}, {Piacentini}, {Piat}, {Pierpaoli}, {Pietrobon},
  {Plaszczynski}, {Pointecouteau}, {Polenta}, {Pratt}, {Pr{\'e}zeau}, {Prunet},
  {Puget}, {Rachen}, {Reach}, {Rebolo}, {Reinecke}, {Remazeilles}, {Renault},
  {Renzi}, {Ristorcelli}, {Rocha}, {Rosset}, {Rossetti}, {Roudier}, {Rozo},
  {Rubi{\~n}o-Mart{\'\i}n}, {Rumsey}, {Rusholme}, {Rykoff}, {Sandri}, {Santos},
  {Saunders}, {Savelainen}, {Savini}, {Schammel}, {Scott}, {Seiffert},
  {Shellard}, {Shimwell}, {Spencer}, {Stanford}, {Stern}, {Stolyarov},
  {Stompor}, {Streblyanska}, {Sudiwala}, {Sunyaev}, {Sutton}, {Suur-Uski},
  {Sygnet}, {Tauber}, {Terenzi}, {Toffolatti}, {Tomasi}, {Tramonte},
  {Tristram}, {Tucci}, {Tuovinen}, {Umana}, {Valenziano}, {Valiviita}, {Van
  Tent}, {Vielva}, {Villa}, {Wade}, {Wandelt}, {Wehus}, {White}, {Wright},
  {Yvon}, {Zacchei}, \& {Zonca}}]{Planck2016XXVII}
{Planck Collaboration}, {Ade}, P.~A.~R., {Aghanim}, N., {et~al.}
  2016{\natexlab{a}}, \aap, 594, A27

\bibitem[{{Planck Collaboration} {et~al.}(2016{\natexlab{b}}){Planck
  Collaboration}, {Aghanim}, {Arnaud}, {Ashdown}, {Aumont}, {Baccigalupi},
  {Band ay}, {Barreiro}, {Bartlett}, {Bartolo}, {Battaner}, {Battye},
  {Benabed}, {Beno{\^\i}t}, {Benoit-L{\'e}vy}, {Bernard}, {Bersanelli},
  {Bielewicz}, {Bock}, {Bonaldi}, {Bonavera}, {Bond}, {Borrill}, {Bouchet},
  {Burigana}, {Butler}, {Calabrese}, {Cardoso}, {Catalano}, {Challinor},
  {Chiang}, {Christensen}, {Churazov}, {Clements}, {Colombo}, {Combet},
  {Comis}, {Coulais}, {Crill}, {Curto}, {Cuttaia}, {Danese}, {Davies}, {Davis},
  {de Bernardis}, {de Rosa}, {de Zotti}, {Delabrouille}, {D{\'e}sert},
  {Dickinson}, {Diego}, {Dolag}, {Dole}, {Donzelli}, {Dor{\'e}}, {Douspis},
  {Ducout}, {Dupac}, {Efstathiou}, {Elsner}, {En{\ss}lin}, {Eriksen},
  {Fergusson}, {Finelli}, {Forni}, {Frailis}, {Fraisse}, {Franceschi},
  {Frejsel}, {Galeotta}, {Galli}, {Ganga}, {G{\'e}nova-Santos}, {Giard},
  {Gonz{\'a}lez-Nuevo}, {G{\'o}rski}, {Gregorio}, {Gruppuso}, {Gudmundsson},
  {Hansen}, {Harrison}, {Henrot-Versill{\'e}}, {Hern{\'a}ndez-Monteagudo},
  {Herranz}, {Hildebrand t}, {Hivon}, {Holmes}, {Hornstrup}, {Huffenberger},
  {Hurier}, {Jaffe}, {Jones}, {Juvela}, {Keih{\"a}nen}, {Keskitalo}, {Kneissl},
  {Knoche}, {Kunz}, {Kurki-Suonio}, {Lacasa}, {Lagache}, {L{\"a}hteenm{\"a}ki},
  {Lamarre}, {Lasenby}, {Lattanzi}, {Leonardi}, {Lesgourgues}, {Levrier},
  {Liguori}, {Lilje}, {Linden-V{\o}rnle}, {L{\'o}pez-Caniego},
  {Mac{\'\i}as-P{\'e}rez}, {Maffei}, {Maggio}, {Maino}, {Mandolesi},
  {Mangilli}, {Maris}, {Martin}, {Mart{\'\i}nez-Gonz{\'a}lez}, {Masi},
  {Matarrese}, {Melchiorri}, {Melin}, {Migliaccio}, {Miville-Desch{\^e}nes},
  {Moneti}, {Montier}, {Morgante}, {Mortlock}, {Munshi}, {Murphy}, {Naselsky},
  {Nati}, {Natoli}, {Noviello}, {Novikov}, {Novikov}, {Paci}, {Pagano},
  {Pajot}, {Paoletti}, {Pasian}, {Patanchon}, {Perdereau}, {Perotto},
  {Pettorino}, {Piacentini}, {Piat}, {Pierpaoli}, {Pietrobon}, {Plaszczynski},
  {Pointecouteau}, {Polenta}, {Ponthieu}, {Pratt}, {Prunet}, {Puget}, {Rachen},
  {Reinecke}, {Remazeilles}, {Renault}, {Renzi}, {Ristorcelli}, {Rocha},
  {Rossetti}, {Roudier}, {Rubi{\~n}o-Mart{\'\i}n}, {Rusholme}, {Sandri},
  {Santos}, {Sauv{\'e}}, {Savelainen}, {Savini}, {Scott}, {Spencer},
  {Stolyarov}, {Stompor}, {Sunyaev}, {Sutton}, {Suur-Uski}, {Sygnet}, {Tauber},
  {Terenzi}, {Toffolatti}, {Tomasi}, {Tramonte}, {Tristram}, {Tucci},
  {Tuovinen}, {Valenziano}, {Valiviita}, {Van Tent}, {Vielva}, {Villa}, {Wade},
  {Wandelt}, {Wehus}, {Yvon}, {Zacchei}, \& {Zonca}}]{Planck2015_XXII}
{Planck Collaboration}, {Aghanim}, N., {Arnaud}, M., {et~al.}
  2016{\natexlab{b}}, \aap, 594, A22

\bibitem[{{Rettura} {et~al.}(2018){Rettura}, {Chary}, {Krick}, \&
  {Ettori}}]{Rettura2018}
{Rettura}, A., {Chary}, R., {Krick}, J., \& {Ettori}, S. 2018, \apj, 867, 12

\bibitem[{{Ruppin} {et~al.}(2019{\natexlab{a}}){Ruppin}, {Mayet},
  {Mac{\'\i}as-P{\'e}rez}, \& {Perotto}}]{Ruppin2019b}
{Ruppin}, F., {Mayet}, F., {Mac{\'\i}as-P{\'e}rez}, J.~F., \& {Perotto}, L.
  2019{\natexlab{a}}, \mnras, 490, 784

\bibitem[{{Ruppin} {et~al.}(2020){Ruppin}, {McDonald}, {Brodwin}, {Adam},
  {Ade}, {Andr{\'e}}, {Andrianasolo}, {Arnaud}, {Aussel}, {Bartalucci},
  {Bautz}, {Beelen}, {Beno{\^\i}t}, {Bideaud}, {Bourrion}, {Calvo}, {Catalano},
  {Comis}, {Decker}, {De Petris}, {D{\'e}sert}, {Doyle}, {Driessen},
  {Eisenhardt}, {Gomez}, {Gonzalez}, {Goupy}, {K{\'e}ruzor{\'e}}, {Kramer},
  {Ladjelate}, {Lagache}, {Leclercq}, {Lestrade}, {Mac{\'\i}as-P{\'e}rez},
  {Mauskopf}, {Mayet}, {Monfardini}, {Moravec}, {Perotto}, {Pisano},
  {Pointecouteau}, {Ponthieu}, {Pratt}, {Rev{\'e}ret}, {Ritacco}, {Romero},
  {Roussel}, {Schuster}, {Shu}, {Sievers}, {Stanford}, {Stern}, {Tucker}, \&
  {Zylka}}]{Ruppin2020}
{Ruppin}, F., {McDonald}, M., {Brodwin}, M., {et~al.} 2020, \apj, 893, 74

\bibitem[{{Ruppin} {et~al.}(2019{\natexlab{b}}){Ruppin}, {Sembolini}, {De
  Petris}, {Adam}, {Cialone}, {Mac{\'\i}as-P{\'e}rez}, {Mayet}, {Perotto}, \&
  {Yepes}}]{Ruppin2019c}
{Ruppin}, F., {Sembolini}, F., {De Petris}, M., {et~al.} 2019{\natexlab{b}},
  \aap, 631, A21

\bibitem[{{Rykoff} {et~al.}(2012){Rykoff}, {Koester}, {Rozo}, {Annis},
  {Evrard}, {Hansen}, {Hao}, {Johnston}, {McKay}, \& {Wechsler}}]{Rykoff2012}
{Rykoff}, E.~S., {Koester}, B.~P., {Rozo}, E., {et~al.} 2012, \apj, 746, 178

\bibitem[{{Saro} {et~al.}(2015){Saro}, {Bocquet}, {Rozo}, {Benson}, {Mohr},
  {Rykoff}, {Soares-Santos}, {Bleem}, {Dodelson}, {Melchior}, {Sobreira},
  {Upadhyay}, {Weller}, {Abbott}, {Abdalla}, {Allam}, {Armstrong}, {Banerji},
  {Bauer}, {Bayliss}, {Benoit-L{\'e}vy}, {Bernstein}, {Bertin}, {Brodwin},
  {Brooks}, {Buckley-Geer}, {Burke}, {Carlstrom}, {Capasso}, {Capozzi},
  {Carnero Rosell}, {Carrasco Kind}, {Chiu}, {Covarrubias}, {Crawford},
  {Crocce}, {D'Andrea}, {da Costa}, {DePoy}, {Desai}, {de Haan}, {Diehl},
  {Dietrich}, {Doel}, {Cunha}, {Eifler}, {Evrard}, {Fausti Neto}, {Fernand ez},
  {Flaugher}, {Fosalba}, {Frieman}, {Gangkofner}, {Gaztanaga}, {Gerdes},
  {Gruen}, {Gruendl}, {Gupta}, {Hennig}, {Holzapfel}, {Honscheid}, {Jain},
  {James}, {Kuehn}, {Kuropatkin}, {Lahav}, {Li}, {Lin}, {Maia}, {March},
  {Marshall}, {Martini}, {McDonald}, {Miller}, {Miquel}, {Nord}, {Ogando},
  {Plazas}, {Reichardt}, {Romer}, {Roodman}, {Sako}, {Sanchez}, {Schubnell},
  {Sevilla}, {Smith}, {Stalder}, {Stark}, {Strazzullo}, {Suchyta}, {Swanson},
  {Tarle}, {Thaler}, {Thomas}, {Tucker}, {Vikram}, {von der Linden}, {Walker},
  {Wechsler}, {Wester}, {Zenteno}, \& {Ziegler}}]{Saro2015}
{Saro}, A., {Bocquet}, S., {Rozo}, E., {et~al.} 2015, \mnras, 454, 2305

\bibitem[{{Sazonov} \& {Sunyaev}(1998)}]{Sazonov1998}
{Sazonov}, S.~Y. \& {Sunyaev}, R.~A. 1998, Astronomy Letters, 24, 553

\bibitem[{{Sembolini} {et~al.}(2014){Sembolini}, {De Petris}, {Yepes},
  {Foschi}, {Lamagna}, \& {Gottl{\"o}ber}}]{Sembolini2014}
{Sembolini}, F., {De Petris}, M., {Yepes}, G., {et~al.} 2014, \mnras, 440, 3520

\bibitem[{{Shi} {et~al.}(2015){Shi}, {Komatsu}, {Nelson}, \& {Nagai}}]{Shi2015}
{Shi}, X., {Komatsu}, E., {Nelson}, K., \& {Nagai}, D. 2015, \mnras, 448, 1020

\bibitem[{{Sunyaev} \& {Zeldovich}(1980)}]{Sunyaev1980}
{Sunyaev}, R.~A. \& {Zeldovich}, I.~B. 1980, \mnras, 190, 413

\bibitem[{{Sunyaev} \& {Zeldovich}(1972)}]{Sunyaev1972}
{Sunyaev}, R.~A. \& {Zeldovich}, Y.~B. 1972, Comments on Astrophysics and Space
  Physics, 4, 173

\bibitem[{{Thompson} {et~al.}(1986){Thompson}, {Moran}, \&
  {Swenson}}]{Thompson1986}
{Thompson}, A.~R., {Moran}, J.~M., \& {Swenson}, G.~W. 1986, {Interferometry
  and synthesis in radio astronomy}

\bibitem[{{Trotta}(2008)}]{Trotta2008}
{Trotta}, R. 2008, Contemporary Physics, 49, 71

\bibitem[{{van Velzen} {et~al.}(2015){van Velzen}, {Falcke}, \&
  {K{\"o}rding}}]{vanVelzen2015}
{van Velzen}, S., {Falcke}, H., \& {K{\"o}rding}, E. 2015, \mnras, 446, 2985

\bibitem[{Wallis(2014)}]{Wallis2014}
Wallis, K.~F. 2014, Statist. Sci., 29, 106

\bibitem[{{Wik} {et~al.}(2008){Wik}, {Sarazin}, {Ricker}, \& {Rand
  all}}]{Wik2008}
{Wik}, D.~R., {Sarazin}, C.~L., {Ricker}, P.~M., \& {Rand all}, S.~W. 2008,
  \apj, 680, 17

\bibitem[{{Wright} {et~al.}(2010){Wright}, {Eisenhardt}, {Mainzer}, {Ressler},
  {Cutri}, {Jarrett}, {Kirkpatrick}, {Padgett}, {McMillan}, {Skrutskie},
  {Stanford}, {Cohen}, {Walker}, {Mather}, {Leisawitz}, {Gautier}, {McLean},
  {Benford}, {Lonsdale}, {Blain}, {Mendez}, {Irace}, {Duval}, {Liu}, {Royer},
  {Heinrichsen}, {Howard}, {Shannon}, {Kendall}, {Walsh}, {Larsen}, {Cardon},
  {Schick}, {Schwalm}, {Abid}, {Fabinsky}, {Naes}, \& {Tsai}}]{Wright2010}
{Wright}, E.~L., {Eisenhardt}, P. R.~M., {Mainzer}, A.~K., {et~al.} 2010, \aj,
  140, 1868

\bibitem[{{Wylezalek} {et~al.}(2013){Wylezalek}, {Galametz}, {Stern}, {Vernet},
  {De Breuck}, {Seymour}, {Brodwin}, {Eisenhardt}, {Gonzalez}, {Hatch},
  {Jarvis}, {Rettura}, {Stanford}, \& {Stevens}}]{Wylezalek2013}
{Wylezalek}, D., {Galametz}, A., {Stern}, D., {et~al.} 2013, \apj, 769, 79

\bibitem[{{Wylezalek} {et~al.}(2014){Wylezalek}, {Vernet}, {De Breuck},
  {Stern}, {Brodwin}, {Galametz}, {Gonzalez}, {Jarvis}, {Hatch}, {Seymour}, \&
  {Stanford}}]{Wylezalek2014}
{Wylezalek}, D., {Vernet}, J., {De Breuck}, C., {et~al.} 2014, \apj, 786, 17

\bibitem[{{Yagoubov} {et~al.}(2020){Yagoubov}, {Mroczkowski}, {Belitsky},
  {Cuadrado-Calle}, {Cuttaia}, {Fuller}, {Gallego}, {Gonzalez}, {Kaneko},
  {Mena}, {Molina}, {Nesti}, {Tapia}, {Villa}, {Beltr{\'a}n}, {Cavaliere},
  {Ceru}, {Chesmore}, {Coughlin}, {De Breuck}, {Fredrixon}, {George}, {Gibson},
  {Golec}, {Josaitis}, {Kemper}, {Kotiranta}, {Lapkin},
  {L{\'o}pez-Fern{\'a}ndez}, {Marconi}, {Mariotti}, {McGenn}, {McMahon},
  {Murk}, {Pezzotta}, {Phillips}, {Reyes}, {Ricciardi}, {Sandri}, {Strandberg},
  {Terenzi}, {Testi}, {Thomas}, {Uzawa}, {Vigan{\`o}}, \&
  {Wadefalk}}]{Yagoubov2020}
{Yagoubov}, P., {Mroczkowski}, T., {Belitsky}, V., {et~al.} 2020, \aap, 634,
  A46

\bibitem[{{Yu} {et~al.}(2015){Yu}, {Nelson}, \& {Nagai}}]{Yu2015}
{Yu}, L., {Nelson}, K., \& {Nagai}, D. 2015, \apj, 807, 12

\end{thebibliography}

\appendix
\section{Mass estimates}
Table~\ref{tab:app:masses} reports the inferred cluster masses when employing the different gNFW models proposed in \citet{Arnaud2010}, \citet{Planck2013V}, and \citet{McDonald2014}. A summary of the profile parameters is provided in Table~\ref{tab:gnfw}.

\begin{sidewaystable*}
    \centering
    \begin{tabular}{ccccccccccccc}
        \hline\hline
        \noalign{\smallskip}
        Cluster ID & $z_{\mathrm{phot}}$ & $\lambda$ &       $M_{500}$       & $\sigma_{\mathrm{eff}}$ &       $M_{500}$       & $\sigma_{\mathrm{eff}}$ &       $M_{500}$       & $\sigma_{\mathrm{eff}}$ &       $M_{500}$       & $\sigma_{\mathrm{eff}}$ &       $M_{500}$       & $\sigma_{\mathrm{eff}}$\\
                   &          --         &       --       & $(10^{14}~M_{\odot})$ &           --            & $(10^{14}~M_{\odot})$ &           --            & $(10^{14}~M_{\odot})$ &           --            & $(10^{14}~M_{\odot})$ &           --            & $(10^{14}~M_{\odot})$ &           --           \\
        \noalign{\vspace{1pt}}
        \hline
        \noalign{\smallskip}
        &&&\multicolumn{2}{c}{universal}&\multicolumn{2}{c}{cool-core}&\multicolumn{2}{c}{disturbed}&\multicolumn{2}{c}{\textit{Planck}}&\multicolumn{2}{c}{\citealt{McDonald2014}}\\
        \noalign{\smallskip}
        \hline
        \noalign{\smallskip}
        \multicolumn{13}{l}{\textit{Significant detection}}\\\noalign{\vspace{1pt}}
        \object{MOO~J0129$-$1640} & $1.05\substack{+0.04\\-0.05}$ & $49\pm7$ & $2.57\substack{+0.30\\-0.30}$ & $7.77$ & $2.09\substack{+0.24\\-0.29}$ & $8.07$ & $3.47\substack{+0.32\\-0.40}$ & $7.63$ & $2.75\substack{+0.32\\-0.38}$ & $7.85$ & $3.72\substack{+1.28\\-1.11}$ & $7.12$ \\\noalign{\vspace{1pt}}
        \object{MOO~J0345$-$2913} & $1.08\substack{+0.03\\-0.04}$ & $53\pm7$ & $1.78\substack{+0.20\\-0.29}$ & $5.32$ & $1.51\substack{+0.21\\-0.24}$ & $5.71$ & $2.09\substack{+0.24\\-0.38}$ & $5.35$ & $1.66\substack{+0.31\\-0.31}$ & $5.58$ & $2.14\substack{+0.94\\-0.74}$ & $5.24$ \\\noalign{\vspace{1pt}}
        \object{MOO~J0917$-$0700} & $1.10\substack{+0.05\\-0.05}$ & $58\pm7$ & $1.66\substack{+0.31\\-0.38}$ & $4.26$ & $1.48\substack{+0.22\\-0.32}$ & $4.49$ & $1.93\substack{+0.49\\-0.67}$ & $3.68$ & $1.55\substack{+0.38\\-0.45}$ & $4.20$ & $1.77\substack{+0.85\\-0.93}$ & $3.41$  \\\noalign{\vspace{1pt}}
                         &                               &          & $2.13\substack{+0.40\\-0.49}$ &        & $1.83\substack{+0.26\\-0.38}$ &        & $2.67\substack{+0.56\\-0.84}$ &        & $2.13\substack{+0.49\\-0.60}$ &        & $2.67\substack{+1.39\\-1.47}$ &  \\\noalign{\vspace{1pt}}
        \object{MOO~J1139$-$1706} & $1.31\substack{+0.03\\-0.05}$ & $53\pm7$ & $2.24\substack{+0.36\\-0.52}$ & $3.81$ & $1.74\substack{+0.24\\-0.36}$ & $3.82$ & $3.16\substack{+0.66\\-0.87}$ & $3.40$ & $2.24\substack{+0.62\\-0.57}$ & $3.78$ & $6.76\substack{+3.89\\-3.42}$ & $4.35$ \\\noalign{\vspace{1pt}}
        \object{MOO~J1342$-$1913} & $1.08\substack{+0.04\\-0.05}$ & $41\pm6$ & $1.95\substack{+0.31\\-0.31}$ & $4.53$ & $1.58\substack{+0.22\\-0.22}$ & $5.42$ & $2.29\substack{+0.37\\-0.37}$ & $5.09$ & $1.91\substack{+0.35\\-0.39}$ & $5.27$ & $2.40\substack{+0.94\\-0.83}$ & $4.25$ \\\noalign{\vspace{1pt}}
        \object{MOO~J1414$+$0227} & $1.02\substack{+0.07\\-0.06}$ & $41\pm7$ & $2.75\substack{+0.32\\-0.32}$ & $6.99$ & $2.24\substack{+0.31\\-0.26}$ & $6.80$ & $3.55\substack{+0.74\\-0.41}$ & $6.98$ & $2.88\substack{+0.66\\-0.53}$ & $7.04$ & $4.37\substack{+1.91\\-1.81}$ & $6.94$ \\\noalign{\vspace{1pt}}
        \object{MOO~J2146$-$0320} & $1.16\substack{+0.05\\-0.05}$ & $50\pm7$ & $1.86\substack{+0.34\\-0.52}$ & $5.35$ & $1.58\substack{+0.27\\-0.34}$ & $5.55$ & $2.58\substack{+0.56\\-0.66}$ & $5.73$ & $1.95\substack{+0.47\\-0.61}$ & $5.62$ & $5.13\substack{+4.57\\-3.29}$ & $5.94$ \\\noalign{\vspace{1pt}}
                         &                               &          & $2.04\substack{+0.42\\-0.56}$ &        & $1.67\substack{+0.31\\-0.40}$ &        & $3.49\substack{+0.78\\-1.14}$ &        & $2.29\substack{+0.69\\-0.79}$ &        & $4.44\substack{+2.52\\-3.39}$ & \\\noalign{\medskip}
        \multicolumn{13}{l}{\textit{Non-detection}}\\\noalign{\vspace{1pt}}
        \object{MOO~J0903$+$1310} & $1.26\substack{+0.05\\-0.08}$ & $29\pm5$ & $0.30^{+0.18}$ & -- & $0.29\substack{+0.17\\-0.20}$ & $0.83$ & $0.27\substack{+0.16\\-0.16}$ & $0.88$ & $0.27\substack{+0.18\\-0.18}$ & $0.61$ & $0.32^{+0.22}$ &  --  \\\noalign{\vspace{1pt}}
        \object{MOO~J1223$+$2420} & $1.09\substack{+0.04\\-0.04}$ & $51\pm7$ & $1.17\substack{+0.27\\-0.38}$ & $2.40$ & $0.94\substack{+0.21\\-0.35}$ & $2.54$ & $1.33\substack{+0.34\\-0.61}$ & $2.54$ & $0.90\substack{+0.32\\-0.44}$ & $2.53$ & $1.52\substack{+0.86\\-0.92}$ & $2.23$ \\\noalign{\vspace{1pt}}
        \object{MOO~J2147$+$1314} & $1.01\substack{+0.06\\-0.07}$ & $38\pm6$ & $1.82\substack{+0.34\\-0.42}$ & $1.26$ & $1.56\substack{+0.26\\-0.38}$ & $1.34$ & $1.62\substack{+0.43\\-0.52}$ & $1.59$ & $2.06\substack{+0.51\\-0.50}$ & $1.23$ & $2.27\substack{+1.10\\-1.26}$ & $1.35$ 
        \\\noalign{\vspace{1pt}}
        \noalign{\smallskip}
        \hline
        \noalign{\medskip}
    \end{tabular}
    \caption{Estimated masses of the VACA LoCA sample clusters. See Sect.~\ref{sec:res} for more details about the effective significance estimate $\sigma_{\mathrm{eff}}$. The two mass values provided for \object{MOO~J0917$-$0700} and \object{MOO~J2146$-$0320} correspond to the masses of each of the individual SZ components detected.}
    \label{tab:app:masses}
\end{sidewaystable*}

\section{$uv$ plots and dirty images}
\begin{figure*}
    \centering
    \includegraphics[clip,trim=0 0mm 0mm 0mm,width=\textwidth]{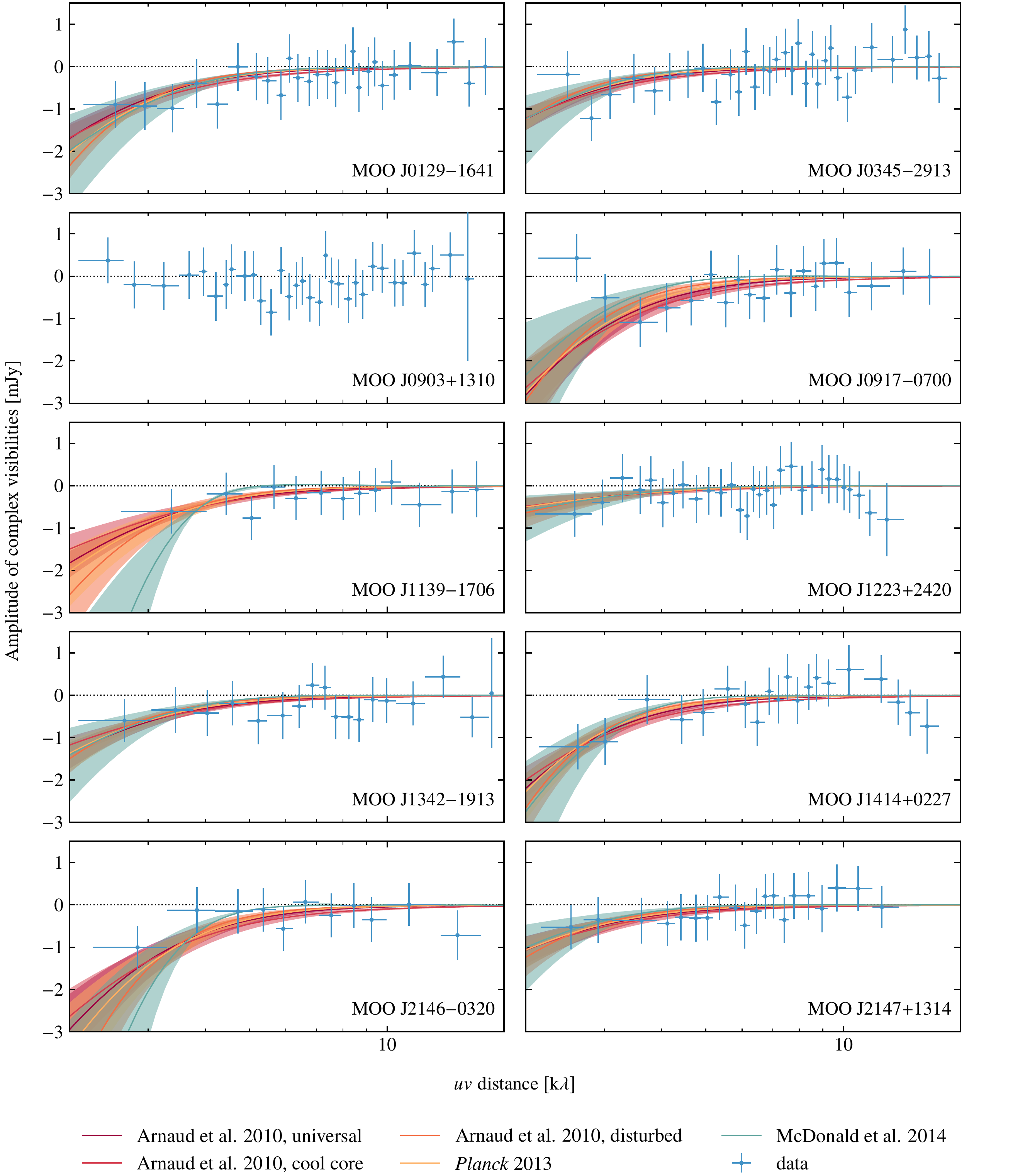}
    \caption{Comparison of the $uv$ profiles of all the gNFW flavors adopted in our work. The data are binned so that each bin contains the same number of visibilities (here set to 2500 for plotting  purposes). Before being averaged,  the phase center was shifted to the position of cluster centroid to minimize the ringing effect due to non-zero phases. A  model profile was not plotted for \object{MOO~J0903$+$1310} as the SZ signal is not detected.}
    \label{fig:app:uvprof}
\end{figure*}

We  provide here the $uv$ radial plots of the data and respective gNFW models for all the VACA LoCA clusters (Fig.~\ref{fig:app:uvprof}). As discussed in Sect.~\ref{sec:res}, all the model profiles show a fairly good agreement with data over the range of probed angular scales, while being affected by a large scatter at short baselines due to the lack of large-scale information. \object{MOO~J0345$-$2913} and \object{MOO~J0917$-$0700} clearly manifest positive modes on small $uv$ scales, while the data points for \object{MOO~J2146$-$0320} are positively offset with respect to the models. These may arise due to off-center SZ components unaccounted for by the model, or due to extended (positive) emission. However, in the case of \object{MOO~J0345$-$2913} the discrepancy is on the order of 1$\sigma$. On the other hand, the SZ signal from both \object{MOO~J0917$-$0700} and \object{MOO~J2146$-$0320} show a complex structure (see Sect.~\ref{sec:res:multi}), and deviations from a gNFW model are not unexpected.

\begin{figure*}
    \centering
    \includegraphics[clip,trim=0 16mm 5mm 26mm,width=\textwidth]{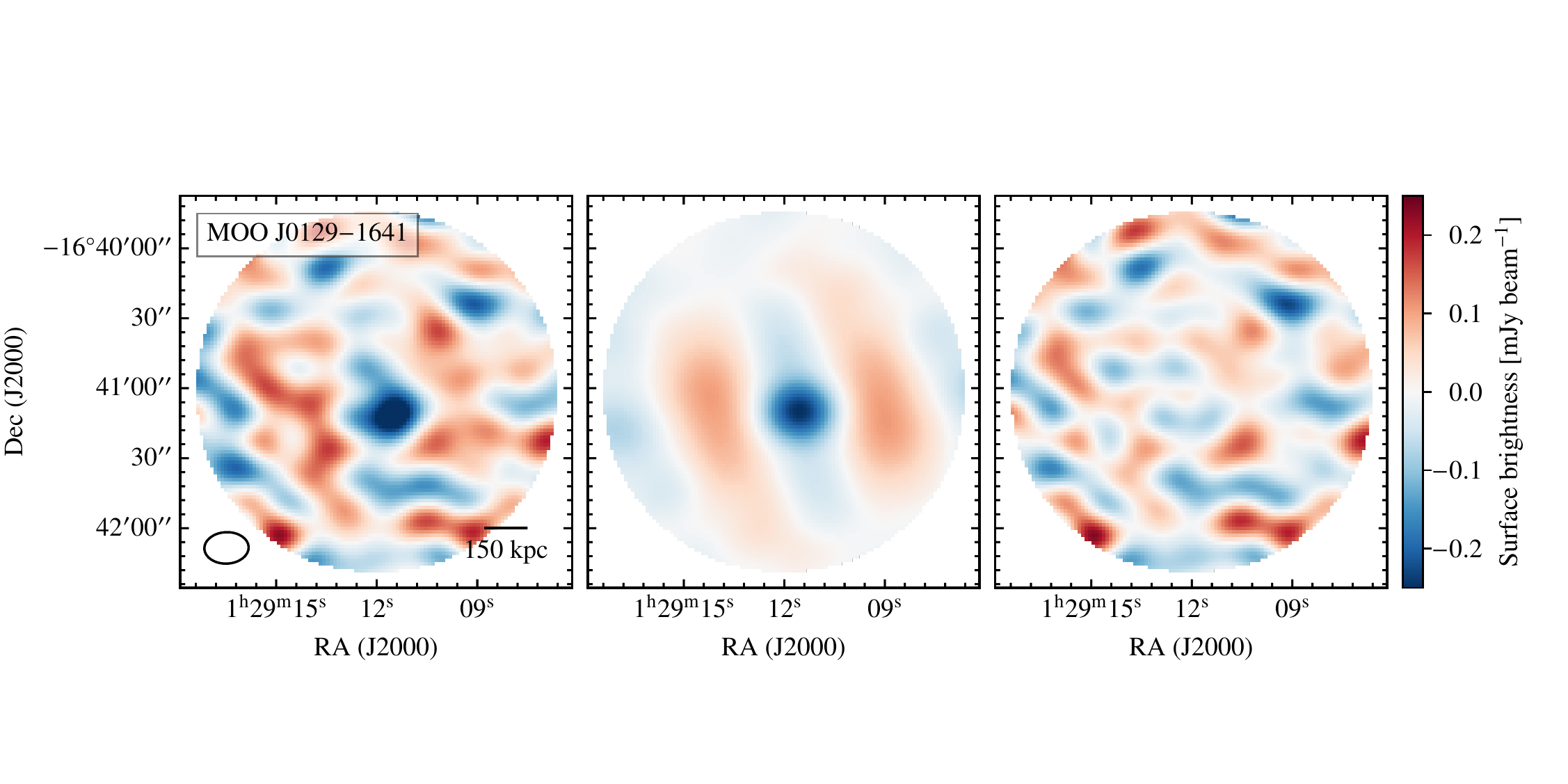}\\
    \includegraphics[clip,trim=0 16mm 5mm 26mm,width=\textwidth]{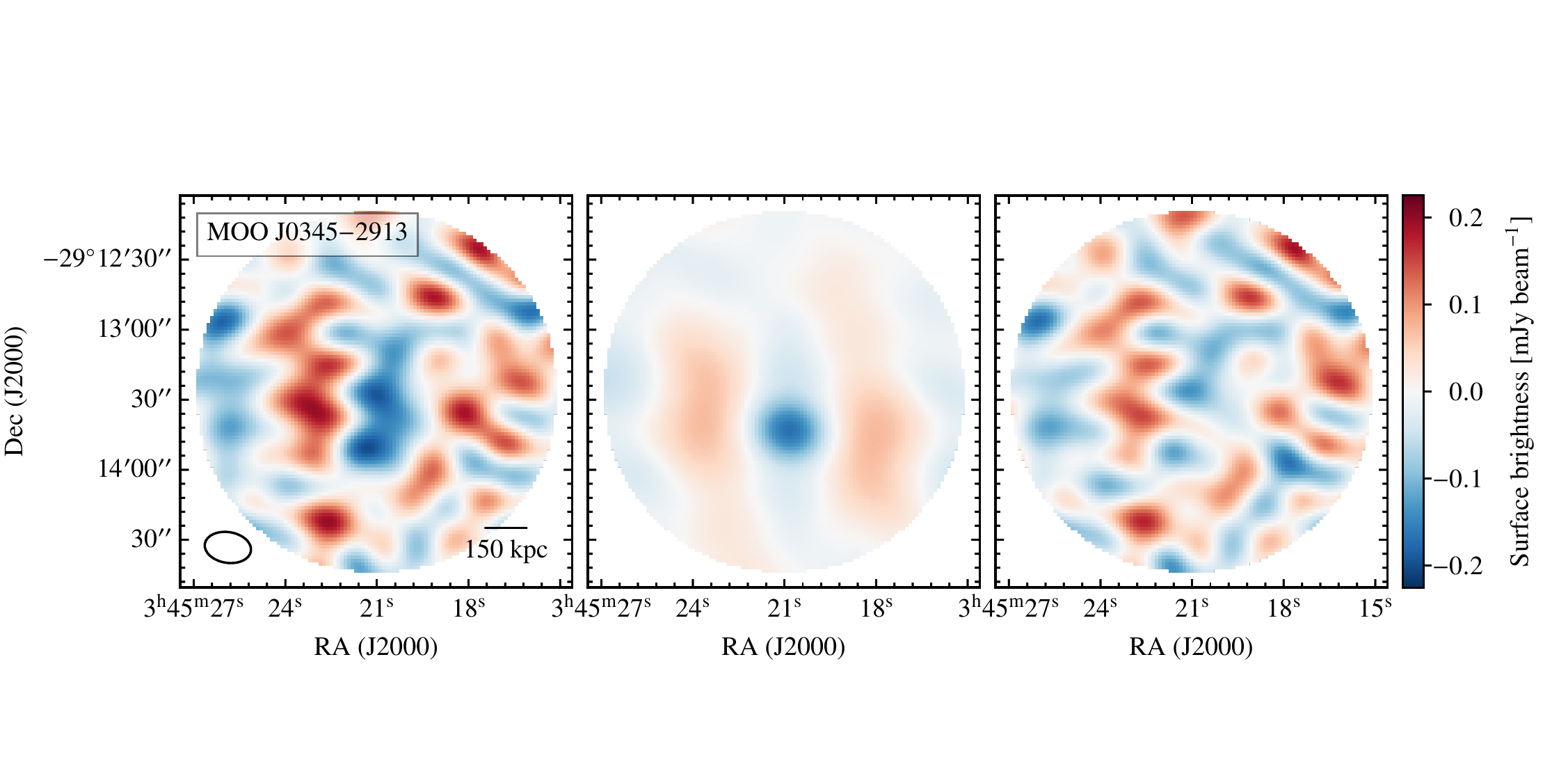}\\
    \includegraphics[clip,trim=0 16mm 5mm 26mm,width=\textwidth]{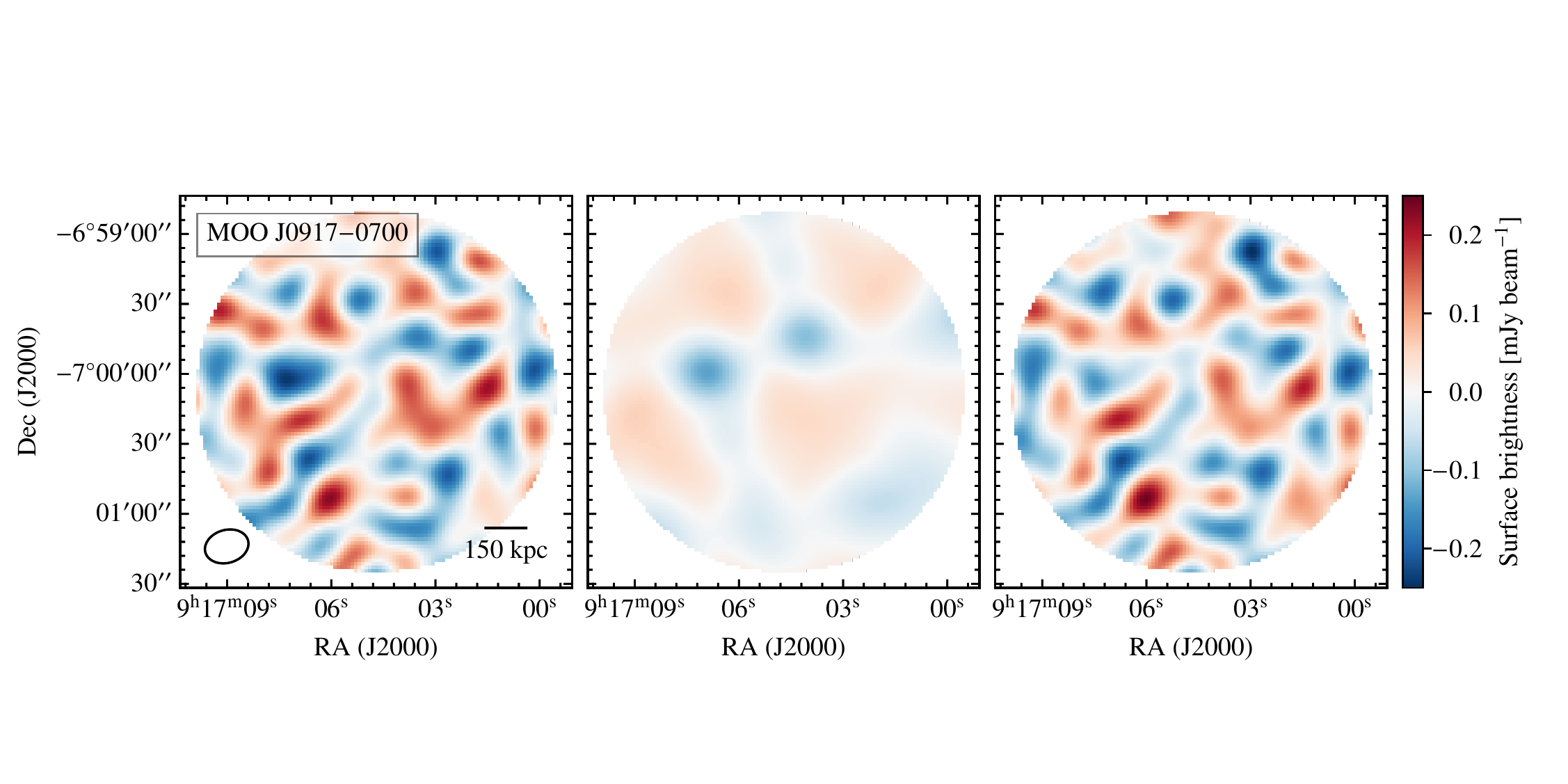}
    \caption{Dirty images of the raw (left), model (center), and residual (right) data of VACA LoCA observations. All the images are generated by applying a multi-frequency naturally weighted, imaging scheme, and extend out to where the ACA primary beam reaches 20\% of its peak amplitude. To better highlight the SZ features in each field  a $10~\mathrm{k\lambda}$ taper is applied, but  without correction for the primary beam attenuation. Furthermore, as in Fig.~\ref{fig:imdirt:mooj0129},   the most significant point-like sources from the raw interferometric data are removed.}
    \label{fig:app:imdirty}
\end{figure*}

To get a more immediate sense of the reconstructed models, we show in Fig.~\ref{fig:app:imdirty} the dirty images of VACA LoCA observations. As there are no sensible differences between the model and residual dirty images generated with different gNFW models, we   consider here only the case of a universal pressure profile \citep{Arnaud2010}. We again note that all the images reported here are   for illustrative purposes only, and were not used in our analysis. 

\section{\textit{Spitzer}/IRAC galaxy overdensities}\label{app:irac}
\begin{figure*}
    \centering
    \includegraphics[clip,trim=0.15cm 0.50cm 2.60cm 1.40cm,height=0.28\textwidth]{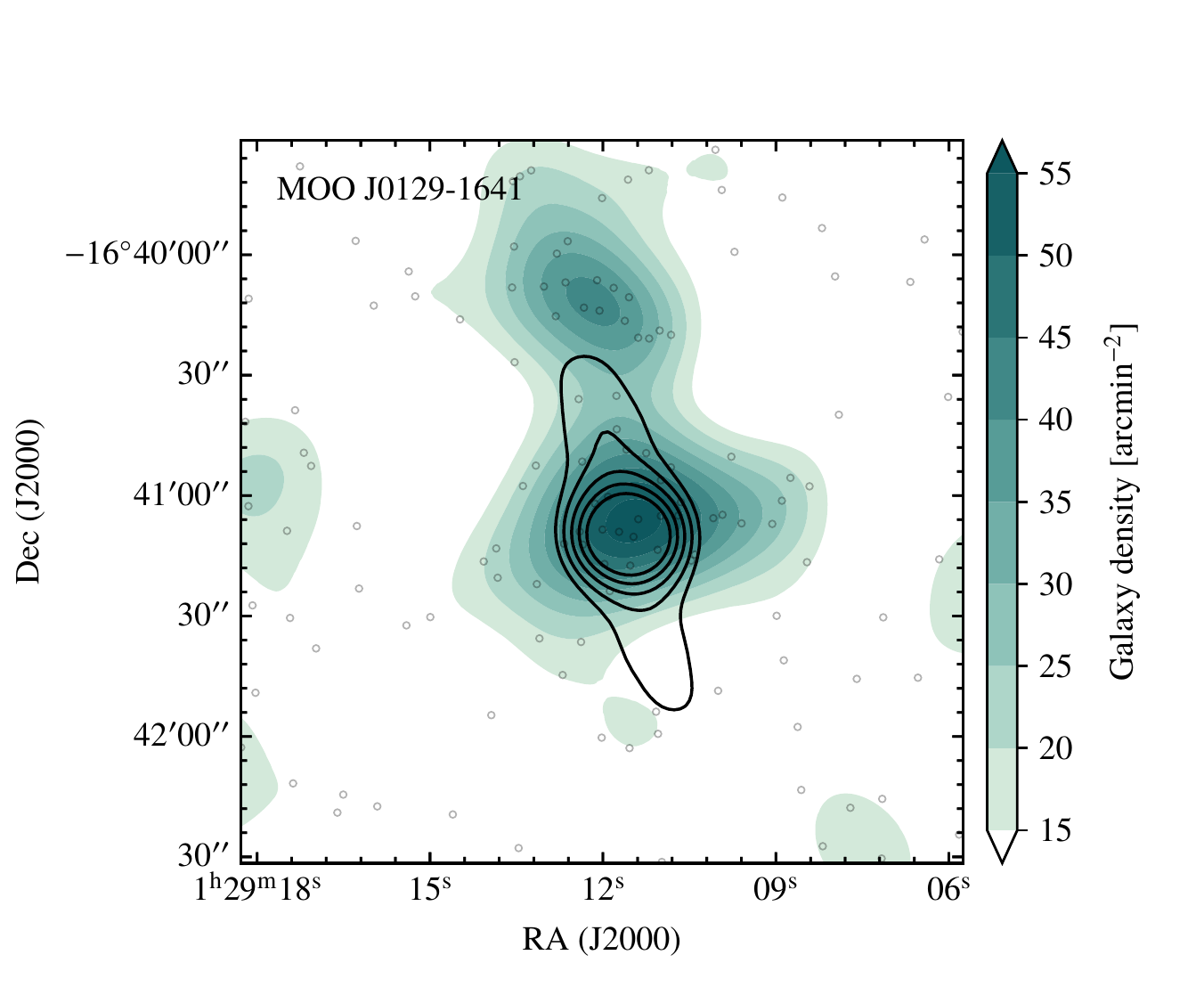}
    \includegraphics[clip,trim=0.80cm 0.50cm 2.60cm 1.40cm,height=0.28\textwidth]{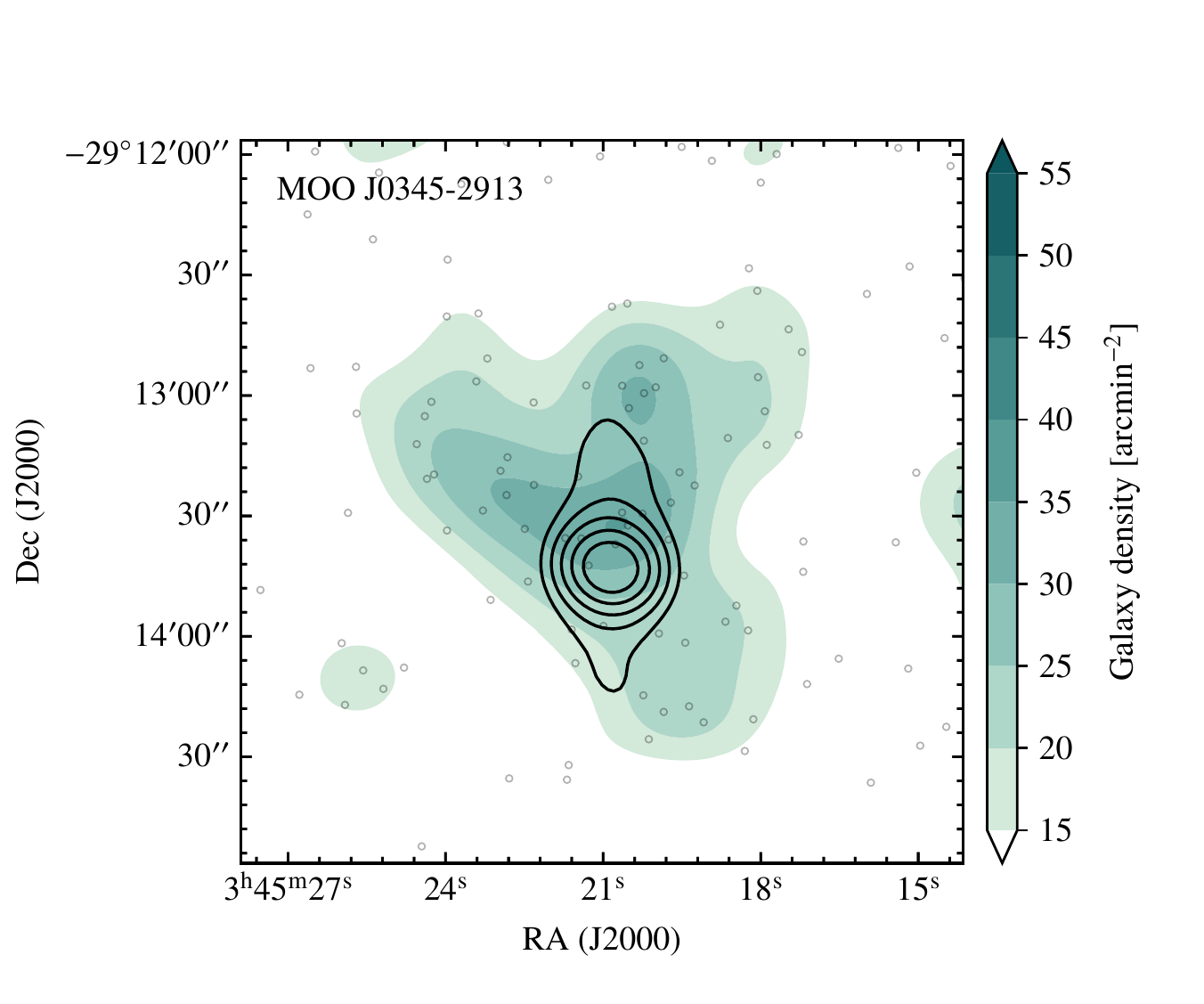}
    \includegraphics[clip,trim=0.80cm 0.50cm 0.60cm 1.40cm,height=0.28\textwidth]{figures/figureA3_MOO_J0917-0700.pdf}\vspace{10pt}\\
    \includegraphics[clip,trim=0.15cm 0.50cm 2.60cm 1.40cm,height=0.28\textwidth]{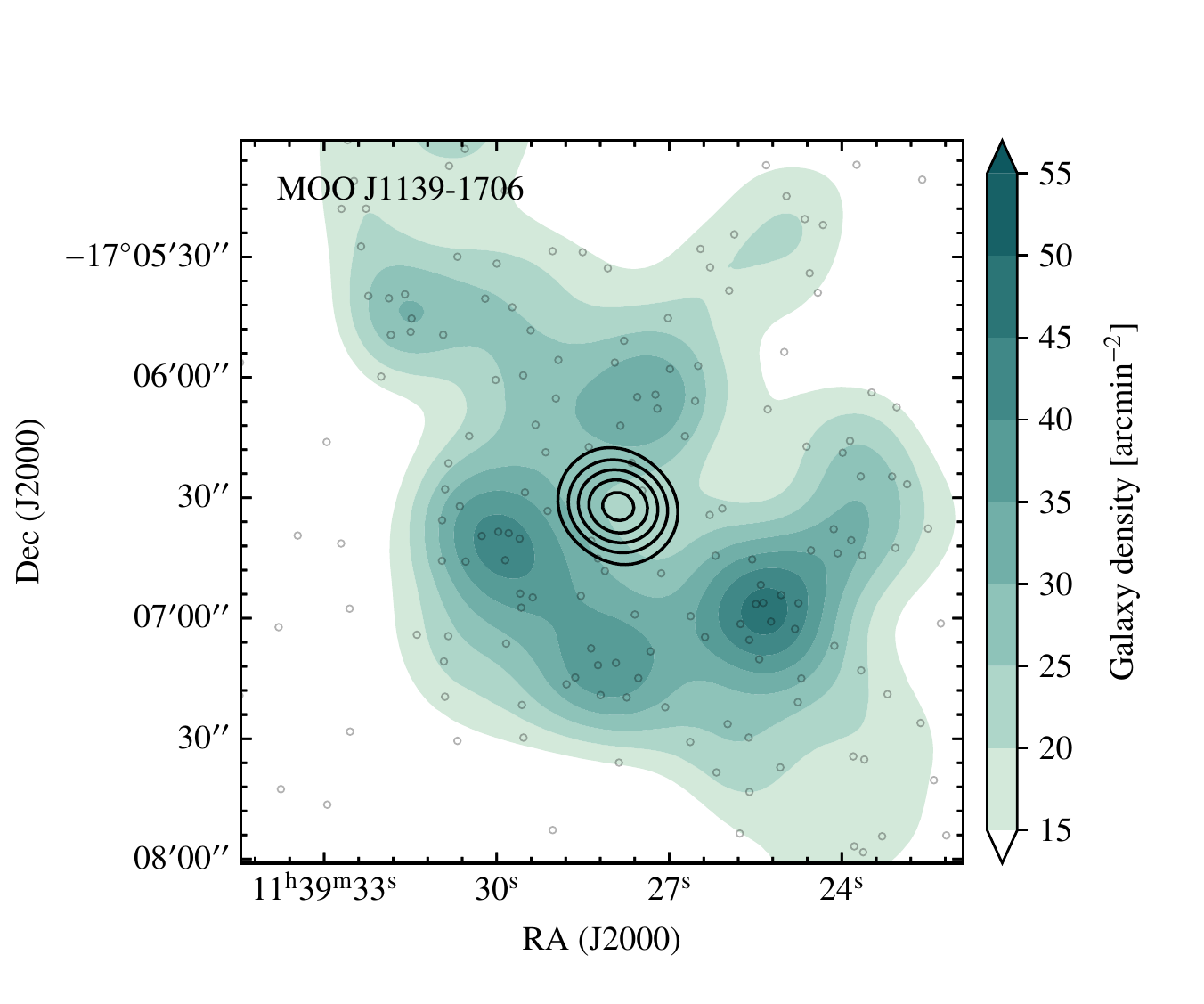}
    \includegraphics[clip,trim=0.80cm 0.50cm 2.60cm 1.40cm,height=0.28\textwidth]{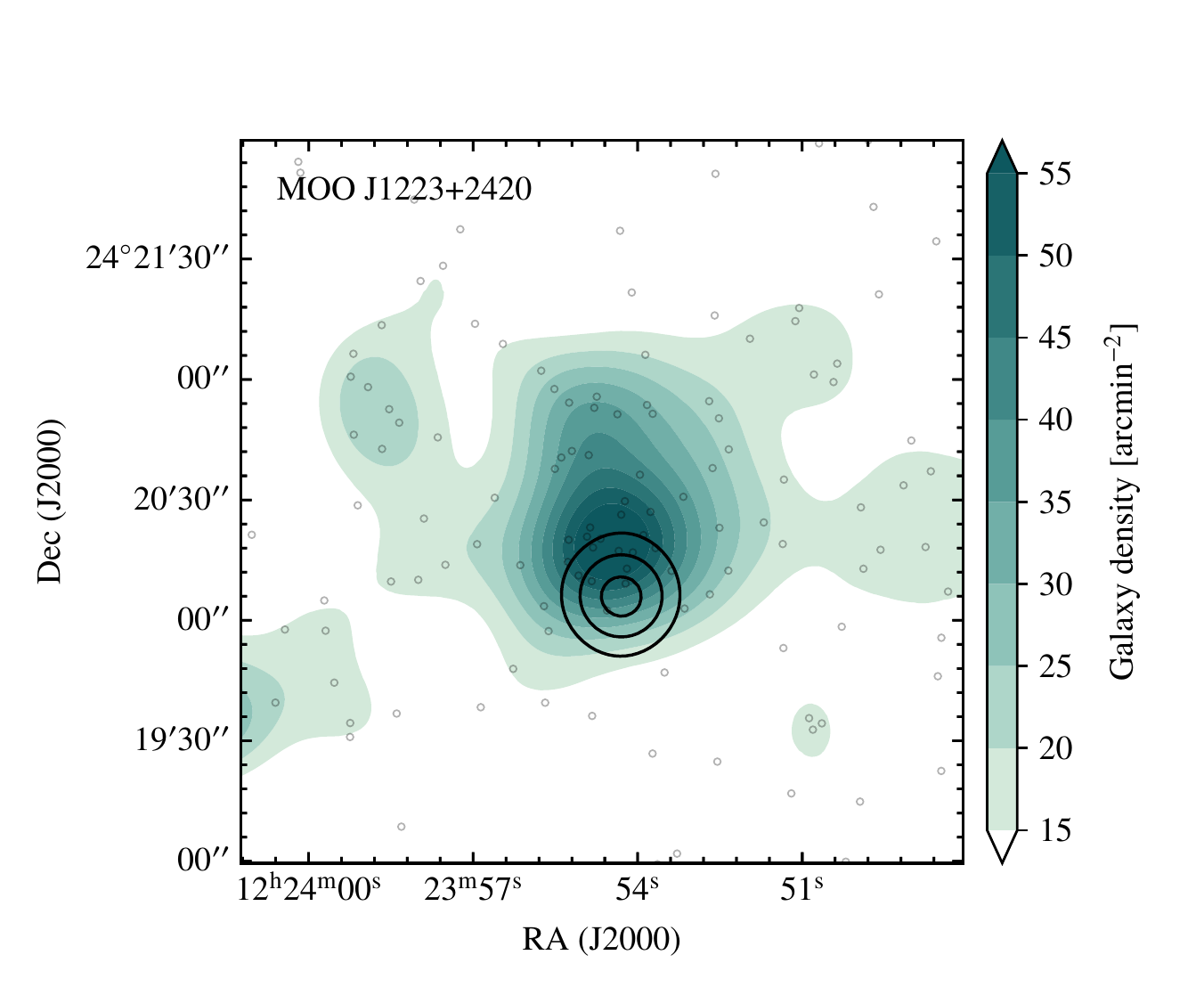}
    \includegraphics[clip,trim=0.80cm 0.50cm 0.60cm 1.40cm,height=0.28\textwidth]{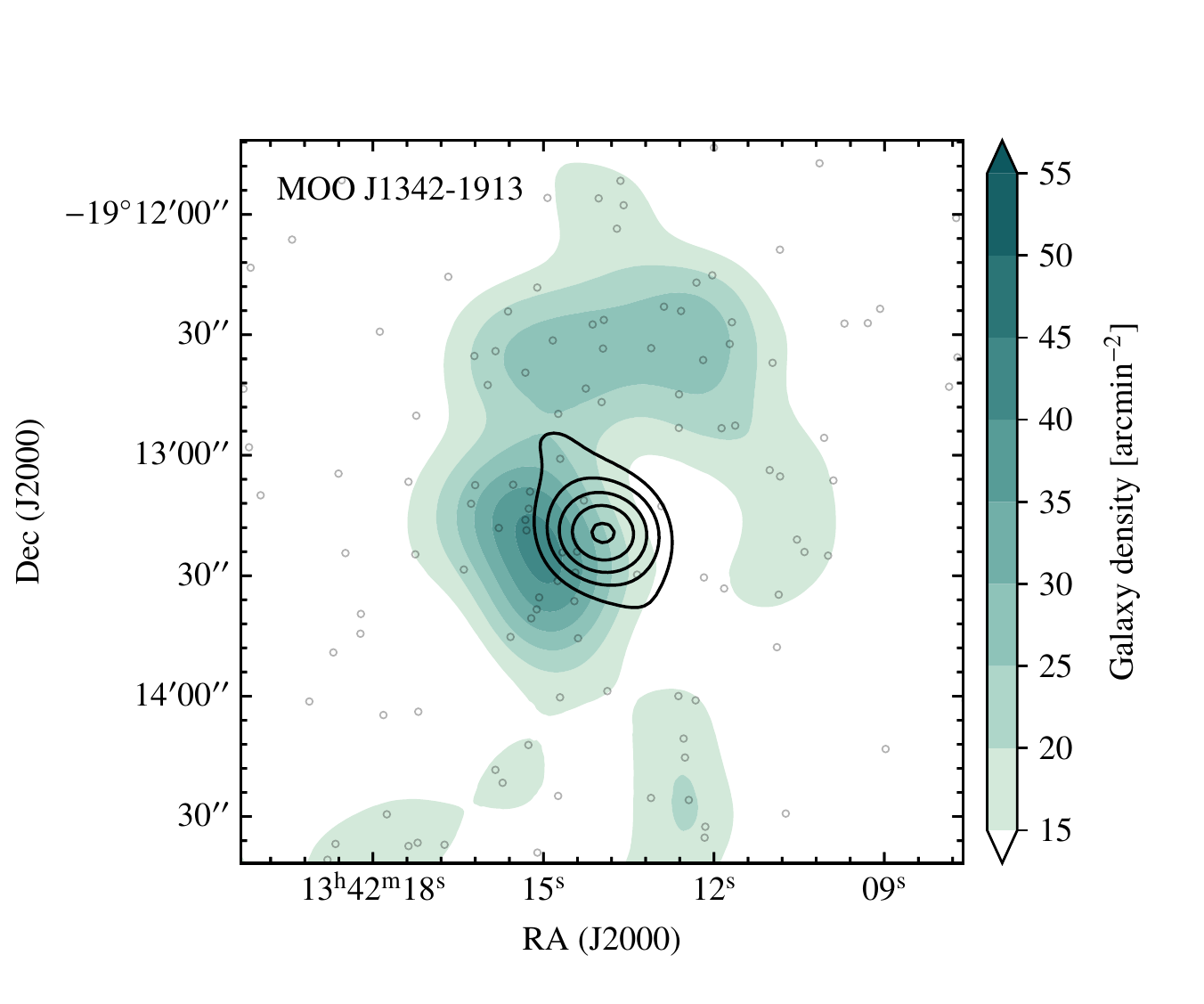}\vspace{10pt}\\
    \includegraphics[clip,trim=0.15cm 0.50cm 2.60cm 1.40cm,height=0.28\textwidth]{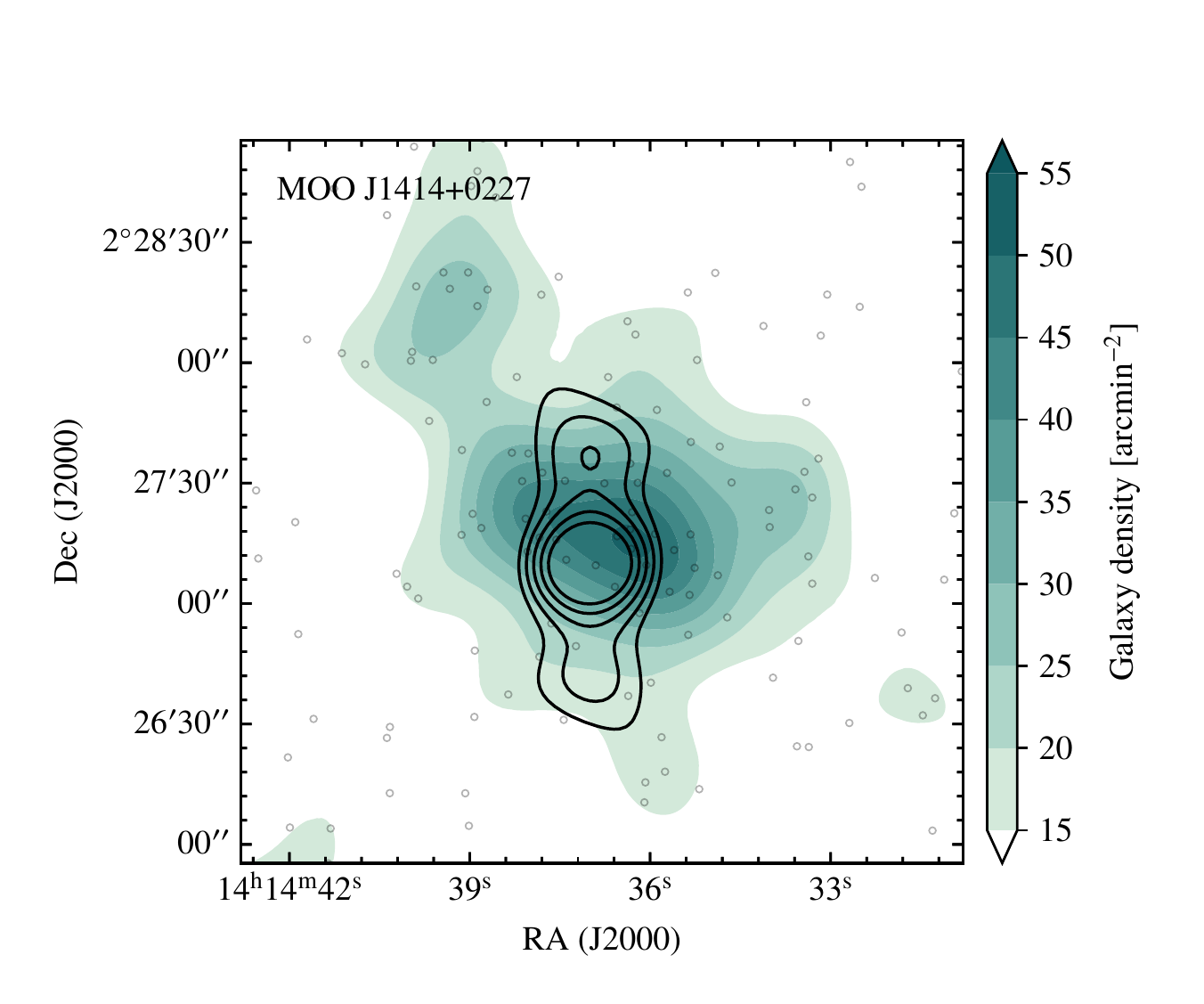}
    \includegraphics[clip,trim=0.80cm 0.50cm 2.60cm 1.40cm,height=0.28\textwidth]{figures/figureA3_MOO_J2146-0320.pdf}
    \includegraphics[clip,trim=0.80cm 0.50cm 0.60cm 1.40cm,height=0.28\textwidth]{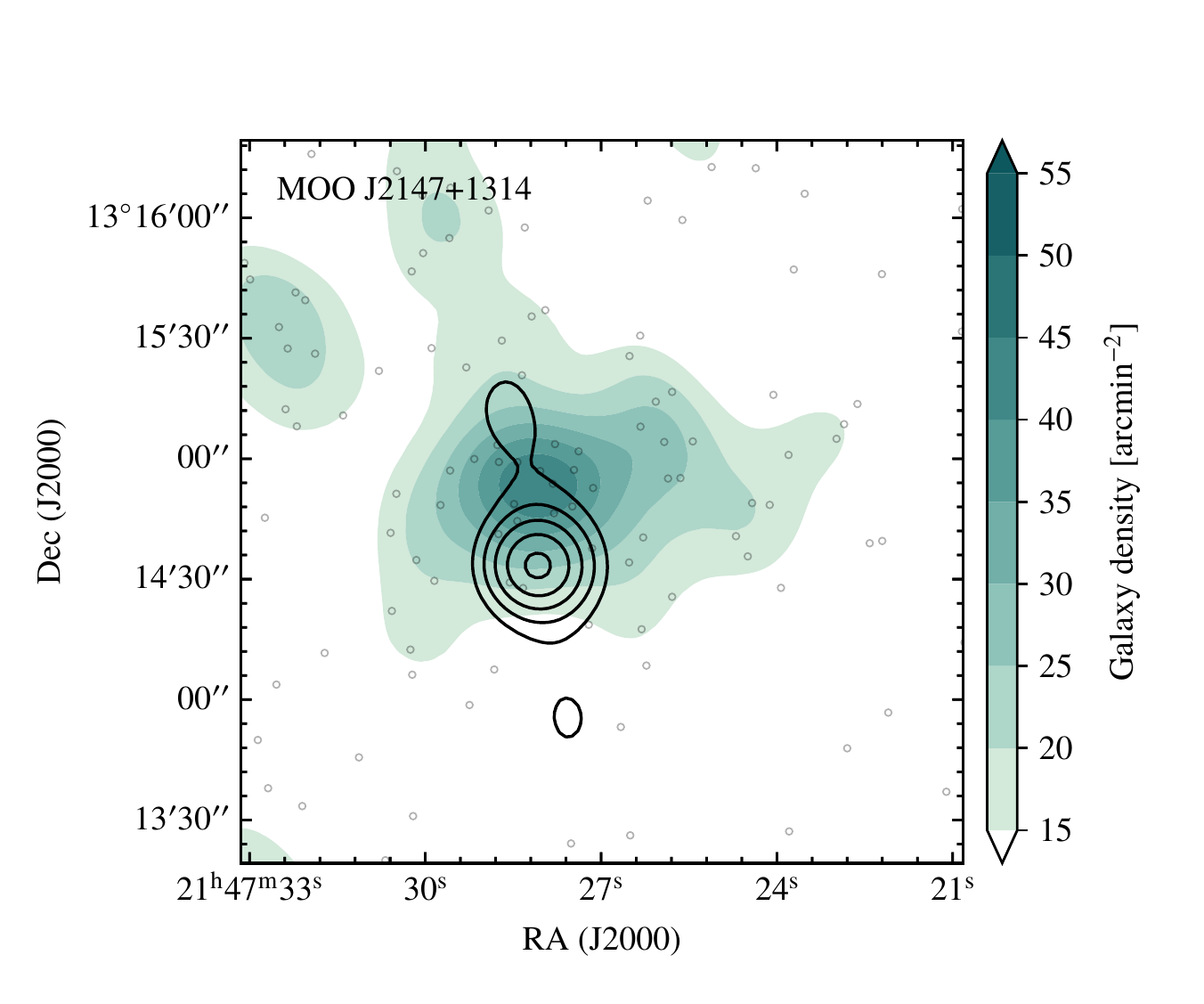}\vspace{10pt}\\
    \caption{\textit{Spitzer}/IRAC galaxy densities in the direction of the VACA LoCA galaxy clusters and contours of the filtered SZ models. See Sect.~\ref{sec:res:multi} and Figure~\ref{fig:irac} for details.}
    \label{fig:app:overlays}
\end{figure*}

For   comparison, we show in Fig.~\ref{fig:app:overlays} the SZ model contours for all the VACA LoCA clusters with either a firm or a marginal detection of their SZ signature overlaid on the maps of their respective color-selected galaxy density distributions from \textit{Spitzer}/IRAC (see Sect.~\ref{sec:res:multi} and Figure~\ref{fig:irac}). For the sake of clarity, we do not plot the secondary negative lobes arising due to the non-Gaussian pattern of the interferometric beam.

\begin{figure*}
    \centering
    \includegraphics[clip,trim=0 16mm 5mm 26mm,width=\textwidth]{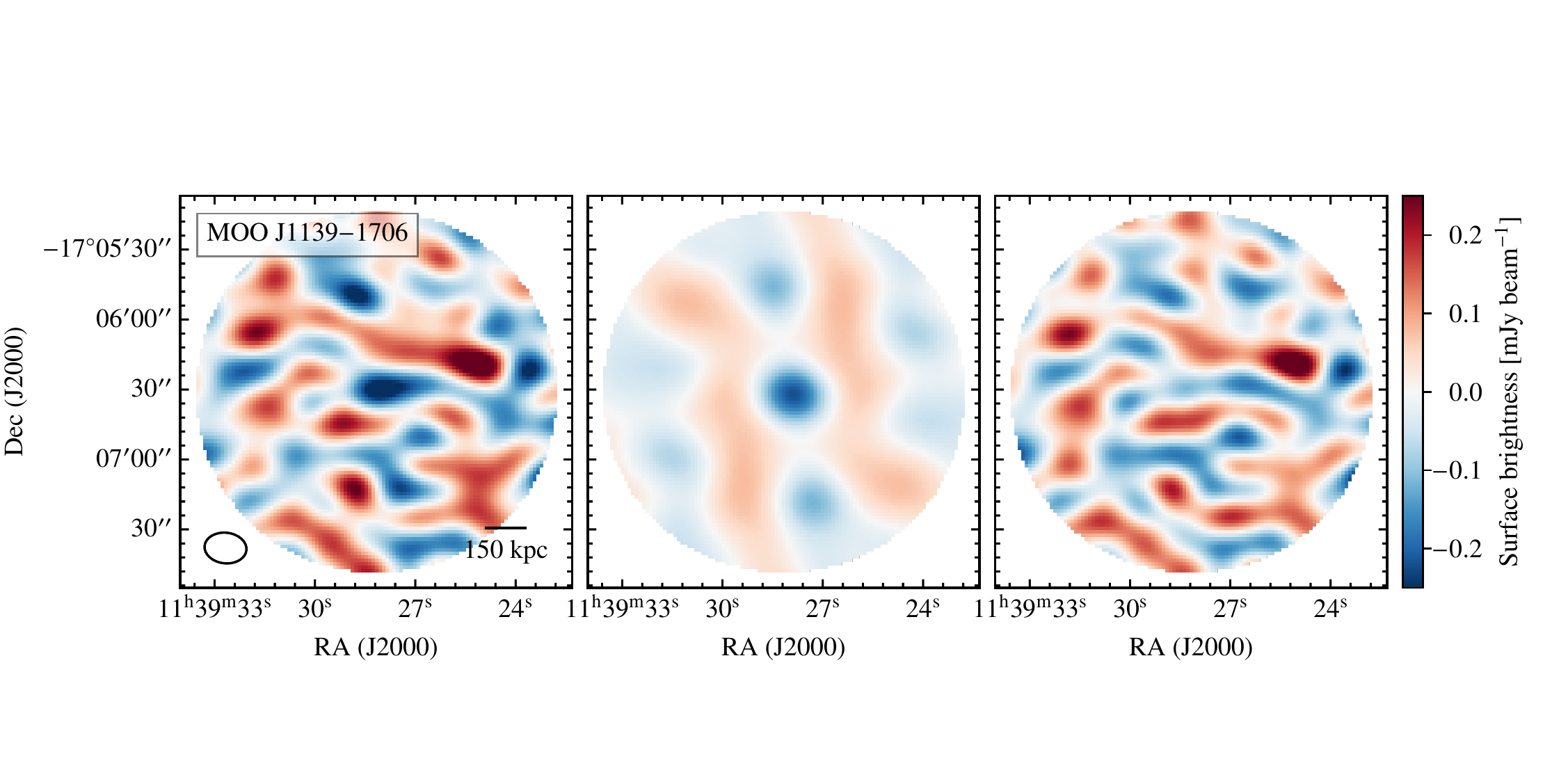}\\
    \includegraphics[clip,trim=0 16mm 5mm 26mm,width=\textwidth]{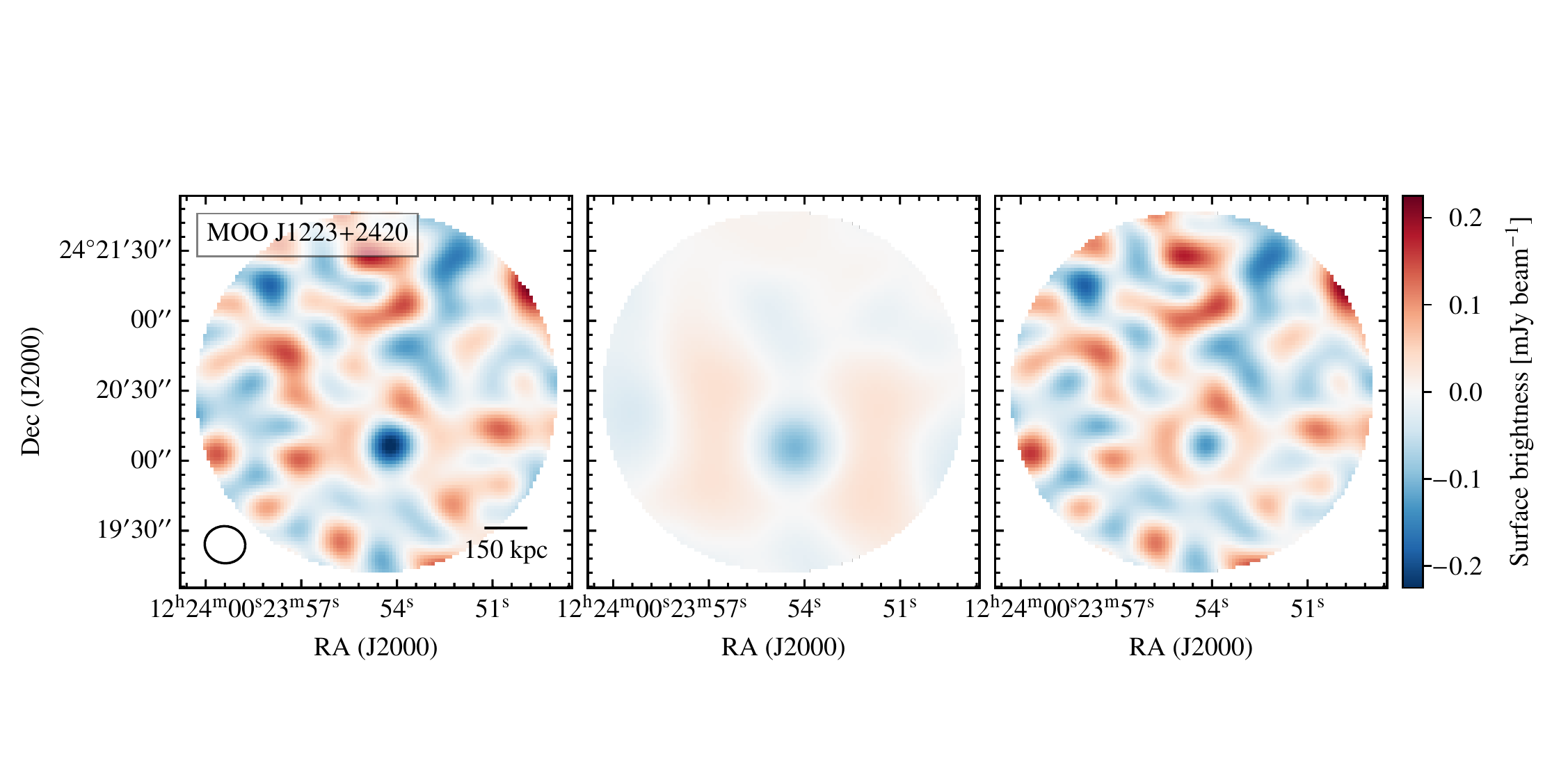}\\
    \includegraphics[clip,trim=0 16mm 5mm 26mm,width=\textwidth]{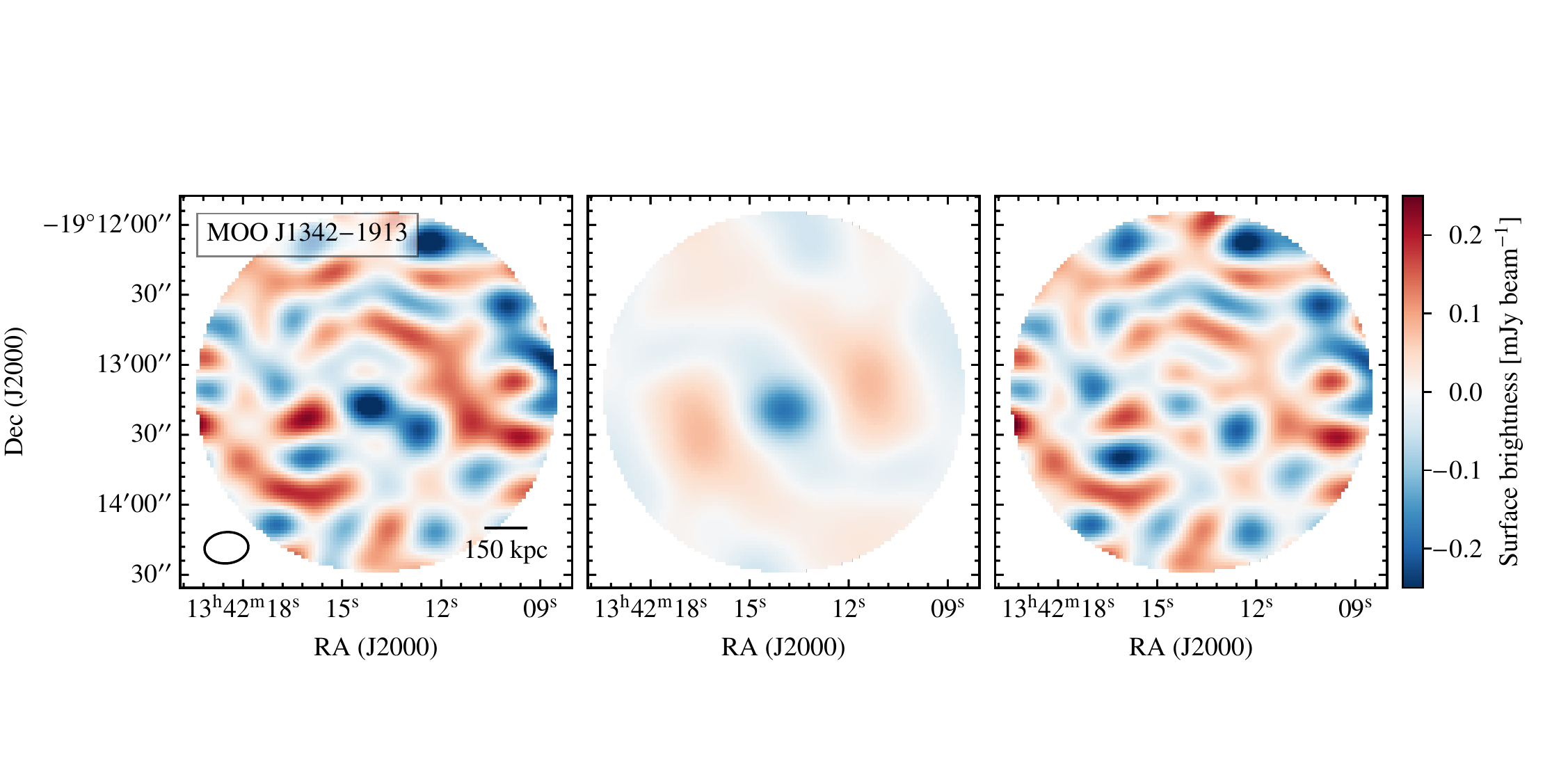}
    \begin{center}\vbox{(continued from Fig.~\ref{fig:app:imdirty})}\end{center}
\end{figure*}
\begin{figure*}
    \includegraphics[clip,trim=0 16mm 5mm 26mm,width=\textwidth]{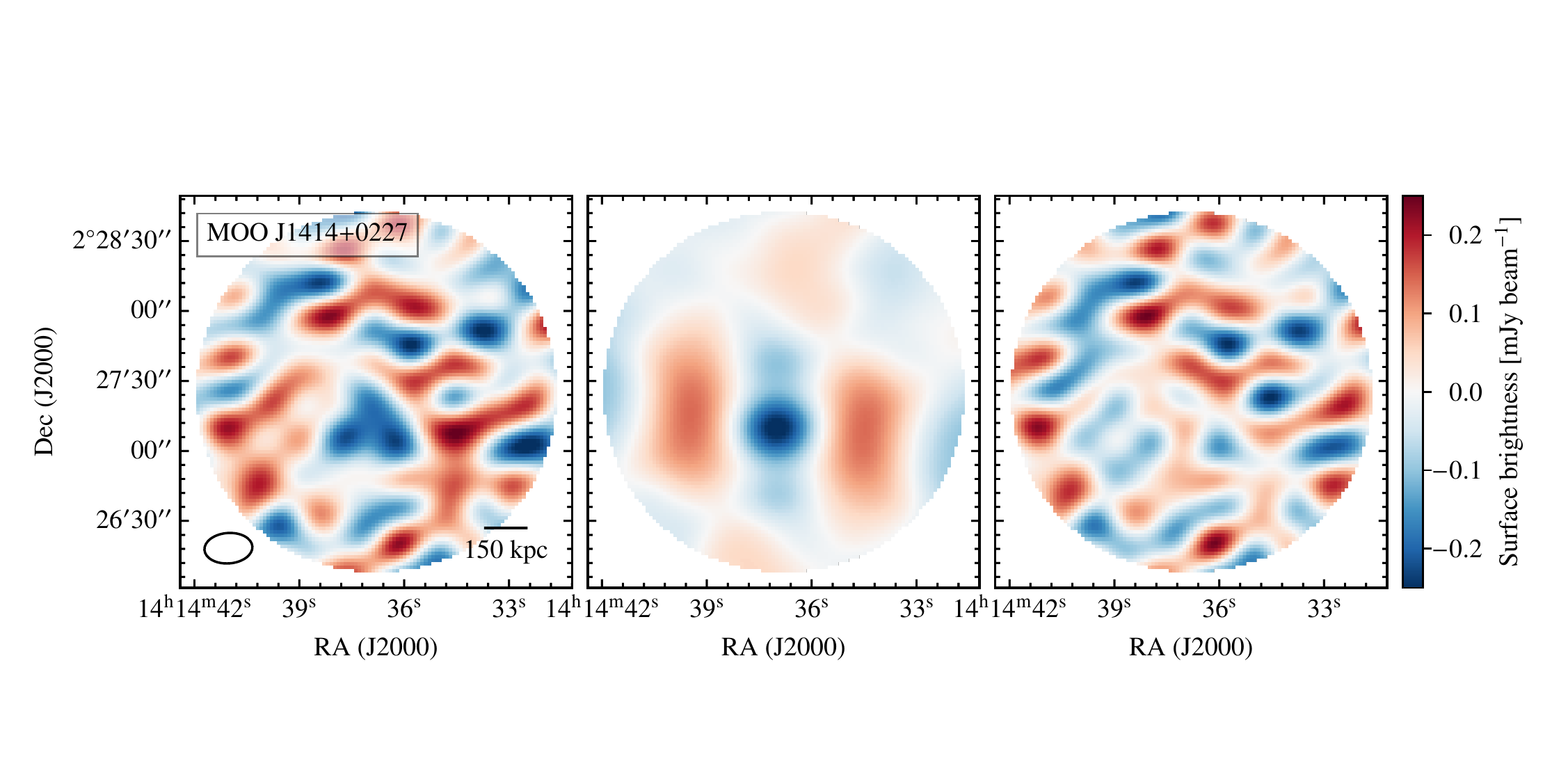}\\
    \includegraphics[clip,trim=0 16mm 5mm 26mm,width=\textwidth]{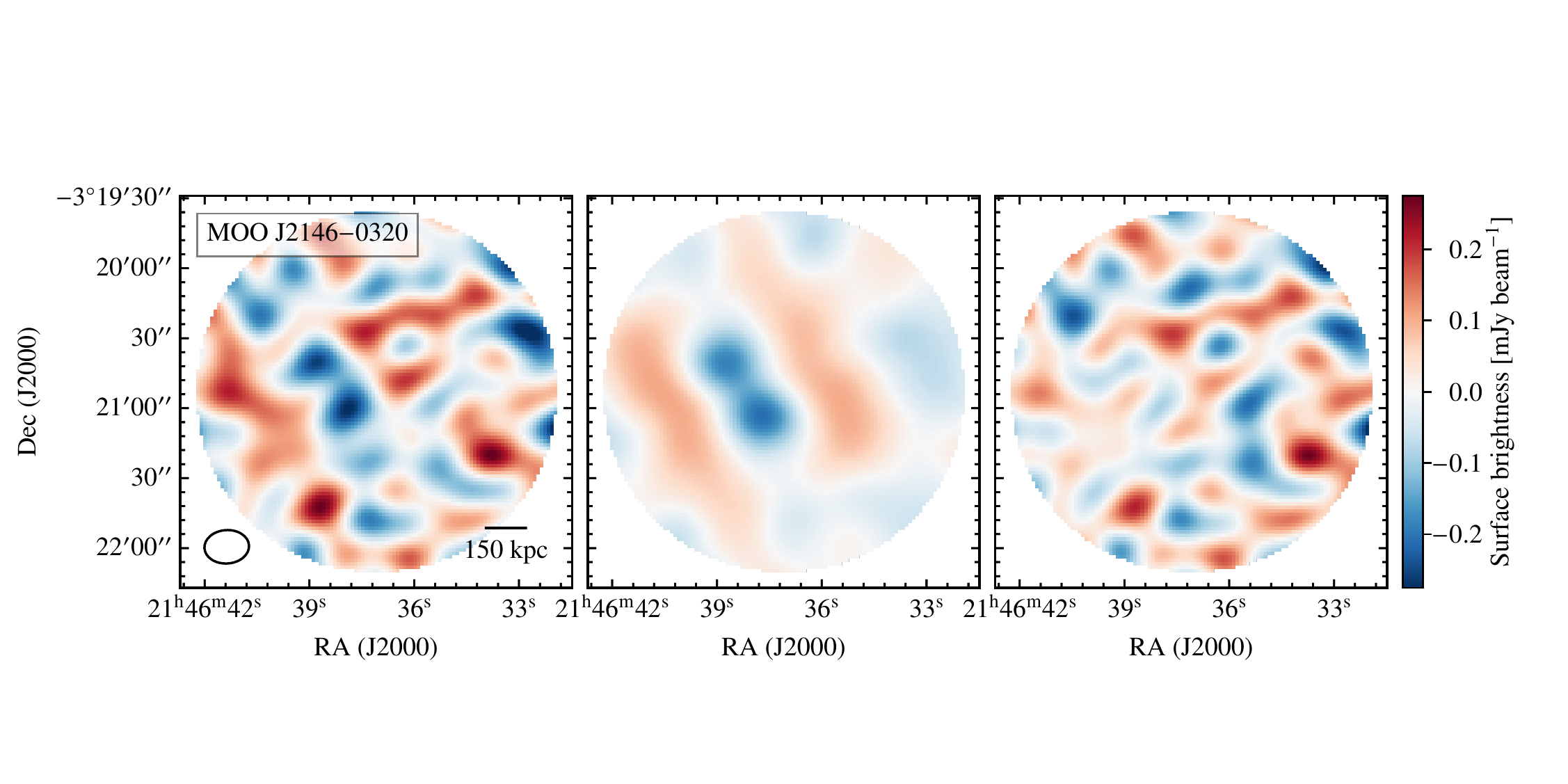}\\
    \includegraphics[clip,trim=0 16mm 5mm 26mm,width=\textwidth]{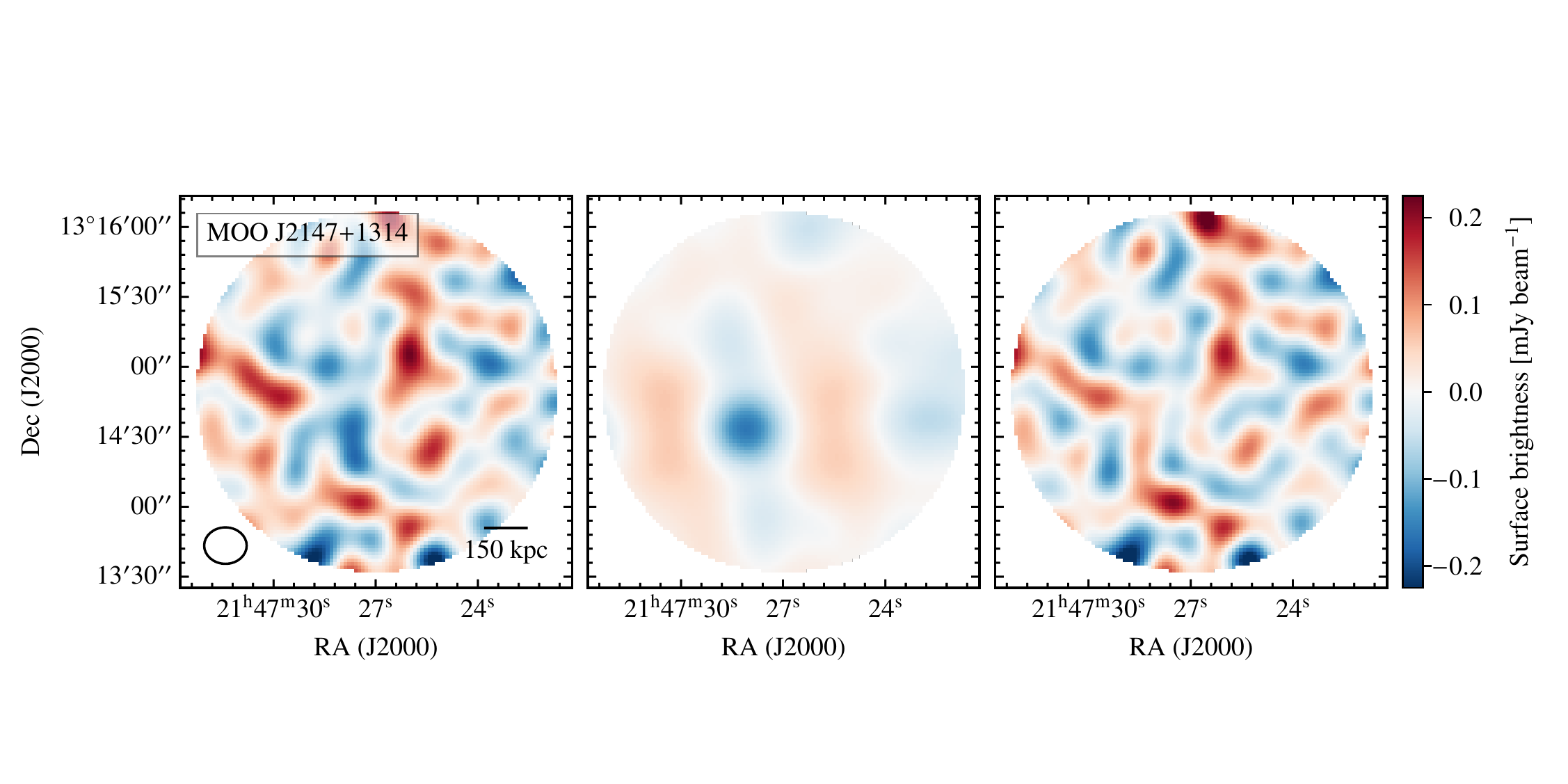}    
    \begin{center}\vbox{(continued from Fig.~\ref{fig:app:imdirty})}\end{center}
\end{figure*}

\end{document}